\documentclass[epj,nopacs]{svjour}
\pdfoutput=1
\usepackage[utf8]{inputenc}
\usepackage{amsmath,amssymb,amsfonts,graphicx,cite,multirow,tikz,enumitem}
\usepackage{pstricks,pdflscape,morefloats,algorithmicx,algpseudocode,slashed}
\usepackage[section]{algorithm}

\graphicspath{{graphics/}}
\makeatletter
\ifx\input@path\@undefined
\def\input@path{{graphics/}}
\else
\g@addto@macro\input@path{{graphics/}}
\fi
\makeatother

\usepackage{color}
\usepackage{hyperref}

\newcommand{\dd}{{\mathrm{d}}}

\newcommand{\PS}{\mathcal{PS}}
\newcommand{\PSV}{\mathcal{PSV}}

\newcommand{\tPS}{\mathcal{\widetilde{PS}}}
\newcommand{\tPSV}{\mathcal{\widetilde{PSV}}}
\usepackage{mathtools}

\preprint{IPPP/17/39\\KA-TP-20-2017\\MAN/HEP/2017/07\\UWTHPH-2017-9\\HERWIG-2017-01\\MCnet-17-07}

\title{Merging NLO Multi-jet Calculations with Improved Unitarization}

\author{Johannes Bellm\inst{1} \and 
Stefan Gieseke\inst{2} \and
Simon Pl\"atzer\inst{1,3,4}}

\institute{ IPPP, Department of Physics, Durham University\and Institute for
  Theoretical Physics, Karlsruhe Institute of Technology\and Particle Physics
  Group, School of Physics and Astronomy, University of Manchester\and
  Particle Physics, Faculty of Physics, University of Vienna}

\date{May 18, 2017}

\abstract{ We present an algorithm to combine multiple matrix elements
  at LO and NLO with a parton shower.  We build on the unitarized
  merging paradigm. The inclusion of higher orders and multiplicities
  reduce the scale uncertainties for observables sensitive to hard
  emissions, while preserving the features of inclusive quantities.  The
  combination allows further soft and collinear emissions to be
  predicted by the all-order parton shower approximation.  We inspect
  the impact of terms that are formally but not parametrically
  negligible.  We present results for a number of collider observables
  where multiple jets are observed, either on their own or in the
  presence of additional uncoloured particles.  The algorithm is
  implemented in the event generator Herwig.  \PACS{{xx.yy.zz}{Xx Yy
      Zz}} }

\begin{document}

\maketitle

%% start contents %%

%% Introduction %%

\section{Introduction}

Multi purpose Monte Carlo event generators \cite{Herwig, Herwig7, Pythia8.2, 
Sherpa} are used in most LHC analyses to obtain predictions for a multitude
of observables at the level of final-state particles.  The outstanding
accuracy of the LHC experiments calls for predictions at the highest possible
theoretical accuracy, where next-to-leading order (NLO) predictions in the
perturbative expansion of quantum chromodynamics (QCD) in the strong coupling
constant $\alpha_S$ have become the de-facto standard during the last decade.

NLO corrections are available for many standard model processes with
a moderate number of additional parton emissions.  Once a higher jet
multiplicity is of interest, it will be increasingly important to simulate
also processes with higher multiplicities at the NLO level.  It is clear that
the computational effort for corrections to processes with higher and higher
multiplicities increases enormously with the number of emitted partons.  With
the help of a parton shower, it is, however, possible to consistently merge
computations with different multiplicities into one inclusive sample that
contains full final states at different jet multiplicity.

The first successful attempts at correcting the results of parton showers
integrated a matrix element (ME) for the emission of one additional particle into
the shower evolution, resulting in so-called matrix-element corrections
\cite{Mike,Norrbin:2000uu}.
The next step from there has been the development of systematic merging
prescriptions, either literally based on a jet definition as the so-called MLM
\footnote{MultiLeg Merging in \cite{MLM}} method \cite{MLM}, or as an approach 
from the analytic formulation of emission probabilities \cite{CKKW}, known as the 
CKKW\footnote{Catani, Krauss, Kuhn and Webber\cite{CKKW}} method.  An alternative
formulation, \mbox{CKKW-L}\footnote{L\"onnblad \cite{CKKW-L}} \cite{CKKW-L}, 
was based on very similar ideas, although here the no-emission probability was not 
computed based on assumptions on the parton shower formulation but rather literally 
taken from so-called trial parton showers, carried out in the very parton shower that
was used for the merging of MEs itself.  The latter approaches show that a residual
merging scale dependence is beyond the level of logarithmic approximation of
the parton shower.  Following the first approaches for $e^+e^-$ annihilation
final states, the CKKW algorithm has been generalized to hadronic collisions
as well \cite{Krauss:2002up}.  A systematic comparison of the different
approaches has been carried out in \cite{alwall2008comparative}.

As opposed to merging several tree-level MEs of different
multiplicity, the development of the last 15 years has led to the fact that
it has become standard to simulate collider processes at NLO.  Two
methods have been pioneered on which all of the following work is based.
In the \textsc{mc@nlo} method \cite{MCatNLO} the parton shower
contribution to a partonic final state is expanded in $\alpha_S$ and
subtracted from the corresponding contributions of the ME, such
that a consistent matching of NLO MEs and parton showers
becomes possible.  The program package of the same name comes with a
multitude of built-in processes that can be simulated with different
parton shower programs.  The other method \textsc{powheg} \cite{POWHEG}
rather changes the Sudakov form factor for the first emissions of a
parton shower in a way that up to this perturbative order the
parton-level answer of the computation is consistent with an NLO
calculation.  The program package \textsc{powhegbox} \cite{POWHEGBOX}
provides a large number of processes that can afterwards be showered
with different parton shower programs.  A number of processes have been
implemented into \textsc{herwig}
\cite{Hamilton:2009ne,Hamilton:2009za,Hamilton:2008pd}. In
\cite{Alioli:2012fc} the soft region of the parton shower is singled out
and resummed separately.  NLO MEs have also been matched to
antenna like showers
\cite{Hartgring:2013jma,Fischer:2016vfv}. First attempts
at matching NLO calculations with an NLO parton shower have been made in
\cite{Jadach:2015mza}.

While the two aforementioned packages address each process separately,
the enormous technical progress of the last decade made possible to simulate
practically any standard model process at colliders completely
automatically. A multitude of programs are capable of providing matrix
elements at NLO \cite{Nagy:2003tz, Actis:2016mpe, GoSam, GoSam2, OpenLoops, 
Alwall:2014hca}.  The generated MEs from these programs provide
information according to a common standard \cite{BLHA2}, which in turn will
then be interfaced to the general purpose event generators to simulate full
hadron level events at NLO \cite{Herwig7,Hoche:2010pf,Hoche:2010kg}.  In
\textsc{herwig~7} \cite{Herwig7} the interface is merely used to compute the
MEs, while the phase space sampling, subtraction terms, and
matching algorithms are performed within \textsc{herwig}.  A newer development
is the possibility to provide theoretical uncertainties, based on scale
variations within these programs and within the parton showers
\cite{LOShowerUncertainties,MEPSatNLOUncertanties,Hoeche:2011fd}.  While a lot
of experience with NLO matching has been accumulated, the first processes have
been combined with NNLO MEs \cite{MiNLO, 
MiNLOprime,Hoeche:2014aia,Hoche:2014dla}.

At the same time, the merging method has evolved into a standard for the
simulation of final states with a large number of jets.  It has been found
that for a consistent matching it is very important to understand the clustering
and subsequent shower in great detail.  In the end the phase space that is
covered by the parton shower has to be matched to the ME phase
space.  In order to achieve this, so-called truncated showers have been
introduced \cite{NL3, CKKW2trunk}.

With the advent of more and more partly automatically generated NLO matrix
elements it has become possible to add virtual corrections to the merging as
well \cite{MEPSatNLOLEP, MEPSatNLO, FxFx}. At first, only the corrections to
the process with lowest multiplicity have been added, such that the overall
normalization of the inclusive cross section has been stabilized, while
multiple jet emissions have still been described using tree-level matrix
elements.  This method can be seen as the unification of the previous matching
and merging approaches.  However, the approach has not been limited to the
lowest multiplicity MEs but rather allowed the introduction of virtual
corrections to higher multiplicity MEs as well, thereby reducing
also the scale uncertainties for observables that previously have only been
described at LO.  In \cite{MEPSatNLOUncertanties} the systematic uncertainties
from perturbative and non-perturbative sources have been addressed.

While most of the work is concerned with NLO QCD corrections and merging of
MEs with additional parton emission, the emission of weak bosons
has been studied in \cite{MergingWeakwithQCD}.  A merged sample of $V$+jets
was obtained by either using the electroweak process as a starting point and
adding further hard emissions to it or, alternatively, starting from a
multi-jet process and then producing the weak boson as a parton shower
emission.  Here, particularly the harder parts of the QCD phase space can be
addressed consistently.

As outlined above, the merging of QCD MEs has matured
significantly over the last years.  While it is possible to improve on the
independence on unphysical scales using the NLO merging, one shortcoming still
are uncontrolled terms of higher orders within inclusive cross sections.  This
point has been addressed conceptually in \cite{HerwigMerging}, where, based on
the parton shower formulation of higher order contributions in \cite{LoopSim},
the problem of unitarity violations has been addressed. A formulation of the
merging has been made that inherently preserves unitarity and thereby also
preserves the inclusive cross section and its given accuracy.  First
implementations of this method are now known as the \textsc{ulops} and
\textsc{unlops} approach \cite{ULOPS, UNLOPS,Lonnblad:2011xx}.

In this paper we present a new implementation of the unitary merging algorithm
that was outlined in \cite{HerwigMerging}, based on the dipole shower module 
\cite{Platzer:2009jq,Platzer:2011bc} of Herwig 7 \cite{Herwig7}.
We address all aspects of the
merging algorithm from clustering to the assignment of subsequent parton
shower phase space as well as a detailed discussion of perturbative scales in
the merging algorithm.  The presented algorithm is built upon a very detailed
formal description of the parton shower contribution at any given order.  This
allows us to discuss not only the merging but also an in-depth analysis of terms
that are beyond the targeted approximation on the perturbative and logarithmic
level.  We test the sensitivity of our merging against variations of higher
order terms and choices of scales.  This allows, in addition to a
fully-realistic simulation, also a somewhat reliable estimate of theoretical
uncertainties.  In this paper we study the canonical examples of $e^+e^-$
annihilation and single vector boson production, accompanied by a number of
jets, at hadron colliders.  Furthermore, we consider Higgs production, as here
the higher order corrections are known to be numerically large. Finally, we
consider pure jet production, which, due to the complexity of colour
structures and the ambiguous definition of a hard object, is the most
difficult process from the perspective of merging.

The paper is organized as follows.  In Sec.~\ref{sec:notation} we introduce a
formalism and notation in order to describe the parton shower analytically and
to formulate all aspects of the subsequent subtractions.  In
Sec.~\ref{sec:lomerging} we describe the unitary merging algorithm for LO
MEs first, including details of the clustering and scale settings.
Later, in Sec.~\ref{sec:NLOmerging}, we extend the algorithm to the unitary
merging with NLO MEs.  After a discussion of the scales we are
using in Sec.~\ref{sec:scales} we present some validation of our approach in
Sec.~\ref{sec:sanitychecks}.  Finally we present results obtained with our
approach in Sec.~\ref{sec:results}.

\section{Notations and Definitions}
\label{sec:notation}

In order to set the scene for describing the complex algorithms involved in
merging higher order cross sections in a detailed way, we start with a review
of fixed order cross sections and parton showers to fix our notation of
basic quantities and definitions.

\subsection{Basic Notation}

We denote the differential cross section for a given process with $n+1$
particles in the final state as
\begin{equation}
  \dd \sigma_{n+1}(\phi_{n+1})=\mathcal{M}(\eta_{n+1},\phi_{n+1}) \dd
  \phi_{n+1}\ ,
\label{eq:crosssection}
\end{equation}
where $\phi_{n+1}$ identifies the phase space point in question and is given
by a set of final state momenta $\{p\}_{n+1}$ and the momentum fractions of
the incoming partons $\eta_{n+1}^{a/b}$. We use the shorthand
$\eta_{n+1}=\{\eta_{n+1}^{a},\eta_{n+1}^{b}\}$ for the pair of the momentum
fractions, and define a phase space configuration to include the momentum
fractions $\eta$
\begin{eqnarray}
  \phi_{n+1}&=&\{\eta_{n+1},\{p\}_{n+1}\} \;\; 
                {\rm and}\;\; \hat{s}(\eta_{n+1})=\eta_{n+1}^{a}\eta_{n+1}^{b}S\;,
\end{eqnarray}
where $S$ is the squared centre of mass energy of the collider in
consideration. The phase space measure is taken to be
\begin{equation}
\dd \phi_{n+1}=\dd\eta_{n+1}\prod_i [\dd p_i](2\pi)^4 \delta \Big(\sum_i p_i
-\eta^a_{n+1} p_a -\eta^b_{n+1} p_b \Big)\ .
\end{equation}
The $\delta$-function implements the constraint of energy and momentum
conservation and the product over the final state particle momenta requires
on-shell particles,
\begin{equation}
  [\dd p_i] = \dd^4p_i\delta^+(p_i^2-m_i^2)\ .
\end{equation}
The differential cross section here describes the transition of a definite
initial to a definite final state, which we refer to as a subprocess.  In
order to calculate the inclusive cross section for a process $a b \to c d$,
where the labels can refer to groups of partons, {\it e.g.} a jet or a proton,
the sum over all subprocesses has to be taken, which is not implicit unless we
explicitly mention this. The weight of a single phase space point is given by
\begin{multline}
\mathcal{M}(\eta,\phi) = \mathcal{M}_i(\eta,\phi)\\=
f_i(\eta,\mu_F)|M_i(\phi,\mu_F,\mu_R)|^2 \mathcal{F}(\eta) \mathcal{S}_i
\Theta_C(\phi)\ . \label{eq:weight}
\end{multline}
We use the shorthand $f_i(\eta,\mu_F) =
f^a_i(\eta^a,\mu_F)f^b_i(\eta^b,\mu_F)$ for the product of parton distribution
functions (PDF), which depend on the factorization scheme chosen and the
factorization scale $\mu_F$.  The MEs can be expressed as a power
series in the strong coupling constant $\alpha_S$.  They depend on the
renormalization scheme and scale $\mu_R$.  The truncation of this series leads
to the terms 'leading order' (LO) and (next-to)$^n$-leading order (N$^n$LO).
Both $\mu_F$ and $\mu_R$ are usually chosen as functions of the phase space
point $\phi$ in order to reduce the sensitivity to logarithms that appear as a
result of truncating the perturbative series. $\mathcal{F}(\eta)$ denotes the
partonic flux factor providing the dimensions of a cross section,
$\mathcal{S}_i$ accumulates symmetry factors for identical final state
particles and averages over initial state degrees of freedom, and
$\theta_C(\phi)$ encodes {\it generation} cuts applied to the hard process
which are either required to obtain a finite cross section or otherwise
increase the efficiency of a subsequent final state analysis.

The higher order contributions taken into account in Eq.~\ref{eq:weight} are
typically a combination of individually divergent contributions. The ultra
violet (UV) divergences are regularized and then removed by renormalization,
introducing the dependence on the unphysical scale $\mu_R$.  Besides
the UV divergences, which stem from large momentum components in loop diagrams,
also the region of small components or collinear momentum configurations can
produce divergences, the latter specifically for massless particles.  These
infrared and collinear (IRC) divergences cancel in IRC safe
observables, when the higher multiplicity, real emission, contributions with
the same coupling power are included in the calculation.

In order to perform next-to-leading order calculations numerically,
specifically using Monte Carlo (MC) methods, several schemes have been
developed \cite{Frixione:1995ms,Catani:1996vz,Kosower:1997zr,Nagy:2003qn}.
One of the key ingredients is the projection of a phase space point
$\phi_{n+1}$ for a $n+1$ particle final state onto a phase space point
$\phi_{n,\alpha}$ of a final state with $n$ particles, {\it i.e.} one particle
less per singular limit $\alpha$,
\begin{equation}
   \phi_{n,\alpha}(\phi_{n+1})
   =\{\eta_{n,\alpha}(\phi_{n+1}),
   \{\tilde{p}\}_{n,\alpha}(\phi_{n+1})\}\ .
\end{equation}
We call this projection a {\it clustering} and in a typical subtraction
procedure multiple of these mappings corresponding to one or more singular
limits need to be taken into account. Associated with each clustering $\alpha$
we can construct variables that describe the degrees of freedom of the
emission,
\begin{equation}
  K_{\alpha}(\phi_{n+1})
  =\{p_{T,\alpha}(\phi_{n+1}),z_{\alpha}(\phi_{n+1}),
  \varphi_\alpha(\phi_{n+1})\}\ .
\end{equation}

The inverse of the clustering, of the $n+1$ particle
phase space point onto a reduced $n$ particle phase space point describes the
splitting or emission from a phase space point $\phi_{n}$ to
$\phi_{n+1}^{\alpha}$ as
\begin{eqnarray}
\phi_{n+1}^\alpha(\phi_n, K)&=&\{\eta_{n+1}^\alpha(\phi_n, K),\{p\}_{n+1}^\alpha(\phi_n,K)\}\;.
\end{eqnarray}
$K$ denotes the three independent (splitting) variables of the additional
momentum in $\phi_{n+1}^\alpha$.  For initial state emissions, the mapping is
accompanied by a change in the momentum fractions and we write
\begin{equation}
\eta_{n+1}^{a,\alpha}(\phi_n,K)\eta_{n+1}^{b,\alpha}(\phi_n,K) =
\frac{1}{x^\alpha(\underbrace{\{\tilde{p}\}_n,\eta_n}_{\phi_n} ,K)}\eta^a_n
\eta^b_n \ ,
\end{equation}
with $x^\alpha$ being a function of $\phi_n$ and $K$. While not strictly
required for a subtraction algorithm nor a parton shower emission, the
mappings are typically constructed to span the entire emission phase space
$\phi_{n+1}$ starting from an underlying Born configuration $\phi_n$.  With
this we can write the cross section in Eq.~\ref{eq:crosssection} as
\begin{multline}
\dd \sigma_{n+1}(\phi_{n+1})|_{\phi_{n+1}=\phi_{n+1}^\alpha(\phi_n, K)}
=\\ \mathcal{M}\left(\frac{\eta_{n}}{x^\alpha(\phi_n,K)},\{p^\alpha(\phi_n,K)\}_{n+1}\right)
\mathcal{J}^\alpha(\eta_n,\phi_n;K)\\ \times
\dd \eta_n \prod_i[\dd p_i]\delta\left(\sum_ip_i-q(\eta_n)\right)\dd K \ ,
\end{multline}
where the Jacobian of the clustering mapping can formally be written as
\begin{equation}
\mathcal{J}^\alpha(\eta_n,\phi_n,K) =\left|\frac{\partial
  \phi_{n+1}^\alpha(\phi_n,K)}{\partial \phi_n\partial K}\right| \ .
\end{equation}

\subsection{Subtraction}
\label{sec:subtraction}

In subtraction formalisms, like the approach by Catani and Seymour (CS)
\cite{Catani:1996vz,Catani:2002hc}, expressions for subtraction terms, dipole
terms in this particular example, are constructed to match the real
emission contributions in the IRC limit. The dipole terms approximate the
ME as
\begin{eqnarray}
&&\mathcal{M}_i^A(\eta_{n+1},\phi_{n+1},q(\eta_{n+1}))=\sum_{\alpha\in \{C\}} \mathcal{D}_i^\alpha(\phi_{n+1})\;.\label{eq:dipoles}
\end{eqnarray}
Here $\alpha\in\{C\}$ contains all possible clusterings of the subprocess
described by $\mathcal{M}_i$ to an underlying Born configuration which has
been defined as the leading order process to which corrections are being
calculated.  The matching of the divergences in the IRC limit renders the
difference $\mathcal{M}_i-\mathcal{M}_i^A$ integrable finite.  Infrared and
collinear safe observables $u(\phi_{n+1})$ are insensitive to IRC emissions,
allowing the construction of a subtraction cross section with observables
evaluated at the reduced kinematics $u(\phi^\alpha_{n})$, such that the
integrated counterpart of the subtraction cross section can be added back to
cancel the explicit poles stemming from loop integrals.  The crucial point is
that $\mathcal{D}^\alpha(\phi_{n+1})$ can be factorized into parts that only
depend on the reduced kinematics $\phi^\alpha_{n}$ and parts that contain the
process independent, divergent behaviour. The latter parts are simple enough to
be integrated analytically in $d$-dimensions and can then be reused for the
subtraction of arbitrary processes.

\subsection{LO $\oplus$ PS}

The divergences in higher order calculations cancel between virtual and real
emission contributions for IRC safe observables. Parton showers employ the
fact that the leading behaviour of exclusive or infrared sensitive observables
is given by the divergent and factorisable parts of the scattering amplitude,
provided that the scale of subsequent emissions are strongly ordered. Parton
showers are constructed as stochastic processes, a consequence of strong
ordering, and build on no-emission probabilities based on unitarity:
\begin{align}
1=w_N(q_a|q_b) \label{eq:identity} 
+ \int_{q_b}^{q_a}\hspace{-2mm}\dd q \,w_E(q) w_N(q_a|q) \;.
\end{align}
Here $q_{a/b}$ parametrize in general IRC divergent regions of the phase space
{\it e.g.} a transverse momentum (approaching both soft and collinear limits)
or an angle (approaching collinear emission only), referred to its evolution
scale or parameter. Here $w_N(q_a|q_b)$ is the probability to not emit between
the scales $q_a$ and $q_b$ and $w_E(q)$ is the rate to instantaneously emit at
a scale $q$. We have so far ignored any further variables required to describe
the full emission, with an integration over these variables implicit unless
stated otherwise.

The form of $w_E(q)$ is derived from the differential cross sections in the
IRC limits.  Soft and collinear limits are typically combined into dipole-type
splitting functions, and Eq.~\ref{eq:dipoles} gives rise to
\begin{multline}
  w_E(q) = \sum_{\alpha\in \{E\}} \int \dd K
  \mathcal{J}^\alpha(\eta_n,\phi_n,K) \\ \times
  w_C \frac{\tilde{\mathcal{D}}_i^\alpha(\phi^\alpha_{n+1}(\phi_n,K))}
       {\mathcal{M}_i(\eta_n,\phi_n,q(\eta_n))}\delta(q-q^\alpha(K))\ \label{eq:splittingprob}.
\end{multline}
Here, $\alpha\in \{E\}$ contains splittings from the state that defines the
subprocess $\mathcal{M}_i$ in the denominator. $\tilde{\mathcal{D}}_i^\alpha$
are the colour averaged, dipole expressions in the large $N_C$ limit. To be
more precise, the set $\{E\}$ of possible splittings is directed by a
probabilistic choice of colour lines/dipoles, denoted as $w_C$.
E.g. an emission from a gluon in the subtraction formalism is regularized by a
sum over all possible spectators weighted with the (spin-)colour correlated
MEs.  In the probabilistic formulation of a parton shower the
gluon carries two (anti-)colour lines connecting the gluon to two
partners. These colour partners can be chosen according to the weight of
the squared large $N$ amplitude in the colour flow representation.  Once the
colour lines for the hard process are fixed, colour lines for subsequent
emissions are then generated through the large-$N_c$ limit of splittings in
the parton shower. The no-emission probability solving Eq.~\ref{eq:identity}
can be written as
\begin{equation}
w_N(q_a|q_b)=\Pi^{(1,2)}(q_a|q_b)  \prod_f \Delta^f(q_a|q_b)\;.
\end{equation}
Here $\Pi^{(1,2)}(q_a|q_b)$ is related to splittings from initial state
particles and $\Delta^f(q_a|q_b)$ solves Eq.~\ref{eq:identity} for final state
emissions; in both cases we refer to the no-emission probabilities as Sudakov
form factors, which relate the emission rate to the no-emission probability
\begin{equation}
\Delta^f(q_a|q_b)=\exp\left(-\int_{q_b}^{q_a}\hspace{-2mm}\dd qw^f_E(q)  \right)\;,
\end{equation}
where $w^f_E(q)$ includes all possible splittings originating from the
configuration $f$. A calculational formalism for parton showers can be set out
by defining the iterative process of adding emissions to a hard state $\phi_n$
starting from a scale $Q$ as a functional on test functions $u(\phi_n,Q)$
describing a generic observable at this state of the evolution,
\begin{multline}\label{eq:emission}
\PS_\mu[u(\phi_n,Q){\rm d}\sigma^j_n(\phi_{n}) ] = u (\phi_n,\mu){\rm
  d}\sigma^j_n(\phi_{n}) w_N(Q|\mu) \\ +\int\hspace{-2mm}
\int_\mu^Q\hspace{-2mm} dq \sum_{i} \PS_\mu\left[u(\phi^i_{n+1}(\phi_n),q)
\right.\\\left.  {\rm d} \sigma_n^j(..)w_N(Q|q) \dd w^i_E(q)\right]
\end{multline}
Integrations over the phase space degrees of freedom contained in ${\rm
  d}\sigma_n^j$ are implicit, and in a MC implementation we associate
the weight ${\rm d}\sigma^j_n(\phi_{n})$ to the state $u(\phi_n,Q)$.  The
operator $\PS_\mu[.]$ generates two contributions: Either the state
$u(\phi_n,Q)$ evolves with probability $w_N(Q|\mu)$ to $u(\phi_n,\mu)$ without
any further radiation or an emission of type $\alpha$ can be generated at any
of the intermediate scales $q$ between the starting scale $Q$ and the cut-off
$\mu$. The state then changes to become $u(\phi_{n+1}^\alpha(\phi_n),q)$. The
latter emission happens with a rate $w_N(Q|q) \dd w^\alpha_E(q)$, where
$w_N(Q|q)$ is the probability of having no emission before $q$ and $\dd
w^\alpha_E(q)$ is the full differential splitting rate for type $\alpha$ (see
Eq.~\ref{eq:splittingprob} for a single $\alpha$).  The integration and the
sum in Eq.~\ref{eq:emission} include all possible emissions at
the intermediate scales.  When an emission is generated, the parton shower
operator acts iteratively on the new state $u(\phi_{n+1}^\alpha(\phi_n),q)$
with starting scale $q$.

In the following we will define an actual implementation of the algorithms in
terms of pseudo code.  We start at this basic level for the sake of a complete
reference. Parton shower algorithms are usually implemented as variations of
the algorithms outlined in Alg.~\ref{alg:partonshower}.  Here the starting
conditions are given by the initial state $u(\phi,Q)$, for which a set of
colour dipoles are chosen from the incoming and outgoing coloured partons
according to the colour-subamplitudes described above.  The maximum scale for
emissions from each of this dipoles is set to the {\it shower starting scale}.
As shown in App.~\ref{app:Example}, parton shower algorithms after traversing
$k$ emissions change the event weights to become
\begin{multline}
X^{\PS}_{k}=
\\\Bigg[\prod_{i=1}^k \alpha_S(q^i)\frac{f_i(\eta_{i},q^i)}{f_{i-1}(\eta_{i-1},q^i)} w_N(q^{i-1}|q^i) 
 P^{\alpha_i}_{i,i-1}(\phi^{\alpha_i}_{i})\Bigg]X_{0}  \ ,
\end{multline}

\begin{algorithm}
\begin{algorithmic}
\Function{PartonShower}{$u(\phi_n, Q) ,\mu$}  
\If{ not Colourlines }
\State Find colour lines
\EndIf
\ForAll{ColourDipole $D_{ij}$ in  $u(\phi_n, Q)$} 
\State $\color{black}\rm Solve$ rnd()=$\Delta^f_{ij}(Q,p^{ij}_T)$ or \Comment{FS emitter}
\State $\color{white}\rm Solve$ rnd()=$\Pi_{ij}(Q,p^{ij}_T)$ for $p^{ij}_T$ \Comment{IS emitter}
\EndFor
\If{$\max(p^{ij}_T)>\mu$}  \Comment{Competition Alg.}
\State Solve $rnd()=\int^z_{z_-}dw_E(p^{\tilde{ij}}_T)/\int^{z_+}_{z_-}dw_E(p^{\tilde{ij}}_T) $  for $z$
\State and $rnd()=\varphi/2\pi $ for $\varphi$ for $\tilde{ij}$ with $p^{\tilde{ij}}_T=\max(p^{ij}_T)$ 
\State Produce $u(\phi^{\tilde{ij}}_{n+1}(\phi_n), p^{\tilde{ij}}_T)$ by splitting $D_{\tilde{ij}}$ 
\State with $K=\{p^{\tilde{ij}}_T,z,\varphi \}$
\State \Return \Call{PartonShower}{$u(\phi^{\tilde{ij}}_{n+1}(\phi_n), p^{\tilde{ij}}_T) ,\mu$}  
\EndIf
\State \Return $u(\phi_n,\mu)$
\EndFunction
\end{algorithmic}
\caption{Parton shower algorithm}
\label{alg:partonshower}
\end{algorithm}

starting from an initial weight $X_0$. Here
$P^{\alpha_i}_{i,i-1}(\phi^{\alpha_i}_{i})$ are the splitting probabilities
without the scale dependent functions $\alpha_S(q^i)$ and the ratio of parton
distributions functions (PDF) accounting for changes in the initial state,
while $\eta_i$ is the pair of momentum fractions after the $i$'th emission.

\subsection{Dividing and filling the phase space}
\label{sec:devideThePhasespace}

In order to include corrections to the parton shower it is possible to match
the parton shower results with higher order MEs. Within these approaches, the
double counting occurring beyond the leading order contributions is subtracted
at the desired, typically next-to-leading, order in $\alpha_S$. Contrary to
this, merging approaches aim at combining parton showers and multijet final
states into one inclusive sample which can describe observables across
different jet bins. These algorithms typically build on dividing the phase
space accessible to emissions into regions of jet production (hard, matrix
element region) and jet evolution (parton shower region).

There are multiple ways how the phase space can be partitioned into regions of
soft (IRC divergent) and regions of hard, large-angle emission. Jet algorithms
like the $k_\perp$ algorithm might be a tool of choice to achieve such a
separation, based on clustering partons according to an interparton separation
measuring how soft and/or collinear an emission has been. Each clustering
scale can then be subjected to a cut classifying it to either belong to the
jet production or jet evolution regime. It proves useful to design such an
iterative jet algorithm to correspond precisely to the inverse action of the
parton shower emission, and taking the actual shower emission scales to
separate phase space into regions of hard and soft emission.  In our default
implementation, we use the simple algorithm presented in
Alg.~\ref{alg:MEregion} to define the ME region. In this region all scales
encountered in any possible clustering configuration need to be larger than a
{\it merging scale} $\rho$.

We now aim to separate the parton shower algorithm into a two step evolution with
divided phase space regions, one for the ME and one for the parton shower. 
The first attempt of doing so is to split the shower emissions 
at the same scale $\rho$ as
\begin{equation}
  \PS_\mu [u(\phi_n,Q)]= \PSV_\mu
     [\PS_\rho[u(\phi_n,Q)]] \label{eq:vetoedshower}\ .
\end{equation}  
The action of the inner parton shower operator $\PS_\rho$ is limited to generate
emissions above $\rho$, while the subsequent action
$\PSV$ fills the region at lower scales, usually referred to as a vetoed
shower.
\begin{algorithm}[b]
\begin{algorithmic}
\Function{MatrixElementRegion}{$u(\phi_n, Q) ,\rho$}
\ForAll{$\phi_{n-1,\alpha}(\phi_n)$  } 
\If{$q^{\alpha}<\rho$} 
\State \Return False
\EndIf 
\EndFor
\State \Return True
\EndFunction
\end{algorithmic}
\caption{}
\label{alg:MEregion}
\end{algorithm}

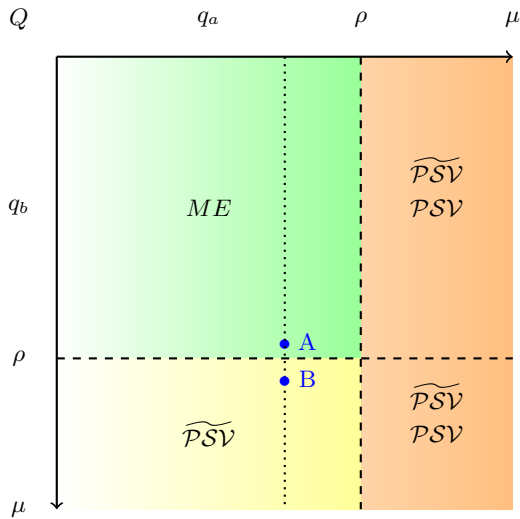
\begin{figure}
  \centering
       \begin{tikzpicture}
     \fill[left color=white,right color=orange!50] (0,0) rectangle (6,-6); 
     \fill[left color=white,right color=green!40] (0,0) rectangle (4,-6);  
     \fill[left color=white,right color=yellow!40] (0,-4) rectangle (4,-6);   
           \node  at (-0.5,0.5) {$Q$};
      \node  at (-0.5,-6) {$\mu$};
      \node  at (6,0.5) {$\mu$};
         \draw[thick,->] (0,0) -- (0,-6);
         \draw[thick,->] (0,0) -- (6,0);
         \draw[thick, style=dashed] (4,0) -- (4,-6);
         \draw[thick, style=dashed] (0,-4) -- (6,-4);
    \node  at (2,0.5) {$q_a$};
    \node  at (-0.5,-2) {$q_b$};
    \node  at (4,0.5) {$\rho$};
    \node  at (-0.5,-4) {$\rho$};
    \node  at (2,-2) {$ME$};
    \draw (3,-4) node[blue,anchor=south] {\textbullet  };
    \draw (3.3,-4) node[blue,anchor=south] {A};
    \node  at (5,-5) {$\PSV$};
    \node  at (5,-2) {$\PSV$};
    \node  at (5,-4.5) {$\tPSV$};
    \draw[thick, style=dotted] (3,0) -- (3,-6);
    \draw (3,-4.5) node[blue,anchor=south] {\textbullet };
    \draw (3.3,-4.5) node[blue,anchor=south] {B};
    \node  at (5,-1.5) {$\tPSV$};
    \node  at (2,-5) {$\tPSV$};
  \end{tikzpicture}
\caption{Simplified example for an emission phase space with two possibilities
  $D_a$, $D_b$ to fill. The ME region is defined, when the scales of both
  clusterings are above a given scale $\rho$. The modified veto shower $\tPSV$ needs
  to fill the remaining region as it would have been filled without division
  of the phase space. Emission by $D_a$ can be created above $\rho$, if the
  scale for $D_b$ is below $\rho$. A simplified vetoed shower $\PSV$ would miss the yellow 
  region for emissions above $\rho$. Further details and an explanation of the emissions $A$ and $B$
   are given in the text.}
  \label{fig:replace6}
\end{figure}

But in a standard shower setup competing channels can radiate within the same
phase space region, {\it e.g.} emissions from the two legs of a dipole system.
Such emissions can, as depicted in Fig.~\ref{fig:replace6}, typically not be
uniquely assigned to one scale and one individual emitter: Phase space points
after emission of one channel could, even though they were generated with a
scale above $\rho$, also be assigned to another splitting with a scale below
$\rho$. Even though the parton shower action was divided in
Eq.~\ref{eq:vetoedshower} into two regions, above and below $\rho$, the region
populated by $\PS_\rho$ is not suited to define the matrix element region,
since an emission above $\rho$ from one dipole leg could exhibit a scale below
$\rho$ w.r.t. to a different dipole leg.

To illustrate this point we will briefly discuss two example emissions $A$ and
$B$ in Fig.~\ref{fig:replace6}.  Dipole $a$ emits at a scale that may or may
not result in a phase space point within the ME region.  The configurations at
this scale are depicted by the dotted line.  In case $A$, the emission is
within the ME phase space and therefore has to be vetoed by the outer vetoed
shower $\PSV$ in Eq.~\ref{eq:vetoedshower}.  In case $B$, the emission would
have also been vetoed by $\PSV$ as an emission from dipole $a$ since the
emission scale is above $\rho$.  However, it is part of the parton shower phase
space of dipole $b$ and is therefore not part of the ME region by the
definition of the cluster algorithm.  To divide the phase space into a ME and
a parton shower region, we have to keep emission $B$ from Dipole $a$ in this
example in order to define the ME region through the cluster algorithm.

\begin{algorithm}[b]
\begin{algorithmic}
\Function{TVPS}{$u(\phi_n, Q) ,p^{\rm hard}_T,\rho$,$N$}  
\State Start shower from state $u(\phi_n,Q)$
\State Veto emissions above $p^{\rm hard}_T$
\If{\Call{MatrixElementRegion}{$\phi^\alpha_{n+1}$,$\rho$}}
\State  \Return \Call{TVPS}{$\phi_n, Q, p^\alpha_T(\phi^\alpha_{n+1}) ,\rho$,$N$} \Comment{Veto}
\EndIf
\State \Return \Call{PartonShower}{$u(\phi^\alpha_n,p^\alpha_T(\phi^\alpha_{n+1}))$,$\mu$}\Comment{Alg.~\ref{alg:partonshower}}
\EndFunction
\end{algorithmic}
\caption{Transparent Vetoed Parton Shower}
\label{alg:tvps}
\end{algorithm}

To solve the problem that arose due to the simplistic splitting of the phase
space in Eq.~\ref{eq:vetoedshower}, we need to modify emissions as follows:
Each emission of the initial, inner showering step needs to fulfil the
condition \mbox{$\prod_i \theta_i(q^i-\rho)=1$}, where $q^i$ are scales obtained from
the possible clusterings. The outer, vetoed shower needs to be allowed to emit
above the scale $\rho$, if the configuration considered might contribute to
the parton shower region as defined through a competing emission
process. Solutions to our aim hence take the form
\begin{equation}
  \PS_\mu [u(\phi_n,Q)]= \tPSV_\mu [\tPS_\rho[u(\phi_n,Q)]]   \label{eq:vetoedshower2}\;,
\end{equation}  
where the $\PS$ and $\tPS$ differ in the emissions at scales below $\rho$ for
any clustering. $\tPS$ emits only in regions which a clustering algorithm
would declare as the ME region, and the counterpart $\tPSV$ fills the
complementary phase space region. 
With this construction we can now replace the inner step of Eq.~\ref{eq:vetoedshower2}
by ME contributions.
An algorithm for $\tPSV$ is given in Alg. \ref{alg:tvps}.

\subsection{Scales}

A number of unphysical scales enter the algorithm. The dependence on these
scales is expected to be minimized upon subsequently including more and more
higher-order corrections in the simulation. This is not always possible, and
we choose to use variations of the scales below as an indication of missing
contributions. In particular, the following scales are considered, possibly as
dynamical quantities depending on the kinematics of the hard process under
consideration:
\begin{itemize}
\item \textbf{Renormalization scale shower $\mu^\PS_R$:} Scale used to
  calculate the value of $\alpha_S(\mu^\PS_R)$ for shower emissions.  It is
  usually related to the ordering variable and/or the respective transverse
  momentum of the shower emission.  The scale can be varied by $\xi_R^\PS$.
\item \textbf{Renormalization scale ME $\mu^{ME}_R(\phi)$:}
  Renormalization scale used to calculate the (N)LO cross section.  In LO or
  matched simulations the scale is applied to the ME calculation
  only.  The scale can be varied by $\xi_R^{ME}$.
\item \textbf{Factorization scale shower $\mu^\PS_F$:} The scale used to
  calculate the value of parton distribution function factors
  $f_{i}(\eta_{i},\mu^\PS_F)/f_{i-1}(\eta_{i-1},\mu^\PS_F)$ for shower
  emissions.  As $\mu^\PS_R$, it is usually related to the ordering variable
  or the respective transverse momentum of the shower emission.  The scale can
  be varied by $\xi_F^\PS$.
\item \textbf{Factorization scale ME $\mu^{ME}_F$:}
  Renormalization scale used to calculate the (N)LO cross section.  In LO or
  matched simulations the scale is applied to the cross section calculation
  only.  The scale can be varied by $\xi_F^{ME}$.
\item \textbf{Shower Starting Scale $Q_S$:} The shower starting scale is used
  to restrict the phase space of shower emissions, both in their transverse
  momenta and, possibly also for their longitudinal momentum fraction. The
  scale can be varied by $\xi_Q$.
\item \textbf{Shower Cut-Off $\mu$:} The infrared cutoff on shower emissions;
  no emissions will be generated with transverse momenta below this
  scale. The scale can be varied by $\xi_{IR}$, and variations here typically
  need to be accompanied by re-fitting the hadronization parameters.
\item \textbf{Merging Scale $\rho$:} The merging scale is a technical
  parameter that divides the phase space into a ME and a shower
  region.  The dependence of the merging scale is non physical and gives an
  indication on the quality of the merging algorithm.
\end{itemize}

\section{Merging multiple LO cross sections}
\label{sec:lomerging}

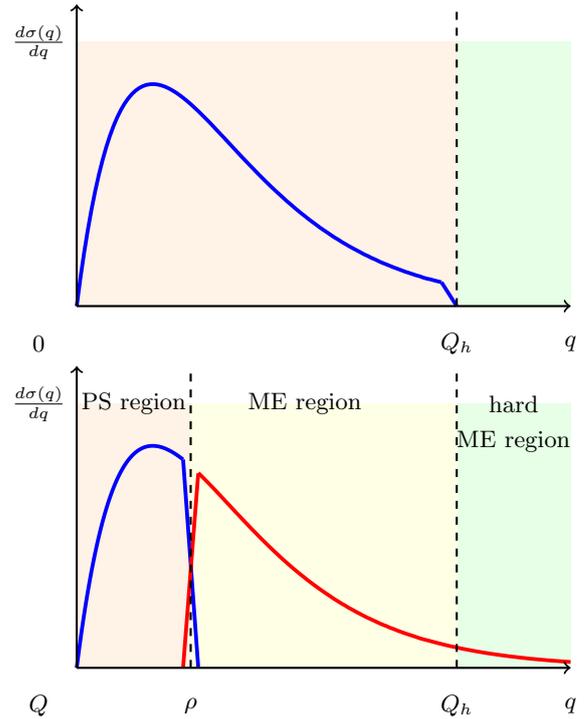
\begin{figure}[t]
  \centering
     \begin{tikzpicture}
     \fill[left color=green!10,right color=green!10] (5,0) rectangle (6.5,3.5);
     \fill[left color=orange!10,right color=orange!10] (0,0) rectangle (5,3.5);  
         \draw [blue,line width=0.5mm] plot [ samples=100, domain=0:4.8 ] 
         (\x,{8*\x*exp(-\x) });
         \draw [blue,line width=0.5mm] plot [ samples=100, domain=4.8:5. ] 
         (\x,{8*4.8*exp(-4.8)*(1-(\x-4.8)/0.2 ) });
           \node  at (-0.5,3.5) {$\frac{d\sigma(q)}{d q}$};
           \node  at (-0.5,-0.5) {$0$};
           \node  at (5,-0.5) {$Q_h$};
           \node  at (6.5,-0.5) {$q$};
         \draw[thick,->] (0,0) -- (0,4);
         \draw[thick,->] (0,0) -- (6.5,0);
         \draw[thick, style=dashed] (5,0) -- (5,4);
  \end{tikzpicture}
  
       \begin{tikzpicture}
     \fill[left color=green!10,right color=green!10] (5,0) rectangle (6.5,3.5);
     \fill[left color=orange!10,right color=orange!10] (0,0) rectangle (1.5,3.5);   
     \fill[left color=yellow!10,right color=yellow!10] (1.5,0) rectangle (5,3.5); 
         \draw [blue,line width=0.5mm] plot [ samples=100, domain=0:1.4 ] 
         (\x,{8*\x*exp(-\x) });
         \draw [blue,line width=0.5mm] plot [ samples=100, domain=1.4:1.6 ] 
         (\x,{8*1.4*exp(-1.4)*(1-(\x-1.4)/0.2 ) });
         \draw [red,line width=0.5mm] plot [ samples=100, domain=1.6:6.5 ]
          (\x,{8*\x*exp(-\x) });
         \draw [red,line width=0.5mm] plot [ samples=100, domain=1.4:1.6 ] 
         (\x,{8*1.6*exp(-1.6)*((\x-1.4)/0.2 ) });
          \node  at (0.75,3.5) {PS region};
         \node  at (3.,3.5) {ME region};
         \node  at (5.75,3.5) {hard };
         \node  at (5.75,3.) {ME region};
           \node  at (-0.5,3.5) {$\frac{d\sigma(q)}{d q}$};
           \node  at (-0.5,-0.5) {$Q$};
           \node  at (1.5,-0.5) {$\rho$};
           \node  at (5,-0.5) {$Q_h$};
           \node  at (6.5,-0.5) {$q$};
         \draw[thick,->] (0,0) -- (0,4);
         \draw[thick,->] (0,0) -- (6.5,0);
         %\draw[brown,thick, style=dashed] (0.75,0) -- (0.75,4);
         \draw[thick, style=dashed] (1.5,0) -- (1.5,4);
         \draw[thick, style=dashed] (5,0) -- (5,4);
  \end{tikzpicture}
  
\caption{An illustration of a schematic, infrared sensitive observable that
  receives contributions by both, parton showers and hard, additional jet
  production, which are reminiscent of a Drell-Yan $p_\perp$
  spectrum. The top panel shows the shower prediction, with no radiation
  produced above a hard scale $Q_h$, and the physical shape of the spectrum in
  the region of low transverse momenta. In the lower panel, we show the
  additional spectrum of a high-$p_\perp$ tail (in red) as populated by a
  cross section with an additional, hard jet.
  This spectrum would diverge
  towards low transverse momenta and so a cut needs to be applied at the
  merging scale $\rho$, below which the phase space is populated by the parton
  shower. Merging algorithms need to consistently combine the two, avoiding
  missing or double-counted contributions around the merging scale.}
  \label{fig:schematicdist}
\end{figure}

We will summarise leading-order (LO) merging in this section, using the
notation we have introduced so far. This will form the basis to detail the
unitarization procedure, as well as the further steps towards NLO
merging. General, qualitative, features of merging algorithms are shown in
Fig.~\ref{fig:schematicdist} using a schematic distribution such as the
transverse momentum spectrum of a colour singlet final state produced in
hadron-hadron collisions.

\subsection{Conventional LO merging}

We briefly establish LO merging algorithms as the basis of the present work,
however the reader is referred to the original publications for specific
details of a particular algorithm. LO merging algorithms like the conventional
CKKW\cite{CKKW}, \mbox{CKKW-L}\cite{CKKW-L} and MLM\cite{MLM} can be summarized to
include cross sections of higher multiplicities in regions of the phase space
where the scales of the extra emissions are above a {\it merging scale}
$\rho$.  The ME region is usually defined in terms of a jet clustering
algorithm, which we discuss in detail in Sec.~\ref{sec:clustering}.  To
merge the cross sections of the different multiplicities the weights of the
phase space configuration are multiplied by additional factors that reflect
the probability that the parton shower would have generated the given
configuration. These factors are built out of Sudakov factors accounting for
the probability of no emission between intermediate scales. The intermediate
scales are obtained by the clustering algorithm, which reflects the
inverse of the parton shower evolution. The shower emission densities are
usually used to select the most probable interpretation in terms of a shower
history, though this procedure is not unique. The re-weighted multiplicities
now enter a modified/vetoed parton shower,  not allowing emissions in
the ME region except if it starts off a state of the highest ME multiplicity
available to the algorithm.

We will detail the ingredients to this generic procedure in the following
subsections. The definition of a ME region and its relation to the
parton shower emission phase space was sketched in
Sec.~\ref{sec:devideThePhasespace} with focus on regions which can be filled
by more than one shower emission process. The reweighting procedure will be
discussed in detail in the sections to follow.

\subsection{Clustering Probabilities}
\label{sec:clusterprob}

An important ingredient for merging multi-jet cross sections is the
reweighting of higher jet multiplicities. This crucially involves which of the
multiple ways a parton shower could have traversed to generate the
configuration at hand is chosen to evaluate the weight factor accompanying
this particular state. These weight factors finally account for the {\it
  exclusiveness}, {\it i.e.} the absence of further resolved radiation on top
of the input jet multiplicity.  The parton shower approximates the next higher
multiplicity as
\begin{equation}
d\sigma_{n+1}(\phi_{n+1},Q)\approx \sum_\alpha d\sigma^\alpha_{n}(\phi_{n,\alpha},Q) w^\alpha_E(Q,\phi_{n+1})\;,
\end{equation}
such that observables obey
\begin{eqnarray}
&&u(\phi_{n+1}) d\sigma_{n+1}(\phi_{n+1},Q)w^{n+1}_N\\
&&\approx u(\phi_{n+1})\sum_\alpha d\sigma^\alpha_{n}(\phi_{n,\alpha},Q) w^\alpha_E(q,\phi_{n+1})w^\alpha_N(Q|q,\phi_{n,\alpha})\;.\nonumber
\end{eqnarray}  
We are therefore led to a recursive construction of the weight $w^{n+1}_N$,
which reads
\begin{align}
&w^{n+1}_N\approx \\
&\sum_\alpha \frac{d\sigma^\alpha_{n}(\phi_{n,\alpha},Q)w^\alpha_E(Q,\phi_{n+1})}{d\sigma_{n+1}(\phi_{n+1},Q)} \frac{w^\alpha_E(q,\phi_{n+1})}{w^\alpha_E(Q,\phi_{n+1})} w^\alpha_N(Q|q,\phi_{n,\alpha})\nonumber
\\
&\approx \sum_\alpha\frac{d\sigma^\alpha_{n} w^\alpha_E}{\sum_\beta d\sigma^\beta_{n}w^\beta_E} \frac{w^\alpha_E(q,\phi_{n+1})}{w^\alpha_E(Q,\phi_{n+1})} w^\alpha_N(Q|q,\phi_{n,\alpha})\;.\nonumber
\end{align} 
We choose to perform the sum as a MC sum by picking one $\alpha$ with
weight $w_C^\alpha=d\sigma^\alpha_{n} w^\alpha_E/\sum_\beta d\sigma^\beta_{n}w^\beta_E$ and include the
remaining part of the expressions via a direct multiplication of the event
weight.

\begin{algorithm}[b]
\begin{algorithmic}
\Function{Cluster Algorithm}{$\phi_n,\rho$}  
\If{{\bf not} \Call{MatrixElementRegion}{$\phi_{n}$,$\rho$}} 
\State\Return $\phi$ with weight 0 \EndIf
\State $\phi \gets\phi_n$,$i \gets n$
\While{$i>0$} 
\State $\Phi\gets $ Empty Selector
\ForAll{$\phi_{i-1,\alpha}$} 
\If{$i = 1$} 
\State $Q\gets Q(\phi_{0,\alpha})$
\Comment{Scale for $\phi_0$}
\Else
\State $Q\gets \max(p^{\alpha,\beta}_T(\phi_{i-2,\alpha,\beta})$ 
\EndIf 
\If{$\phi_n \in PS(\phi_{i-1,\alpha},Q)$} 
\State  add $\phi_{i-1,\alpha}$ to  $\Phi$ with weight $\tilde{D}^\alpha_i(\phi_n )$
\EndIf
\EndFor
\If{$\Phi$ empty} 
\State\Return $\phi$
\EndIf 
\State $\phi \gets $ weighted selection from $\Phi$,$i \gets i-1$
\EndWhile 
\State\Return $\phi$
\EndFunction
\end{algorithmic}
\caption{The clustering algorithm employed to interpret a hard process input
  in terms of the shower action.}
\label{alg:clustering}
\end{algorithm}

\subsection{Clustering}
\label{sec:clustering}

We summarize the clustering algorithm employed in the merging strategy in
algorithmic form in Alg.~\ref{alg:clustering}.  The probability to choose a
given clustering has already been discussed in Sec.~\ref{sec:clusterprob}, and
is based on the parton shower approximation.  At the beginning of the
algorithm we ensure that the phase space point $\phi_n$ is contained inside
the ME region for chosen merging scale $\rho$. We now select all possible
clusterings $\alpha$ that provide scales and splitting variables that could
have been used to perform a shower emission from the underlying kinematics
$\phi_{n-1,\alpha}(\phi_n)$; this includes the restriction imposed by the
shower starting scale $Q$ assigned to the underlying configuration
$\phi_{n-1,\alpha}(\phi_n)$.  $Q$ either is the initial shower scale if no
further clustering is possible ({\it i.e.} we have reached the lowest order
input process) or the maximum scale $p^{T,\alpha\beta...}_{ij,k}$ assigned to
all subsequent clusterings, {\it e.g.} for secondary clusterings
$p_T^{\alpha\beta}(\phi_{n-2,\alpha\beta}(\phi_{n-1,\alpha}(\phi_n)))$.

Configurations $\phi_{n-k,\alpha...}(...(\phi_n))$ that are not ruled out by
the constraints are inserted into a $selector$\footnote{The entries of a
  selector can be accessed randomly according to the weight that has been
  assigned to the entry.} with the weight as it is described in
Sec.~\ref{sec:clusterprob}. From these possible clusterings we choose one
according to the weight of the shower approximation and continue this
procedure iteratively from the state obtained through the clustering.  If no
possible clustering is found the algorithm terminates and the phase space
point is returned to be the {\it{seed}} phase space point to initiate the modified
showering process.

\subsection{Starting Scale}
\label{sec:startingscale}
If the clustering algorithm returns a {\it seed} configuration
$\phi_{0,\alpha...}$ with the lowest order multiplicity for which the
production process is defined, the shower starting scale $Q_S$ is given by the scale $Q_P$
one would assign to the initial process in the absence of any merging
prescription,
\begin{equation}
Q_S(\phi_n)= Q_P(\phi_n)\ .\label{eq:startscale1}
\end{equation}
Otherwise, if the clustering algorithm terminates at a clustered phase space
point $\phi_{n-m}$ with a higher than the lowest multiplicity, this
contribution is treated as an additional hard process. In this case, which can
be encountered either by processes for which no matching shower clustering is
available or for which the emission considered is happening outside of the
shower phase space, we now choose the starting scale $Q_S$ as
\begin{equation}
Q_S(\phi_{n-m})= \max(Q_P(\phi_{n-m}),\min(p^T_{ij,k})),\label{eq:startscale}
\end{equation}
where $Q_P$ is calculated with the same description as for the initial
process, {\it e.g.} $H_T$ or invariant mass of bosons or jets, and
$p^T_{ij,k}$ are the transverse momenta the dipole configuration $(ij,k)$
would have assigned to the showering process. This choice is reminiscent of
what is done by the scales used in the shower phase space itself, which
typically possess an upper kinematic limit and are otherwise driven by choosing
a scale of the same order as the typical hard process scales. Just
as with the algorithmic choice of the clustering procedure, this scale
prescription amounts to a genuine uncertainty of the algorithm, which will be
quantified in detail in future work.

\subsection{LO merging}
\label{sec:historyweight}
\begin{algorithm}[b]
\begin{algorithmic}
\Function{HistoryWeight}{$\phi_B,\phi_n,Q_S(\phi_B)$} 
\State $\phi \gets \phi_B$; $Q \gets Q_S(\phi_B)$;  $w \gets (1+ \frac{\alpha_S(q)}{2\pi}K_g)^{N_{S}}$
\State $\phi_{+} \gets$ next step to  $\phi_n$
\While{$\phi_+$ not $\phi_n $}
\State $q \gets p_T(\phi_{+}(\phi))$\Comment{See Sec.~\ref{sec:CMW} for $K_g$ and $N_{S}$}
\State $w_\alpha\gets \frac{\alpha_S(q)}{\alpha_S(Q_S(\phi_B))}(1+ \frac{\alpha_S(q)}{2\pi}K_g)$
\State $w_f \gets \frac{f^{1,2}(\phi,Q)}{f^{1,2}(\phi,q)}$
\State $w_\Delta \gets w_N(Q|q,\phi)$
\State $\phi \gets \phi_+$; $Q\gets q$; $w \gets w_\alpha w_f w_\Delta w $
\State $\phi_{+} \gets$ next step to  $\phi_n$
\EndWhile
\State \Return $w  \frac{f^{1,2}(\phi_n,q)}{f^{1,2}(\phi_n,Q_S(\phi_B))}$
\EndFunction
\end{algorithmic}
\caption{ }
\label{alg:historyweight}
\end{algorithm}

We are now in the position to write down an expression for the merged cross
section following the spirit of \cite{HerwigMerging,UNLOPS} as
\begin{align}
&\PS[{\rm d}\sigma^{\rm merged} u (\phi_n)]\label{eq:conventionalmerged}\\\nonumber&=
 \tPSV_\mu \left[ u(\phi_0,Q)  {\rm d}\sigma_0(Q)  w^0_L \right] \\\nonumber&+
\sum_{n=1}^{N-1} \tPSV_\mu \left[u(\phi_{n},q^{n-1}){\rm d}\sigma_{n}(Q)w^n_L \frac{f^{(1,2)}_n(\eta_n,q_n)}{f^{(1,2)}_n(\eta_n,Q)} \prod_{k=1}^n w^k_I\right]\\\nonumber&+
\PS_\mu \left[u_i (\phi_{N},q^{N}){\rm d}\sigma_{N}(Q)
  \frac{f^{(1,2)}_N(\eta_N,q_N)}{f^{(1,2)}_N(\eta_N,Q)}\prod_{k=1}^N
  w^k_I\right]\ .
\label{eq:emission2}
\end{align}
For compactness and readability we suppress the integration over the phase
space and the summation of single subprocesses. The higher multiplicity cross
sections ${\rm d}\sigma_{n/N}(Q)$ are constrained to the ME region
and $N$ is the highest multiplicity for which cross sections (in this case at
tree level) are taken into account by the merging algorithm. Higher jet
multiplicities are solely generated by the parton shower. $w_L$ is the
no-emission probability combined with a PDF ratio to adjust the scale used in
the ME calculation to the one chosen for a respective parton shower emission,
\begin{eqnarray}
w^k_L&=&  \Pi^{(1,2)}_{q_{k}\to \rho} \prod_f \Delta^{q_k}_\rho \;.
\end{eqnarray}
Here $q_0=Q$ and $(q_k,\eta_k)$ are the scales and momentum fractions
encountered in the clustering procedure. $x_n$ is the momentum fraction
associated to the phase space point $\phi_n$, for which the ME was
calculated. In App.~\ref{app:Example} we construct an explicit example of the
weight accumulated through a parton shower history obtained by the clustering
algorithm. $w_I^k$ is the weight for no emission off internal lines, combined
with ratios of parton distribution functions and the strong coupling required
to instantiate the intermediate, parton-shower, emission scales,
\begin{align}
w^k_I&= \sum_\alpha \frac{w^k_{C,\alpha}}{\sum_s w_{C,\beta}^k
}\frac{\alpha_S(q_{k})}{\alpha_S(\mu_R)}\\\nonumber
&\cdot\underbrace{\frac{f^{(1,2)}_k(\eta_{k-1},q_{k-1})}{f^{(1,2)}_k(\eta_{k-1},q_{k})}\Pi^{(1,2)}({q_{k-1}|
    q_k})}_{\approx \Delta({q_{k-1}| q_k}) } \prod_f \Delta(q_{k-1}| q_k)
\end{align}
Note that $w_I^k$ also depends on the clustering algorithm through the
probability which has been used to choose a particular parton shower history.
For future reference we also define
\begin{equation}
w^n_H=\frac{f^{(1,2)}_n(\eta_n,q_n)}{f^{(1,2)}_n(\eta_n,\mu_F)} \prod_{k=1}^n
w^k_I \ .
\end{equation}
The full algorithm to calculate the weight for a specific history is presented
in Alg.~\ref{alg:historyweight}.

\subsection{Unitarized LO merging}
\label{sec:ULOM}
\begin{algorithm}[b]
\begin{algorithmic}
\Function{UnitarizedMergingAlgorithm}{$\phi_n,\rho,Q$}  

\State Select subprocess $S_i$ with $w^i_S/\sum_j w^j_S$ 
\State $w\gets \sum_j w^j_S/ w^i_S$
\State Produce $\phi_n$ for $S_i$
\State $\phi_B\gets$\Call{Cluster Algorithm}{$\phi_n,\rho,Q$}   \Comment{Alg.~\ref{alg:clustering}}
\If{$\theta_C(\phi_B)=0$} 

\Return 0\Comment{Sec.~\ref{sec:Cuts}}\EndIf 
\If{$\phi_B=\phi_n$} \Comment{No clustering}

\Return $w \dd\sigma_{n,i}(\phi_n,Q_S(\phi_n))$ \Comment{see Eq.~\ref{eq:startscale} for $Q_S()$} \EndIf 
\State $w_H\gets$\Call{HistoryWeight}{$\phi_B,\phi_n,Q_S(\phi_B)$} \Comment{Alg.~\ref{alg:historyweight}}
\State \Return
\State  \hspace{2.5mm} Rnd($2u(\phi_{n},q_{n})$,$-2u(\phi_{n-1,\alpha H},q_{n-1})$)
\State  \hspace{2.5mm} $ w\cdot w_H \dd\sigma_{n,i}(\phi_n,Q_S(\phi_B))$
\EndFunction
\end{algorithmic}
\caption{ }
\label{alg:unitarizedLOmerging}
\end{algorithm}

Parton showers are built on factorization properties of tree-level amplitudes
in singly-unresolved limits. The corresponding virtual contributions are
derived by imposing a unitarity argument, and inclusive quantities are hence
unchanged after parton shower evolution. Upon replacing the emission
probability of the parton shower through a merging algorithm we expect a
non-unitary action since the virtual contributions, present at leading
logarithmic level to all orders in the Sudakov form factor, have not been
changed such as to retain the correct resummation properties. The amount by
which inclusive cross sections are changed through such a procedure have been
addressed in \cite{HerwigMerging}, and are essentially probing the quality of
the parton shower approximation by integrating the difference between a parton
shower approximated and a full real emission matrix element, weighted by the
Sudakov form factor for the relevant singular limit considered. As such, no
logarithmically enhanced terms are expected to contribute to inclusive cross
sections. However, with NLO merging in reach for which a potentially much more
dangerous mis-cancellation in inclusive quantities is expected, we first
address how inclusive quantities can be constrained through appropriate
subtractions within merged cross sections.

While the first approaches to unitarized merging suggested an exact
restoration of inclusive quantities, we here relax this criterion such as to
capture terms with a logarithmic enhancement only. To this extent, for each
clustering $\phi_n$ to $\tilde{\phi}^\alpha_{n-1}$, we determine the
respective transverse momentum $p_\perp^\alpha$ and longitudinal fraction
$z^\alpha$ of the emission, which are required to be within the phase space
available to shower emissions. If no such configuration is obtained for
potential clustering we assume that a genuine correction of hard jet
production has been found which will then set the initial conditions for
subsequent showering, subject to acceptance of a hard trigger object like a
$Z$ boson or a jet within the generation cuts at the level of this particular
hard process configuration.

Those logarithmically enhanced contributions which are subtracted from the
lower multiplicity in order to constrain inclusive cross sections provide
approximate NLO corrections in the same spirit as the LoopSim \cite{LoopSim}
method has established capturing the most enhanced corrections. Having
identified and being able to control these contributions we are in the
position to establish a unitarized merging of NLO multijet cross sections.

In Sec.~\ref{sec:devideThePhasespace} we have shown how the parton shower
evolution can be factored into a region of jet production (ME region) and jet
evolution, subject to a merging scale $\rho$ and including the fact that
emissions with transverse momenta below or above $\rho$ do not necessarily
respect this ordering when the kinematic mapping is defined with respect to a
different dipole configuration than the emitting one. It is therefore clear
that shower emissions off a configuration which is contained in the ME region
({\it i.e.} one with {\it all} evolution scales above $\rho$) cannot be
subjected to a naive veto on transverse momenta. Only if the full kinematics
of the emission are known can we test if {\it one or more} of the emitting
configurations gave rise to an emission below $\rho$ in case of which the
emission is accepted as being contained in the shower region. We are convinced
that this procedure of a modified vetoed shower is precisely equivalent to the
truncated showering procedures employed elsewhere
\cite{Hamilton:2009ne,Hoche:2010kg}.

\subsection{Generation Cuts}
\label{sec:Cuts}

While generation cuts within a merging algorithm are not required to render
cross section predictions with additional jets finite, they might still be
desirable to enhance populating certain regions of phase space. Even more
than with a fixed jet multiplicity cut migration does become an issue within
this context, specifically as clustered phase space points which have been
identified as seed processes, will or will not pass the generation cut
criteria independently of the acceptance of the unclustered configuration. Just
as with cut migration present in standard shower evolution off a hard seed
process, we indeed require that cuts are solely applied to the seed process
which has been identified after the clustering procedure.

This still requires that care needs to be taken in the region subject to the
migration and results should only be considered away from these
boundaries, just as is the case for showering jetty processes in a standard
setup. Ideally, generation cuts should not be required but efficiency issues
should be addressed through a biasing procedure complemented by a reweighting
such as to ensure that no event which possibly could contribute to an
observable of interest will be discarded. We will address this issue in more
detail in future work.

\section{NLO Merging}
\label{sec:NLOmerging}

In this section we explicitly construct the merging of multiple NLO cross
sections. We will follow the proposal of unitarized merging algorithms, which
specifically focus on potentially problematic terms arising in NLO multi jet
merging. Based on the combination of multiple LO cross sections with the
parton shower, which constitutes an improved, resummed prediction, the task
actually boils down to matching such a calculation consistently to the first
${\cal O}(\alpha_S)$ correction in each jet multiplicity. These corrections can
receive both virtual corrections to the multiplicity at hand and approximate
contributions from the parton shower action on a lower multiplicity.  The
remaining dependence on these corrections in inclusive cross sections can be
traced back to a mismatch of the leading-order parton shower attempting to
approximate a next-to-leading order correction, and will explicitly be removed
by the unitarization procedure.

We first aim at correcting cross sections above the merging scale to be
accurate to NLO. To this extent we consider the expression for the LO merged
sample for multiplicity $n$ in a region where all resolution scales are above
the merging scale,
\begin{align}
d\sigma_n u(\phi_n,q_n)  w^n_H  &- \int_\rho^{q_n}\hspace{-3mm}dq \sum_\alpha \frac{w_{C,\alpha}}{\sum_\beta w_{C,\beta}} u(\phi^{\alpha}_n,q^{\alpha}_n) d\sigma_{n+1}  w^{n+1}_H\\
& +   d\sigma_{n+1} u(\phi_{n+1},q_{n+1}) w^{n+1}_H \nonumber \ ,
\end{align}
which exhibit the $\mathcal{O}(\alpha_S)$ expansion
\begin{align}
\label{eq:LOexpansion}
d\sigma_n u(\phi_n,q_n) \left.\frac{\partial w^n_H}{\partial \alpha_S}\right|  &- \int_\rho^{q_n}\hspace{-3mm} dq \sum_{\alpha} \frac{w_{C,\alpha}}{\sum_\beta w_{C,\beta}} u(\phi^{\alpha}_n,q^{\alpha}_n) d\sigma_{n+1}  \\
& +   d\sigma_{n+1} u(\phi_{n+1},q_{n+1}) \nonumber\;.
\end{align}
Here we use $\partial w^n_H/\partial \alpha_S|$, the derivative of the history
weight with respect to $\alpha_S$ evaluated at $\alpha_S = 0$ for all additional powers of $\alpha_S$ with respect to the production process. Such quantities
will be abbreviated as $w_{\partial H}$ in the following. Note that while $ d\sigma_{n+1}$ is already of
the order we required the expansion to cover, the probability to cluster the
unitarization expression is remaining in this expansion.

Besides expanding the shower/merged expressions, the calculation of NLO
corrections with MC techniques requires the introduction of subtraction for the
virtual and real contributions, see Sec.~\ref{sec:subtraction}.  On combining
the different parts in the following, we discuss how the different
contributions interact in order to achieve NLO accuracy above the merging
scale; we will also show that below the merging scale the real emission is
generating corrections to the shower approximation present otherwise, such
that we can achieve NLO accuracy below the merging scale in a standard
matching paradigm.

\subsection{Real emission contributions}
\label{sec:realemissioncontribution}

\begin{algorithm}[b]
\begin{algorithmic}
\Function{RealEmissionWeight}{$\phi_{n+1}$}

\ForAll {$\alpha$ with $p_T^\alpha<\rho$ }
\If{ not \Call{MatrixElementRegion}{$\tilde{\phi}^\alpha_{n}$,$\rho$}} 
\State\Return 0 \EndIf
\EndFor

\State Choose $\alpha'$ randomly
\State $\phi_B\gets$\Call{Cluster Algorithm}{$\phi_{n,\alpha'},\rho$}  
\If{$\theta_C(\phi_B)=0$} 

\Return 0\Comment{Sec.~\ref{sec:Cuts}}\EndIf 

\State $w_H\gets$\Call{HistoryWeight}{$\phi_B,\phi_{n,\alpha'},Q_S(\phi_B)$}  \Comment{Alg.~\ref{alg:historyweight}}

\If{\Call{MatrixElementRegion}{$\phi_{n+1}$,$\rho$}}\Comment{ME-region}
\State $\phi_{n,C}\gets$\Call{ClusterAlgorithm}{$\phi_{n},\rho$,1 step} \Comment{Alg.~\ref{alg:clustering}}
\If{$\phi_{n,C}=\phi_{n,\alpha'}$ } \Comment{See Sec.~\ref{sec:realemissioncontribution} (A)}
\State \Return 
\State Rnd($2u(\phi_{n,\alpha'},q_{n})$,$-2u(\phi_{n-1,\alpha'},q_{n-1})$)\Comment{Sec.~\ref{sec:UnitarisingNLO}}
\State $N_{Dip}w_H(\dd\sigma_R(\phi_{n+1})-D^{\alpha'}(\phi_{n+1}))$
\Else  
\State \Return
\State Rnd($2u(\phi_{n,\alpha'},q_{n})$,$-2u(\phi_{n-1,\alpha'\beta},q_{n-1})$)\Comment{Sec.~\ref{sec:UnitarisingNLO}}
\State $N_{Dip}w_H(-D^{\alpha'}(\phi_{n+1}))$
\EndIf 
\Else \Comment{Shower Region}
\If{rnd()$>0.5$ }
\State \Return \Comment{See Sec.~\ref{sec:realemissioncontribution} (B)}
\State Rnd($2u(\phi_{n+1},q_{n})$,$-2u(\phi_{n,\alpha'},q_{n-1})$) \Comment{Sec.~\ref{sec:UnitarisingNLO}}
\State $2w_H\left[\dd\sigma_R(\phi_{n+1})-\sum_\alpha P^{\alpha}(\phi_{n+1})\theta\left(p^{\alpha\beta}_T-\rho\right)\right]$
\Else
\State \Return \Comment{See Sec.~\ref{sec:realemissioncontribution} (C)}
\State Rnd($2u(\phi_{n,\alpha'},q_{n})$,$-2u(\phi_{n-1,\alpha'H},q_{n-1})$)\Comment{Sec.~\ref{sec:UnitarisingNLO}}
\State  $2N_{Dip}w_H(P^{\alpha'}(\phi_{n+1})-D^{\alpha'}(\phi_{n+1}))\theta\left(p^{\alpha'\beta}_T-\rho\right)$
\EndIf 
\EndIf 
\EndFunction
\end{algorithmic}
\caption{ }
\label{alg:real}
\end{algorithm}

In the NLO merging algorithm the subtraction for the real emission
contributions is more complicated. In the MC@NLO approach the expansions of
the shower expressions up to $\mathcal{O}(\alpha_S)$ generate terms that need
to be subtracted in order to remove double counted contributions.  In the NLO
merging the expansion of the LO merging up to $\mathcal{O}(\alpha_S)$ produces
similar subtraction terms. There are three different phase space regions to
consider:
\begin{enumerate}[label=(\Alph*)]
\setcounter{enumi}{0}
\item {\bf ME Region:} If the phase space point $\phi_{n+1}$ of the real
  emission is contained in the ME region, the LO merging has already populated
  the region with LO corrections to the parton shower. The
  $\alpha_S$-expansion of the LO merged contribution, see expressions
  proportional to $\phi_{n+1}$ in Eq.~\ref{eq:LOexpansion}, need to be
  subtracted in this region in order to solve the double counting of ME
  corrections in this region. The same double counting argument now requires
  that the second expression of Eq.~\ref{eq:LOexpansion}, which is stemming
  from the unitarization expressions of the LO merging needs to be added to
  the expansion of $\alpha_S$. This contribution is proportional to the real
  emission contribution, in the ME region and is clustered to the underlying
  Born phase space points $\tilde{\phi}^\alpha_{n}$ according to the weight
  $w_{C,\alpha}/\sum_\beta w_{C,\beta}$. Note that this weight is implicitly
  generated by the clustering algorithm of the LO merging. The same weight
  needs to be included here, since it is not part of the
  $\alpha_S$-expansion. In addition to the clustered real emission, the
  contributions from subtraction terms need to be constructed. Since in the CS
  dipole subtraction, the subtraction dipoles are evaluated according to the
  real emission phase space points, but observables are evaluated at one of
  the underlying Born phase space points, the subtraction terms can be
  calculated alongside the clustered real emission contribution.  We generate
  this contribution as follows: $\phi_{n+1}$ is clustered randomly to one of
  the underlying $\tilde{\phi}^\alpha_{n}$ (the point at which the dipole term
  $D_{\alpha}$ is calculated) and in addition $\phi_{n+1}$ is clustered with
  the same algorithm used in LO merging.  Only if $\tilde{\phi}^\alpha_{n}$
  coincides with the LO clustering, the real emission point is calculated
  including the subtraction dipole at this point.  Otherwise only the dipole
  $D_\alpha$ is retained. By multiplying the result with the number of dipoles
  we compensate for the random choice of the first clustering.  With this
  strategy we generate the same clustering weights as used in the LO
  merging. Since the dipoles and the integrated counterparts must cancel, the
  at first randomly chosen kinematics $\tilde{\phi}^\alpha_{n}$ now needs to
  be subjected to the definition of the ME region of the process with $n$
  additional legs. If it is contained in this phase space volume, the point is
  accepted and the algorithm proceeds to generate a history for
  $\tilde{\phi}^\alpha_{n}$ and the virtual contributions.
\end{enumerate}
The next region of the real emission phase space to be considered is the
region outside the ME region.  This region is populated in the LO merging by
emissions with at least one parton shower emission. A simple addition of the
real emission contributions in this region would therefore produce a similar
double counting as it is known from the MC@NLO approach. In order to solve
this problem, the parton shower outside the ME region needs to be expanded in
$\alpha_S$ and subtracted from the real emission contribution in this region.
To this extent, we consider two further regions:
\begin{enumerate}[label=(\Alph*)]
\setcounter{enumi}{1}
\item The {\bf differential PS-region}: This region of phase space is
  populated by the transparent veto algorithm $\tPS$. The real emission
  contribution needs to be subtracted by a shower approximation above the
  shower cutoff. While in the fixed order calculation the real emission is
  subtracted with the dipole expressions that are contributing to observables
  that are evaluated at the reduced phase space point
  $\tilde{\phi}^\alpha_{n}$, the expansion of the shower expansion contributes
  to the same real emission phase space point $\phi_{n+1}$.  Compared to the
  dipole expressions, the $\mathcal{O}(\alpha_S)$ expansion of the shower are
  restricted by ordering in the shower evolution scale.  Only these expansions
  contribute, for which the clustering scale $p^\alpha_T$ associated to the
  clustering $ \tilde{\phi}^\alpha_{n}$ is ordered with respect to the shower
  starting scale evaluated for $ \tilde{\phi}^\alpha_{n}$ if $n=0$ or if any
  of the underlying scales $p^{\alpha,\beta}_T$ associated to any of the
  clusterings $ \tilde{\tilde{\phi}}^{\alpha,\beta}_{n-1}$ is larger than
  $p^\alpha_T$.
\item {\bf Clustered PS-region:} The expansion of the parton shower outside
  the ME region, which was calculated with the real emission in the previous
  region has a counterpart in the no emission probability in the shower
  algorithm.  As for matching algorithms this no emission probability needs to
  be expanded and subtracted from the clustered phase space points $
  \tilde{\phi}^\alpha_{n}$.  These expansions are constructed like the
  subtraction expressions of the previous region but opposite in sign and are
  here calculated with the dipole expressions of the CS dipole
  subtraction. While the expansion is ordered in the evolution scale the
  dipole contributions do not require this ordering, as the integrated
  counterpart, which subtract the IRC singularities of the virtual
  corrections, have no restriction on the analytically integrated expressions.
\end{enumerate}
Within the actual implementation, we have full control over the contributions
from these different regions, along with the Born and virtual contributions,
which are somewhat simpler to handle and will be discussed in detail in the
next section.

\subsection{Virtual contributions}

\begin{algorithm}[b]
\begin{algorithmic}
\Function{PartialAlpha}{$\phi_B,\phi_n,Q_S(\phi_B)$}
\State $w_{\partial\alpha} \gets N_{S}\frac{\alpha_S(Q_S(\phi_B))}{2\pi}K_g$\Comment{See Sec.~\ref{sec:CMW} for $K_g$ and $N_S$}
\State $\phi \gets \phi_B$; $\phi_{+} \gets$ next step to  $\phi_n$
\While{$\phi_+$ not $\phi_n $}
\State $q \gets p_T(\phi_{+}(\phi))$
\State $w_{\partial\alpha}\gets w_{\partial\alpha}+\frac{\alpha_S(Q_S(\phi_B))}{2\pi}(b_0\log(\frac{q^2}{Q_S(\phi_B)^2})+ K_g)$
\State $\phi \gets \phi_+$; $\phi_{+} \gets$ next step to  $\phi_n$
\EndWhile
\State \Return $w_{\partial\alpha}$
\EndFunction
\end{algorithmic}
\caption{ }
\label{alg:partialalpha}
\end{algorithm}

\begin{algorithm}[b]
\begin{algorithmic}
\Function{PartialPDF}{$\phi_B,\phi_n,Q_S(\phi_B)$}
\State $w_{\partial f} \gets 0$; $Q\gets Q_S(\phi_B)$
\State $\phi \gets \phi_B$; $\phi_{+} \gets$ next step to  $\phi_n$
\While{$\phi_+$ not $\phi_n $}
\State $q \gets p_T(\phi_{+}(\phi))$
\For{$i \in \{a,b\}$}
\If{$\eta^i(\phi) \neq \eta^i(\phi_+)$}
\State  $w_{\partial f} \gets w_{\partial f}+ \frac{\alpha_S(Q_S(\phi_B))}{\pi}\log (\frac{q}{Q})\times$
\State   $ \sum_j \Big(\int_0^1 \frac{\dd z}{z} \frac{f_j(\eta^i(\phi)/z,Q_S(\phi_B))}{f_i(\eta^i(\phi),Q_S(\phi_B))}P_{ij}(z) \times$
\State  $ \theta(\eta^i(\phi)-z) - P_{ji}(z)\Big) $
\State $Q\gets q$
\EndIf
\EndFor
\State $\phi \gets \phi_+$; $\phi_{+} \gets$ next step to  $\phi_n$
\EndWhile
\For{$i \in \{a,b\}$}
\If{$\eta^i(\phi_n) \neq \eta^i(\phi_B)$}
\State  $w_{\partial f} \gets w_{\partial f}+ \frac{\alpha_S(Q_S(\phi_B))}{\pi}\log (\frac{Q_S(\phi_B)}{q})\times$
\State   $ \sum_j \Big(\int_0^1 \frac{\dd z}{z} \frac{f_j(\eta^i(\phi_n)/z,Q_S(\phi_B))}{f_i(\eta^i(\phi_n),Q_S(\phi_B))}P_{ij}(z) \times$
\State  $ \theta(\eta^i(\phi_n)-z) - P_{ji}(z)\Big) $
\EndIf
\EndFor
\State \Return $w_{\partial f}$
\EndFunction
\end{algorithmic}
\caption{ }
\label{alg:partialpdf}
\end{algorithm}
Besides the real emission contribution, described in the previous section, the
virtual correction including one loop diagrams contribute to the NLO
corrections.  In a fixed order approach, after choosing a renormalization
scheme for UV divergences and a subtraction scheme for IRC divergences the
calculation is unique up to the (presumably dynamical) scale choice. The
argument of the running couplings of the LO approximation to the cross section
needs to be the same as the renormalization scale used in the calculation of
the virtual amplitudes to ensure the proper cancellation of scale variations
at the NLO. However, the choice of the argument of the running coupling for
the virtual and real corrections can be different as the terms induced by this
ambiguity are of relative order $\mathcal{O}(\alpha_S^2)$ to the LO
calculation.  
Combining parton showers and fixed-order corrections at the NLO is a more
complicated task, which is constrained by preserving both the resummation
properties of the shower and the accuracy of the fixed-order calculation, and
the respective ambiguities need to be carefully examined and adjusted to
reflect this aim. While the last section was mainly concerned with the second
and third term in the fixed-order expansion of the unitarized LO merging
described in Eq.~\ref{eq:LOexpansion}, we will now consider exact virtual
contributions to replace the approximate corrections encountered while
unitarizing the LO merging. As in the previous section, where the expansions of
the emission probabilities were subtracted in order to remove double counted
contributions, the aim of this section is to identify the still missing pieces
induced by the parton shower, that need to be subtracted as they are already
covered by the virtual contribution.  In matching to NLO input, also the
dynamic scale choice incorporated through the clustering and history
reweighting procedure, needs to be taken into account. Specifically the first
term of Eq.~\ref{eq:LOexpansion}, corresponding to the expansion of the
history weights, needs to be subtracted in order to ensure scale compensation
at NLO.  This expansion generates terms such as
\begin{equation}
\prod_i\frac{\alpha_S(q_i)}{\alpha_S(\mu)}=1-\sum_i
b_0\frac{\alpha_S(\mu)}{\pi}\log
\left(\frac{q_i}{\mu}\right)+\mathcal{O}(\alpha_S^2)\label{eq:alphasexpansion}
\ ,
\end{equation}
which, without a corresponding subtraction would spoil NLO accuracy. By
subtraction of the expansion of the history weight we map (single) logarithms
of the shower scales to the invariants relevant in the virtual corrections,
which we expect to be small if the history weighting procedure is able to
capture the gross dynamics of the multiscale processes under consideration.

\begin{algorithm}[b]
\begin{algorithmic}
\Function{PartialSudakov }{$\phi_B,\phi_n,Q_S(\phi_B)$}
\State $w_{\partial\Delta} \gets 0$; $Q\gets Q_S(\phi_B)$
\State $\phi \gets \phi_B$; $\phi_{+} \gets$ next step to  $\phi_n$
\While{$\phi_+$ not $\phi_n $}
\State $q \gets p_T(\phi_{+}(\phi))$
\ForAll{ColourDipole $D_{ij}$ in  $u(\phi, Q)$} 
\State $w_{\partial\Delta} \gets w_{\partial\Delta}+\int_q^Q \dd w^{ij}_{E}(Q)$ \Comment{Emis. prob. $w^{ij}_{E}$  } 
\EndFor
\State $Q \gets q$
\State $\phi \gets \phi_+$; $\phi_{+} \gets$ next step to  $\phi_n$
\EndWhile
\State \Return $w_{\partial\Delta}$
\EndFunction
\end{algorithmic}
\caption{}
\label{alg:partialsudakov}
\end{algorithm}

We will now describe algorithms to calculate the $\mathcal{O}(\alpha_S)$
expansion of the history weights. The history weights consist of three
contributions as discussed in Sec.~\ref{sec:historyweight}. First we have the
$\alpha_S$ ratio, for which the expansion is simply given by the running
coupling. Corrections which have been calculated in the $\overline{\text{MS}}$
scheme require an additional compensating term for the CMW\footnote{Catani, Marchesini and Webber in \cite{CMW}} prescription used
in the shower\footnote{In the CMW scheme parts of the NLO corrections for
  simple color singlet production and decays are already covered in the choice
  of $\Lambda_{QCD}$.}, cf. the discussion in Sec.~\ref{sec:CMW}.

Alg.~\ref{alg:partialalpha} is generating the contribution in
Eq.~\ref{eq:alphasexpansion} together with the corrections for using the CMW
scheme in the parton shower evolution. 

The second contribution to the history
weight is the PDF ratio. This contribution can be obtained using the
$\mathcal{P}$ operator of the subtraction formalism,
Alg.~\ref{alg:partialpdf}.

The third contribution in the expansion of the history weights is the
expansion of the Sudakov form factors. As the computation of the Sudakov
exponent is performed by sampling the exponent with MC methods, it is possible
to use the same routines to evaluate the first-order expanded form factor. In
contrast to the Sudakov sampling the scale of $\alpha_S$ is kept fix and
the PDF ratio is evaluated at the scale of the last emission.\footnote{Note
  that this is of the same level of accuracy for an NLO calculation.}

The virtual contribution of the NLO correction $d\sigma^V_n$, contains,
\begin{itemize}
\item the interference of the loop diagrams and the tree level contribution, 
\item the UV-counterterms, 
\item the integrated dipole counter terms together with the collinear
  counterterm from PDF renormalization as contained in $\mathbf{I},\mathbf{P}$
  and $\mathbf{K}$ operators of the CS subtraction formalism
  \cite{Catani:1996vz,Catani:2002hc}.
\end{itemize} 
$d\sigma^V_n$ depends on the unphysical renormalization and factorization
scales $\mu_R$ and $\mu_F$.  For higher multiplicities we however use a scale
given by the shower history {\it e.g.} in $\alpha_S$-ratios, which (see the
arguments above) might not be able to ensure scale compensation through NLO
with an arbitrary choice of $\mu_{R,F}$. We are therefore led to calculate
the virtual contributions at a scale of $Q_S(\phi_B)$ to ensure scale
compensation. In practice we calculate the virtual contributions as outlined
in Alg.~\ref{alg:virtual}.

\begin{algorithm}[b]
\begin{algorithmic}
\Function{VirtualContribution}{$\phi_n$}  

\State $\phi_B\gets$\Call{Cluster Algorithm}{$\phi_n,\rho,Q$}  \Comment{Alg.~\ref{alg:clustering}}
\If{$\theta_C(\phi_B)=0$} 

\Return 0\Comment{Sec.~\ref{sec:Cuts}}\EndIf 

\If{$\phi_B=\phi_n$} \Comment{No clustering}

\Return $\dd\sigma^V_{n,i}(\phi_n,Q_S(\phi_n))$ \Comment{see Eq.~\ref{eq:startscale}} \EndIf 

\State $w_{H\;}\gets$  \Call{HistoryWeight}{$\phi_B,\phi_n,Q_S(\phi_B)$}  \Comment{Alg.~\ref{alg:historyweight}}
\State $w_{\partial \alpha}\gets$ \Call{PartialAlpha}{$\phi_B,\phi_n,Q_S(\phi_B)$}  \Comment{Alg.~\ref{alg:partialalpha}}
\State $w_{\partial \Delta}\hspace{-0.5mm}\gets$ \Call{PartialSudakov}{$\phi_B,\phi_n,Q_S(\phi_B)$}  \Comment{Alg.~\ref{alg:partialsudakov}}
\State $w_{\partial f}\gets$ \Call{PartialPDF}{$\phi_B,\phi_n,Q_S(\phi_B)$} \Comment{Alg.~\ref{alg:partialpdf}}

\State \Return 
\State  \hspace{2.5mm}Rnd($2u(\phi_{n},q_{n})$,$-2u(\phi_{n-1,\alpha H},q_{n-1})$)
\State  \hspace{2.5mm}$w_H \big[\dd\sigma^V_{n}(\phi_n,Q_S(\phi_B))$
\State  \hspace{2.5mm}$ -(w_{\partial \alpha}+w_{\partial \Delta}+w_{\partial f})\dd\sigma^B_{n}(\phi_n,Q_S(\phi_B))\big]$

\EndFunction
\end{algorithmic}

\caption{ }
\label{alg:virtual}
\end{algorithm}

\subsection{Unitarising the NLO corrections}
\label{sec:UnitarisingNLO}

Having the framework for unitarized LO merging at hand, along with the
subtractions required to match NLO calculations for individual multiplicities,
the unitarization of NLO merged cross sections is now straightforward: In
order to ensure that the dipole subtraction terms cancel, their integrated
contribution and the differential counterparts to these need to be subject to
the same reweighting procedure at the underlying Born phase space point
encountered. The virtual contributions are treated by the algorithm the same
as the respective Born contributions.

In the case of real emission contributions, we consider a clustered phase
space point $\phi_{n-1}^\alpha$ just as a Born-like contribution. We choose the
index $\alpha$ randomly and reweight with the number of possibilities, and
therefore effectively integrate the dipole contribution over the full phase
space to match their integrated counter parts.  Secondary clusterings from the
point $\phi^\alpha_{n-1}$ to $\phi^{\alpha,\beta}_{n-2}$ above the merging
scale will be considered as a virtual contribution, and their subtraction
realizes the unitarization of NLO corrections as outlined in
\cite{HerwigMerging}.

\subsection{Ambiguities in $\mathbf{\boldsymbol{\alpha}_S}$ expansions}
\label{sec:varyingexpansion}

Combinations of parton shower calculations and fixed order corrections like
matching and merging are subject to a number of ambiguities. While there are 
both technical and algorithmic details that can turn out to be numerically
significant, the by far most striking ambiguity is in terms which are beyond
the control of the fixed order input at hand. Specifically the weights applied
within the NLO merging algorithm can be adjusted already at one higher order
in the strong coupling, which is beyond both the fixed-order and parton-shower
accuracy. We address this ambiguity in detail and consider a number of
different schemes of expanding the history weights to fixed order:
 
Each history weight is composed as a product of factors $w^i_X$, where $X\in
\{\alpha_S,f,\Delta\}$ for every step $i$ of the history. All of the factors
depend on the scales $\{q^i,q^{i-1},Q_S ... \}$ and can be written as
$w^i_X=1+\alpha_S(\mu)w^i_{\partial_X}(\mu)+\mathcal{O}(\alpha^2_S(\mu))$. The
result of Alg.~\ref{alg:virtual} is calculating this expansion of the history
weights together with the contribution from the LO merging weights to obtain:
\begin{equation}
d\sigma^B_{n} \prod_i \left(\left[1 - \sum_{X } \alpha_S
  w^i_{\partial_X}\right]\prod_{X } w^i_X \right)\label{eq:critical1} \ ,
\end{equation}
where the LO case is reproduced in the first term of each summand.  Expanding
in $\alpha_S$ the expression collapses to
\begin{equation}
 d \sigma^B_{n}(1+\mathcal{O}(\alpha^2_S))\label{eq:critical2} \ .
\end{equation}
With this expansion at hand, we can remove the $\mathcal{O}(\alpha_S)$
contribution of the subsequent showering, which will be replaced by the
respective correction from the NLO calculation.

\begin{figure}
\centering
\scalebox{.75}{\includegraphics{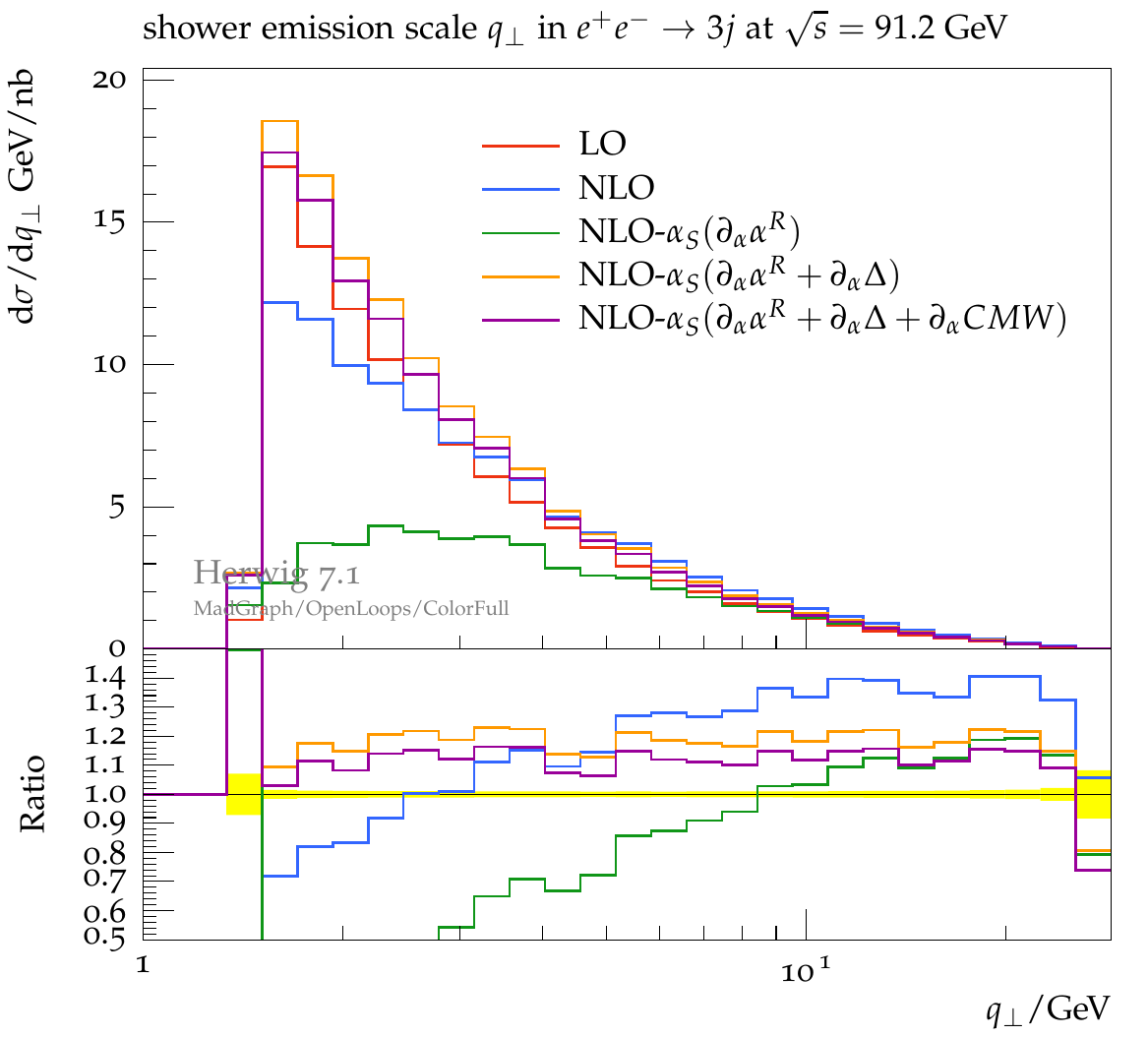}}
\caption{\label{fig:expansions}NLO corrections and shower expansion terms in
  the case of the shower emission scale in $e^+e^-$ to jets events. See text
  for details.}
\end{figure}

Eq.~\ref{eq:critical1} is, however, not the only possibility to achieve a
clean NLO correction (cf. Eq.~\ref{eq:critical2}). A different choice is given
by
\begin{equation}
d\sigma^B_{n} \prod_i \left(\left[\prod_{X } w^i_X  - \sum_{X } \alpha_S
  w^i_{\partial_X}\right] \right) \ .
\end{equation}
This leads us to consider the schemes listed in Tab.~\ref{tab:schemes}.
\begin{table}
  \centering
\begin{center}
  \begin{tabular}{ll}
    \hline\hline
    Scheme & Expression \\ \hline
     0  & $\prod_{X } w^i_X $ (no hist. exp., not NLO correct)\\     
    1 & $\left[1 - \sum_{X } \alpha_S w^i_{\partial_X}\right]\prod_{X } w^i_X $ \\
    2 & $\left[1  -  \alpha_S w^i_{\partial_\alpha}- \alpha_S w^i_{\partial_f}   -   \alpha_S w^i_{\partial_\Delta}/  w^i_\alpha  \right] \prod_{X } w^i_X$ \\ 
    3 & $\left[1 - \sum_{X } \alpha_S w^i_{\partial_X}/  w^i_\alpha\right]\prod_{X } w^i_X $ \\ 
    4 & $\left[1  -  \alpha_S w^i_{\partial_\alpha}/  w^i_\alpha - \alpha_S w^i_{\partial_f}   -   \alpha_S w^i_{\partial_\Delta} \right] \prod_{X } w^i_X$ 
        \\ \hline\hline
  \end{tabular}
\end{center}

\caption{Expansion schemes considered for the treatment of terms beyond
  NLO. \label{tab:schemes}}
\end{table}
Scheme 0 is not NLO accurate as the expansion contains $\mathcal{O}(\alpha_S)$
expressions of the same order as the $\mathcal{O}(\alpha_S)$ corrections of
the NLO contribution. We merely use this for illustrative proposes. Scheme 1
originates from Eq.~\ref{eq:critical1}. Since the history weight $w^i_\alpha$
is in general larger than one, scheme 2 suppresses the contribution which
originates from expanding the Sudakov form factors. We expect these terms, to
first order, to be proportional to $\alpha_S \log^2(q_1/q_2)$ and as such the
dominating part of the history weight expansion for large scale
separations. 
Scheme 3 is similar to scheme 1 but suppresses the expansion by keeping the $\alpha_S$
weight fixed. Scheme 4 was introduced as the opposite to \mbox{scheme 2}. Here we suppress the 
negative  contribution of the $\alpha_S$ ratio expansion, rather than the positive 
Sudakov expansion. Further schemes could be constructed by rearranging the expressions in a
way that keeps the $\alpha_S$ expansion fixed.

 We have illustrated the effect of these contributions in
Fig.~\ref{fig:expansions}, using the relevant quantity of the emissions'
transverse momentum, comparing the subtractions for a fixed order NLO prediction of the
three jet cross section. Taking into account all of the subtractions is
resulting in a $K$-factor close to one from leading order to the adjusted
next-to-leading order. While one would assume that such a prescription is
guaranteeing a somewhat reasonable behaviour it is hard to claim which scheme
should be considered optimal, and such statements would require a (process
specific) comparison to NNLO corrections.

\section{CMW Scheme, Scale Variations and Notation}
\label{sec:scales}

In this section we discuss additional input to the algorithm and the
freedom of choosing schemes beyond the accuracy of our merged cross
section calculation. One of the choices is the running and input value
of the strong coupling constant which we discuss in detail in
Sec.~\ref{sec:CMW}. We then elaborate on scale variations which we have
chosen to assess the theoretical uncertainty of the merged calculation,
followed by a discussion about the functional choice of the merging
scale cut. At the end of this section we introduce a short notation for 
merged simulations.

\subsection{The CMW scheme}
\label{sec:CMW}

The value and the treatment of the strong coupling constant $\alpha_S$ in
parton showers is delicate. In some implementations the value of $\alpha_S$ is
directly chosen from the PDG or from the PDF set in use.  Other approaches use
$\alpha_S$ as a tuning parameter. In \cite{CMW} it is shown that leading parts
of the higher order effects for specific processes can be resummed by using a
modified value of $\alpha_S$,
\begin{equation}
  \alpha'_S(q)=\alpha_S(k_g q)\approx \alpha_S( q)\left(1+K_g \frac{
      \alpha_S( q)}{2\pi}\right)\label{eq:CMWalpha}\ ,
\end{equation} 
where 
\begin{equation}
K_g=C_A\left(\frac{67}{18}-\frac{1}{6}\pi^2\right)-\frac{5}{9}N_F\;
\end{equation}
and $k_g=\exp(-K_g/b_0)$ with $b_0=11-2/3N_F$.  

This modification is usually referred to as the CMW \cite{CMW}
 or Monte Carlo scheme.  Using the modified version of $\alpha_S$ in the
merging algorithm can help to describe data but changes the scheme used
earlier to calculate the MEs. At LO the effect is beyond the
claimed accuracy, but by merging NLO cross sections, the prior estimate
of higher order corrections needs to be subtracted to match the scheme
used in the ME calculation.  For every $\alpha_S$ ratio in the shower
history reweighting of the form $\alpha'_S(q)/\alpha_S(\mu_R)$ the NLO
merged algorithm must subtract the $\alpha_S$ expansion
$\dd \sigma_B K_g\alpha_S/2\pi$.  Note that this additional linear term
and scaling $q$ with the factor $k$ in Eq.~\ref{eq:CMWalpha}
produce the same $\mathcal{O}(\alpha_S)$-expansion. Both schemes are
implemented and can be studied using Herwig.  A detailed study of
uncertainties related to choices in this scheme will be published
elsewhere. 

\subsection{Scale variations }
\label{sec:scalevariations}

All scales that have been used to evaluate the predictions can be varied
to estimate theoretical uncertainties.  Usually constant factors are
used to alter the scales up and down. We include five different factors.
At first we have the variation $\xi_{R/F}^{ME}$ of the production
process. We call the renormalization and factorization scales used here
$\mu_{R/F}^{ME}$, weighted with $\xi_{R/F}^{ME}$.  They apply to the
production process and -- if a full shower history is found by the
clustering algorithm -- the reweighting of the history that restores the
weights for the assumed production process.  Additional emissions are
produced with the scale of the shower splitting.  These scales are
$\mu_{R/F}^{\PS}$ and are altered by factors $\xi_{R/F}^{\PS}$. The last
scale we need is the shower starting scale $Q_{S}$ varied by
$\xi_{Q}$.  While $\xi_{R/F}^{\PS}$ apply to any shower emission the
scaling factor $\xi_{Q}$ is only chosen for the initial emission.

If we assume e.g.\ two emissions and a shower history that reaches to
the production process, the weight needed for the
$\alpha_S$-reweighting is
\begin{equation}
  w_{\alpha_S}^H=\frac{\alpha'_S(\xi_{R}^{\PS}q_2)}{\alpha_S(\mu_R)}\frac{\alpha'_S(\xi_{R}^{\PS}q_1)}{\alpha_S(\mu_R)}\left(\frac{\alpha'_S(\xi_{R}^{ME}\mu_R)}{\alpha_S(\mu_R)}\right)^{N_S} ,
\end{equation}
where $\mu_R$ is the scale used for the ME calculation and $N_S$ is the
order of $\alpha_S$ in the seed process.  $\alpha'_S$ is the possibly
modified version of $\alpha_S$ due to the inclusion of the CMW scheme as
described in above in Sec.~\ref{sec:CMW}.

Accordingly, the PDF ratios read  
\begin{align}
w_{f}^H =&
\frac{f^{1,2}(\phi_{n+2},\xi_{F}^{\PS}q_2)}
		{f^{1,2}(\phi_{n+2},\mu_F)}
\frac{f^{1,2}(\phi^{\alpha}_{n+1},\xi_{F}^{\PS}q_1)}
	   {f^{1,2}(\phi^{\alpha}_{n+1},\xi_{F}^{\PS}q_2)}\nonumber\\
  &\times
\frac{f^{1,2}(\phi^{\alpha\beta}_{n},\xi_{F}^{ME}\mu_F)}
       {f^{1,2}(\phi^{\alpha\beta}_{n},\xi_{F}^{\PS}q_1)}\ .
\end{align}
Here, $f^{1,2}(\phi_{n+2},\mu_F)$ are the weights that have been used to
calculate the ME.

One could also construct a scenario where the shower should make use of
a different PDF set than the ME calculation, e.g.\ LO for the former and
NLO PDFs for the latter.  Then the part in the ratios that are rescaled
with $\xi_{F}^{\PS}$ belong to the shower related PDFs and the one with
$\xi_{F}^{ME}$ needs to be used in the ME related set.  In NLO merged
samples this difference between the two sets needs to be corrected by
subtracting the difference of the two sets in order to preserve NLO
accuracy.

\subsection{Functional form of the merging scale cut}
\label{sec:smearing}

At the phase space boundary of the ME region the difference between
shower and ME corrected contributions can lead to
discontinuities of the order of the difference between the ME and the
shower approximation, further reduced by the Sudakov suppression.  While
the ``jumps'' at the boundaries are expected to be less than about
$10\,\%$, we still include a parameter to smear these effects in order to
reduce the possibility that automated algorithms will misinterpret the
effects as physically relevant.  This smearing is performed by choosing
the merging scale $\rho_s$ on an individual phase space point level
around the central value $\rho_C$ as
\begin{equation}
\rho_s=\rho_C\cdot(1+(2\cdot r-1)\cdot \delta)\ ,
\end{equation} 
where $r$ is a random number in the interval $[0,1]$ and $\delta$ can be
chosen in the range $[0,0.2]$. The chosen scale $\rho_s$ is then fixed for 
the duration of the full event.

\subsection{Notation}
\label{sec:notation2}
The results we present for the merging of different multiplicities are
labeled according to the number of tree- and one-loop MEs
used. Speaking of LO and NLO is inappropriate in most cases, as e.g.\
the notion NLO not only refers to the calculation itself but also very
strongly to the observable under consideration.  Hence, we try to be
very explicit in labeling every single multiplicity of \emph{additional}
parton emissions.  Let us consider the production of a $Z^0$ boson with
additional jets as an example for any final state $\Phi$ of the Born
process under consideration.  In LO-merging we add tree-level matrix
elements for the additional partonic multiplicities, such that we write
\begin{displaymath}
  Z(0,1,2,\ldots n) 
\end{displaymath}
with $n \geq 0$ for the merging of MEs with 
\begin{displaymath}
  Z^0+0, Z^0+1, \ldots, Z^0+n
\end{displaymath}
additional partons. 

When merging higher order MEs as well, we denote every
multiplicity $n$ for which, in addition to the tree-level matrix
elements, we also consider one-loop corrections with an extra $*$ as
$n^*$.  Hence, as a fairly general example, we use
\begin{displaymath}
  Z(0^*, 1^*, 2, 3) 
\end{displaymath}
to denote $Z^0$ production with up to 3 additional parton emissions,
where we also use one-loop MEs for $Z^0+0$ and $Z^0+1$
parton processes.  The special case $Z(0^*, 1)$ of this notation would
describe the ``matching through merging'' limit of our merging approach.
It has the same level of accuracy for any observable as the conventional
ME plus parton shower matching algorithms at NLO.  

It should be clear that the accuracy of the above mentioned example
$Z(0^*, 1^*, 2, 3)$ depends on the observable we consider.  For the
given example the inclusive cross section or differential cross section
of the hardest jet's transverse momentum would be described at NLO
level, while the transverse momentum of the second jet and third jet
would be described at LO only. The fourth and higher jets would
certainly only be described at the LLA level of the parton shower.  We
may wonder about the accuracy of the inclusive cross section in this
case.  Albeit formally at the NLO level, the $Z(0^*, 1^*, 2, 3)$ sample
undoubtfully contains more perturbative information than the
$Z(0^*, 1)$ which is the smallest sample that contains an NLO
description of the inclusive cross section.  The former case contains
almost all ingredients of the NNLO, except the two-loop virtual
contributions, which are only represented by NLO plus leading logarithms
from the parton shower. 

\section{Sanity Checks}
\label{sec:sanitychecks}

To validate the merging algorithm several sanity checks have been
performed.  For simple processes we check the Sudakov suppression from
our implementation against an independent Mathematica
implementation. Further we can check that the subtraction of the
real emission contribution is performed in an IR safe way. To validate
the weights of the shower history, we replace a LO ME by the
corresponding dipole expression and expect similar results as we would
with pure shower emissions.  Since the merged cross section is corrected
by the algorithm for hard emissions in the ME region, soft physics
should be hardly affected by the algorithm. A quantity sensitive to soft physics in Herwig is 
the cluster mass spectrum, which will be discussed.  
The introduction of an
auxiliary cross section can provide a reduction of events with negative
weights.  In the last part of this section we compare the various
schemes introduced in Sec.~\ref{sec:varyingexpansion}.

\begin{figure}[t]
  \centering
  \scalebox{.7}{\input{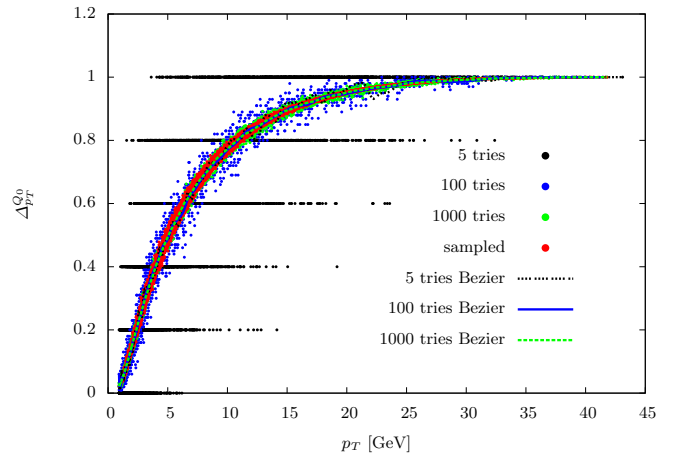}}
  \caption{Comparison of the implementation of a Sudakov form factor for
    a $q\bar{q}$ pair with $Q_0=91.2\,\text{GeV}$. The black, blue and
    green points correspond to \mbox{CKKW-L} type Sudakov suppression, where
    the number of no emissions above the scale $p_T$ is counted. The red
    points are evaluated Sudakov suppressions by directly sampling the
    exponent with a MC error of 5 \%.  The three Bezier lines, which are in perfect 
    agreement are produced by the corresponding gnuplot averaging procedure.}
  \label{fig:SudakovCKKWL}
\end{figure}%

\subsection{Sudakov Sampling}

An important ingredient of merging matrix elements of various jet
multiplicities is the handling of the Sudakov suppression of the higher
jet multiplicities.  Here various schemes have been introduced in the
past.  The original CKKW implementation reweights the individual
contributions by analytical suppression factors. The \mbox{CKKW-L} approach
uses trial emissions to veto events producing emissions into the ME
region of the higher jet multiplicity.  In the implementation
described here the Sudakov suppression is sampled for each step of the
history according to the contribution the shower uses to sample the next
emission. This approach is equivalent to the \mbox{CKKW-L} implementation in
the $N\to \infty$ limit for sampling with precision $\epsilon \to 0$. 
By sampling the suppression factors the
adaptive MC sampling of the cross section has access to the weights
produced by the sampling and can adjust to the weighted
production. Phase space regions with large scale gaps are suppressed
even though the contribution to the cross section without reweighting
would be large.  Fig.~\ref{fig:SudakovCKKWL} shows an example of the
weights produced by the sampling (red dots) compared to a \mbox{CKKW-L} trial
approach with 5/100/1000 trials.  The average of the trials agrees very
well with the sampled contribution.

\subsection{Subtraction Plots}

\begin{figure}[t]
  \centering
 \scalebox{.45}{\includegraphics{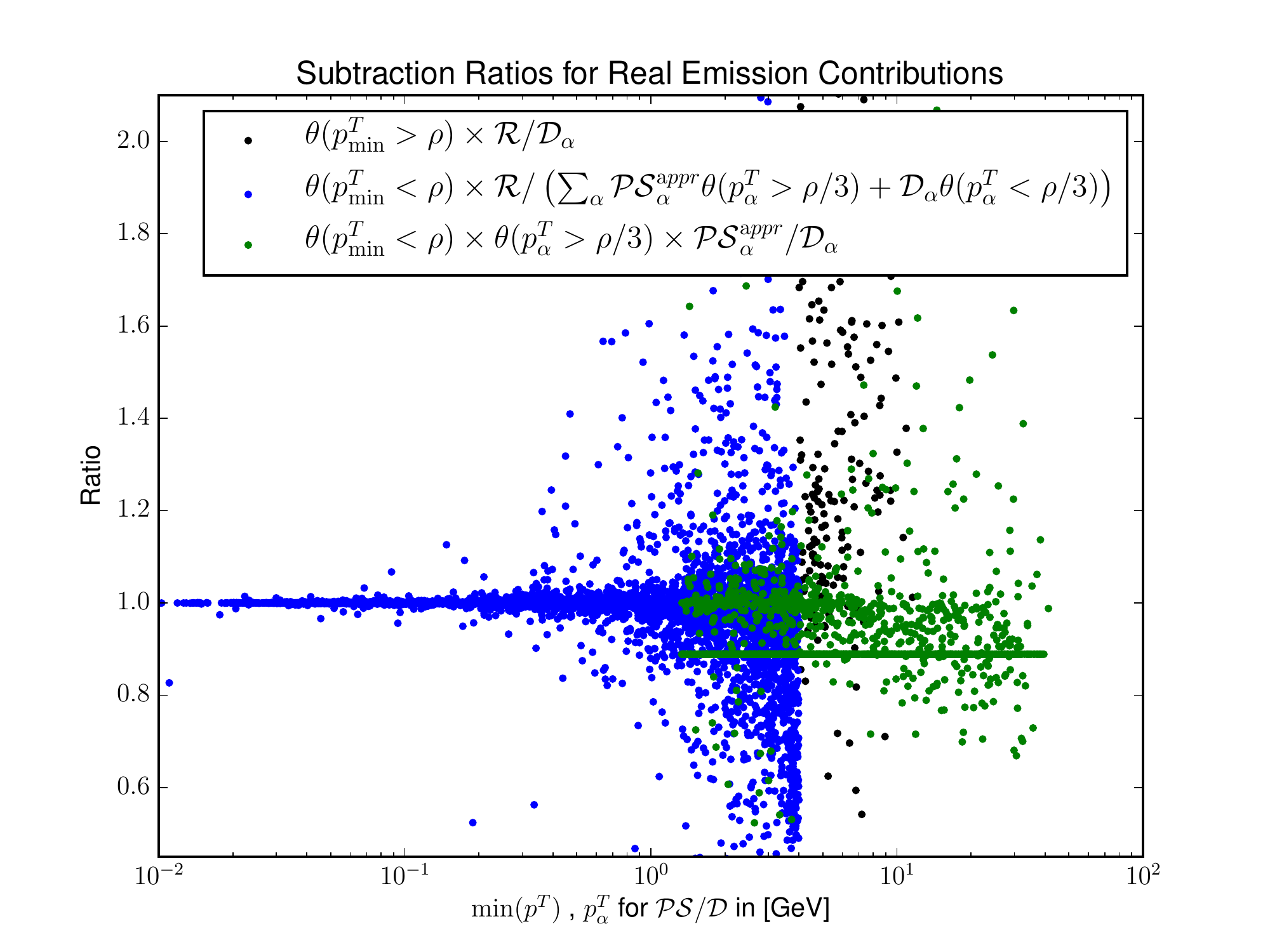}}
 \scalebox{.45}{\includegraphics{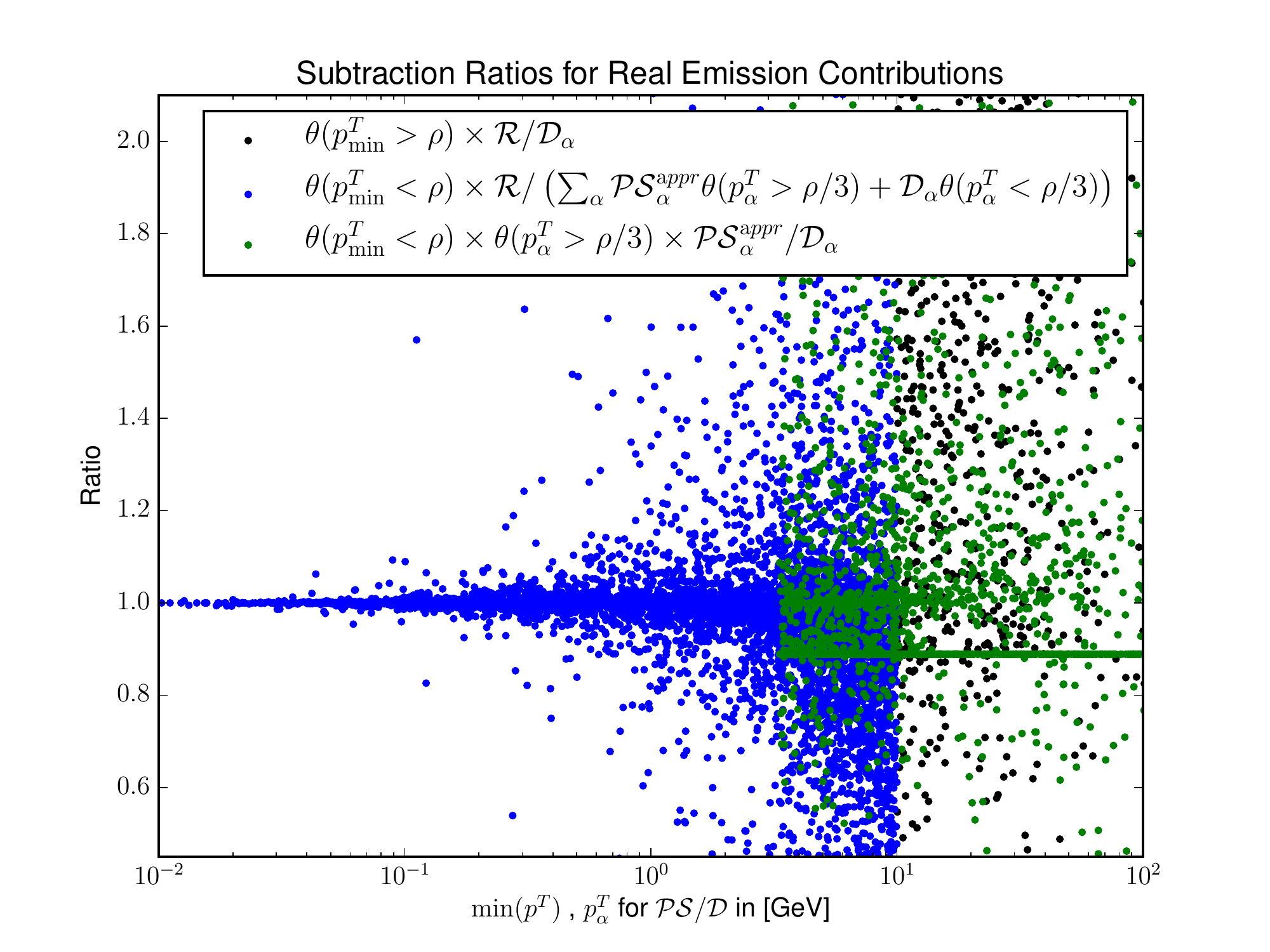}}
   \caption{Subtraction plot for the ``$1^*$'' real emission contribution
     in $ee(0^*,1^*,2)$ and $Z(0^*,1^*,2)$.}
  \label{fig:subtraction}
\end{figure}

\begin{figure}[t]
  \centering
 \scalebox{.7}{\includegraphics{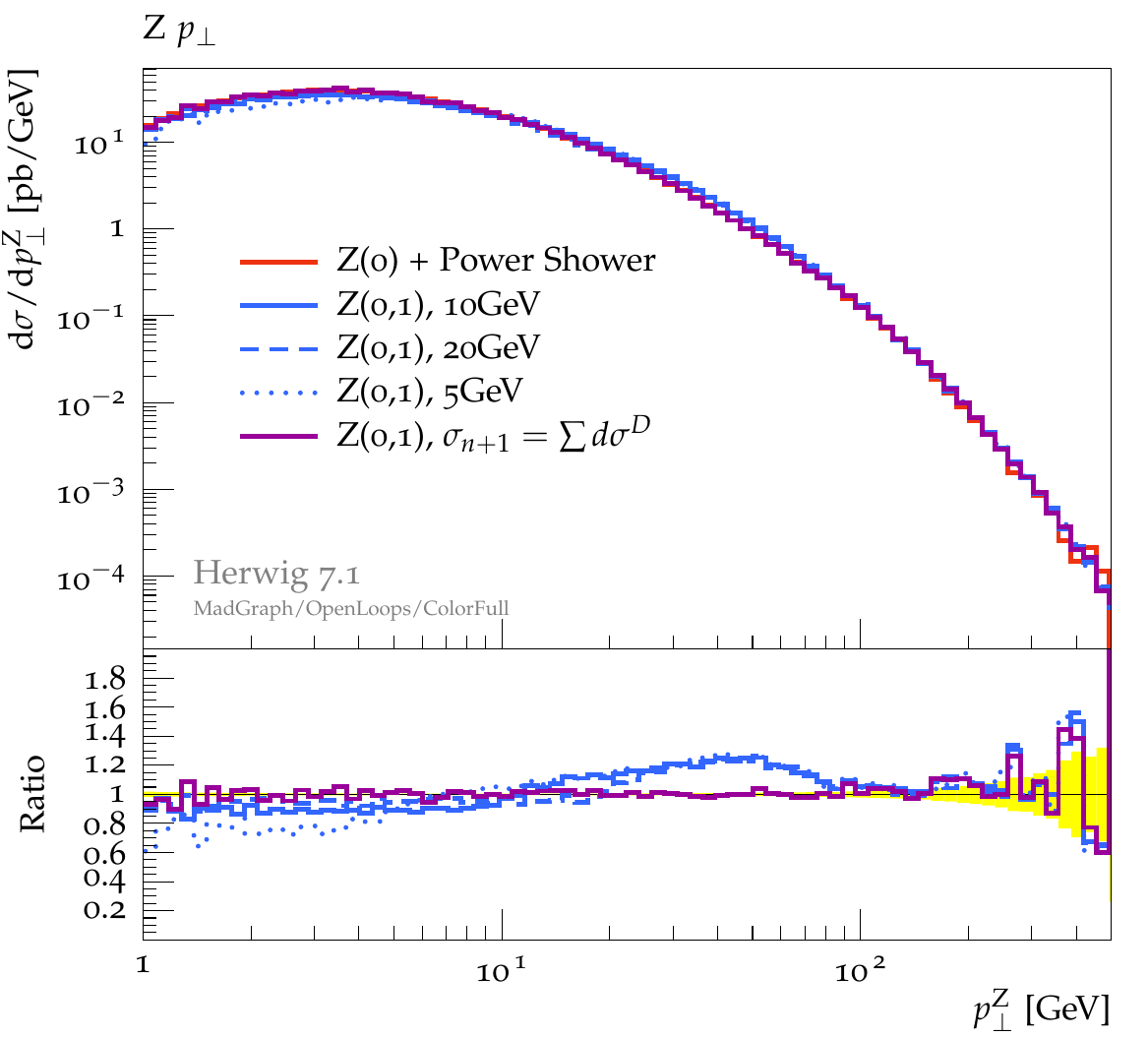}}
 \caption{$Z^0$-production: comparing LO plus power shower (red) with a
   merged sample (blue) and the merged sample where the MEs for the
   $(Z^0+1\,\text{jet})$ samples have been replaced by the sum of the
   corresponding dipoles.}
  \label{fig:DIPforME}
\end{figure}

Performing NLO calculations with MC techniques requires the introduction
of subtraction terms for the virtual and real contributions as described
in Sec.~\ref{sec:subtraction}. In the CS framework the real emission
contributions are subtracted with auxiliary cross sections that cancel
versus the integrated counterparts for the virtual contributions. 
In the NLO merging algorithm the subtraction for the real emission
contributions is more complicated, see
Sec.~\ref{sec:realemissioncontribution}.  
The expansion of the LO merging up to $\mathcal{O}(\alpha_S)$ 
produces similar subtraction terms.  There are three different phase space
regions, the ME-region (A), the differential PS-region (B) and the
clustered PS-region (C) as described in
Sec.~\ref{sec:realemissioncontribution} which can be checked for
subtractive properties.  

In Fig.~\ref{fig:subtraction} we show ratios of real emission
contribution, dipole expressions and shower approximations for the real
process with four partons in the merged simulation of
$e^+e^-\to q\bar q$ merged with two NLO corrections.  The black points
represent the ratios for phase space region (A)  in \ref{sec:realemissioncontribution}, 
namely the phase space region that provides splitting scales that are all above the merging
scale.  In green we plot the ratio of shower approximation over the
dipole contribution for points that are not in the ME region (B) but above a
IR cutoff\footnote{For numerical stability we choose $\rho/3$ as a dynamical cut for the separation between the fully dipole subtracted and shower approximated subtracted region.}. 
The blue points represent the ratio of real emission
contribution with respect to the sum of dipoles if the lowest scale is
below the IR cutoff or the sum of the shower approximation if all
clustering scales are above the IR cutoff. This region is also part of the 
region (B) in \ref{sec:realemissioncontribution}.
For points in region (A) we do not expect the ratio to be one as the ratio consists of 
one of the subtraction contributions that needs to be integrated but is not necessarily 
fully subtractive. The green points should cluster at one or a colour factor that shows 
the difference between the large $N$ shower approximation used to mimic the shower action.
The blue points need to converge to one when the minimum clustering scale approaches
zero as the dipole contributions tend to subtract the real emission contribution.
In Fig.~\ref{fig:subtraction} we see the expected behaviour.

\subsection{Simple check for $\mathbf{\boldsymbol{\alpha}_S}$ and PDF ratios and merging scale variation}
\label{sec:DipforME}

\begin{figure}[t]
  \centering
  \scalebox{.75}{\includegraphics{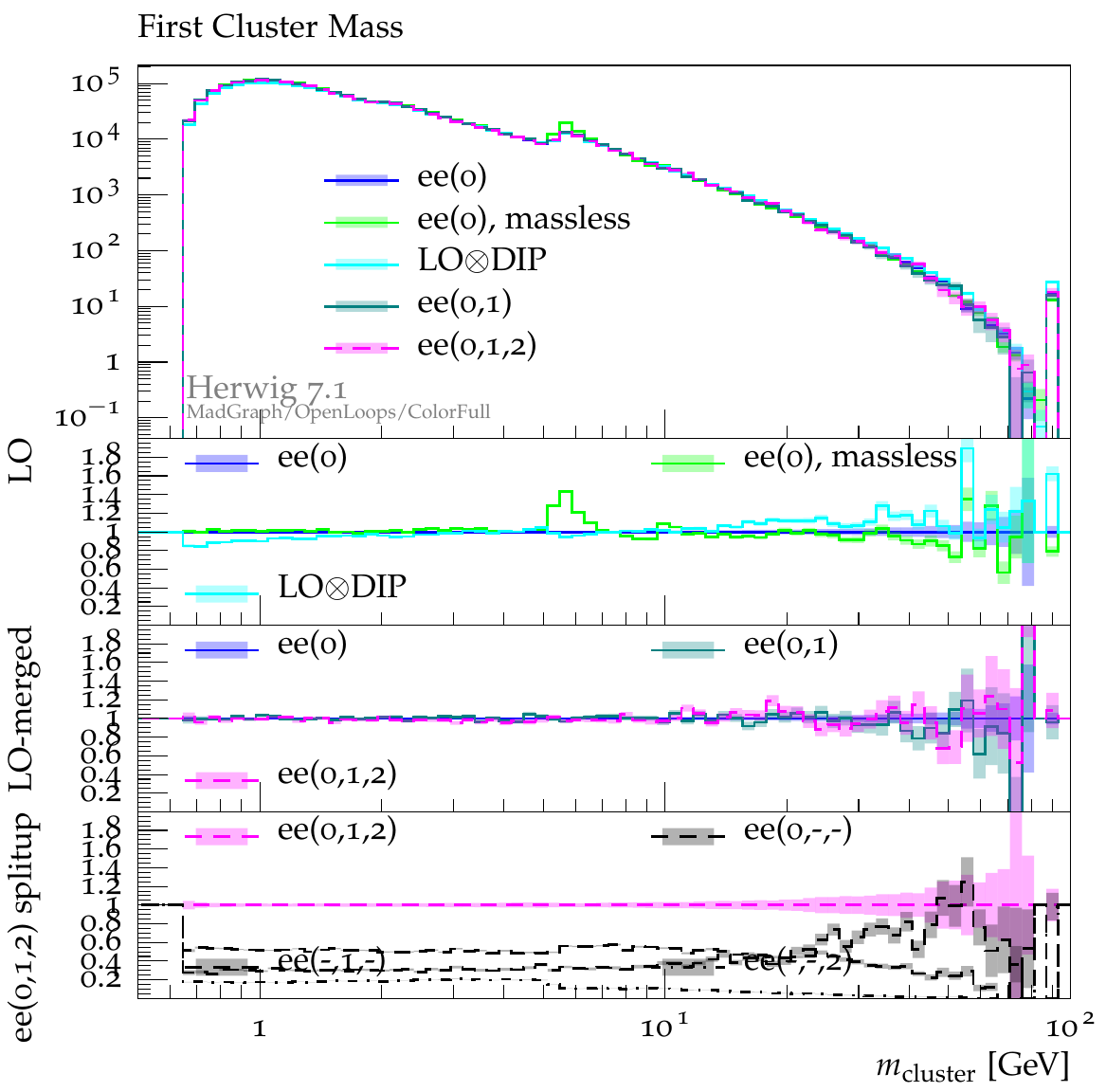}}
  \caption{Mass spectrum of the Herwig cluster hadronization model at
    LEP. We show the mass spectrum of primary clusters. The merging
    hardly modifies the spectrum. In the lowest ratio plot the
    contributions of $ee(1,2,3)$ are split up into multiplicities.
    Note: The mass spectrum of primary colour-singlet clusters is not 
    connected with the clustering of the cluster algorithm described in 
    Sec.~\ref{sec:clustering}. }
  \label{fig:clustermassLO}
\end{figure}

\begin{figure}[t]
  \centering
  \scalebox{.75}{\includegraphics{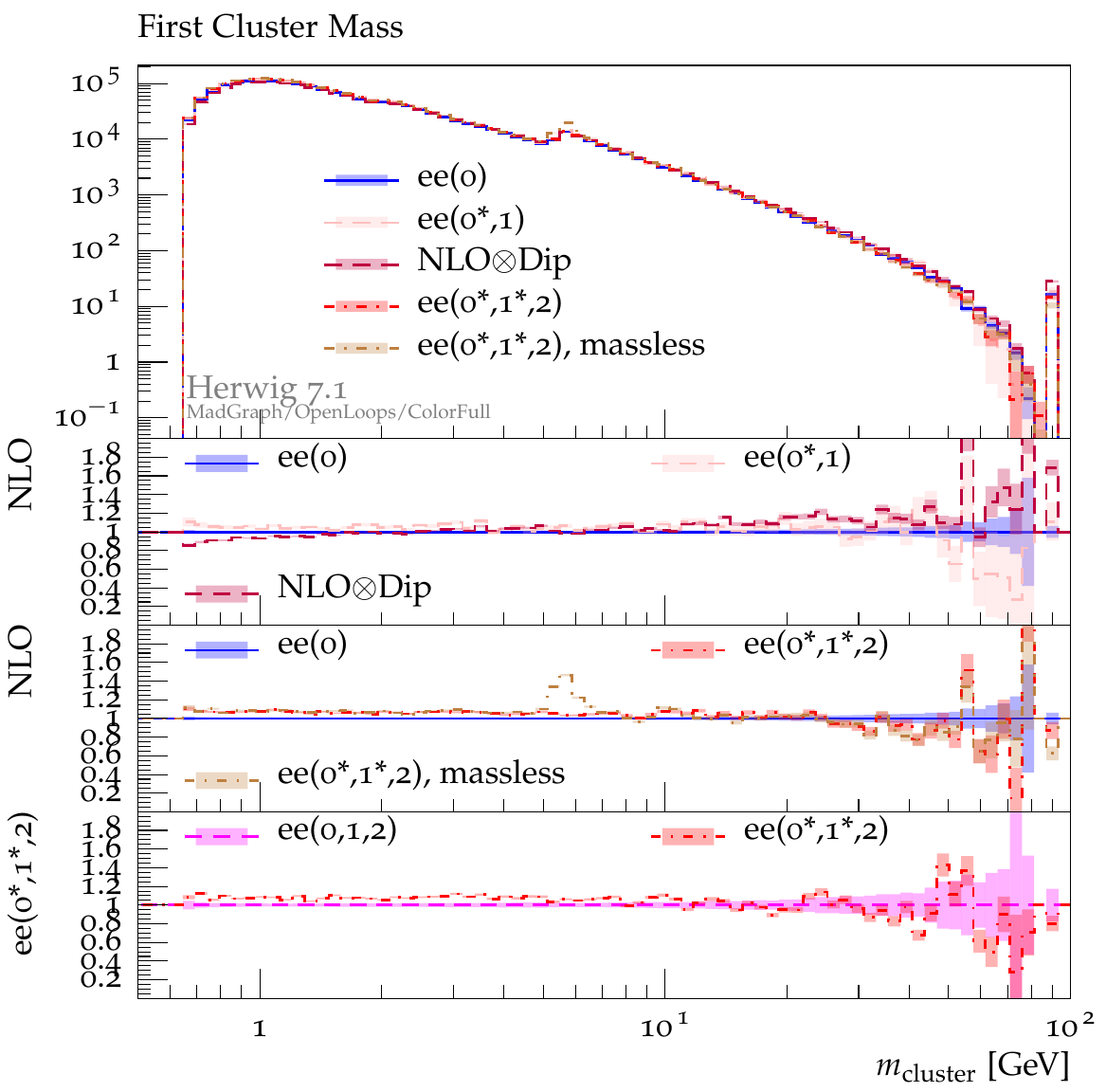}}
  \caption{As Fig.~\ref{fig:clustermassLO} the mass spectrum of primary
    formed clusters is shown. Here the NLO variants in multi jet merging
    and the MC@NLO like contribution are shown.}
  \label{fig:clustermassNLO}
\end{figure}

In order to validate the implementation of $\alpha_S$ and PDF ratios, we
replace the MEs in LO $Z^0$ production, merged with one additional
emission, at the LHC, with the corresponding shower approximation.  In
Fig.~\ref{fig:DIPforME} we show the result with the same parameters as
described in Sec.~\ref{sec:setup}.  The red distribution shows LO $Z^0$
production with a so called power shower, achieved by applying no
restrictions on the emission phase space of the shower and starting the
shower at the highest possible scale $s$.  In purple the same reweights
and clustering properties are used as one would use in merging the two
processes with the difference that we replace the $(Z^0 + 1\,\text{jet})$
sample by the sum of the corresponding dipoles, which corresponds here
to the shower approximation.  The blue lines represent the LO merged
result $Z(0,1)$, see Sec.~\ref{sec:notation2} for notation, with one
additional multiplicity merged to the production process with merging
scales: $10\,\text{GeV}$ (solid), $5\,\text{GeV}$ (dotted) and
$20\,\text{GeV}$ (dashed) and full MEs.  The difference to the pure
shower case or the replacement with the dipole expressions is due to the
MEs.  Similar checks have been performed for light and massive final
state radiation where we find similar agreement between pure parton
shower and merging with shower approximations for the first emission.

\subsection{Cluster Mass Spectrum}

When merging several multiplicities the description of hard emissions is
improved with the information of higher order MEs and approximations
used by the parton shower are rectified in its hard emission
region. Effects of soft physics models are expected to be hardly
affected by the inclusion of these corrections. One of the observables
sensitive to the soft physics is the cluster mass spectrum of Herwig.
We expect the parton shower at the end of the merging to evolve the
final state on the soft end into a cluster spectrum that is almost
unchanged.  The cluster mass spectrum is the essential input for the
hadronization model and the description of e.g.\ hadron multiplicities
would suffer and would require a retuning if this part of the simulation
is strongly affected by the improvements made for hard emissions.  

In Figs.~\ref{fig:clustermassLO} and \ref{fig:clustermassNLO} we compare
the cluster mass spectrum of primary clusters in $e^+e^-$ collisions.
Fig.~\ref{fig:clustermassLO} shows the cluster spectra of LO+PS and LO
merged samples with up to two additional jets. While the massless case
shows a minor spike near the bottom mass, the merging of multiple cross
sections hardy alter the contributions.  In the third ratio plot we
split up the various contributions of the $ee(0,1,2)$, namely the
merging of three multiplicities leading in sum to the full result.
Fig.~\ref{fig:clustermassNLO} compares the same spectrum for NLO merged
contributions and again only the massless showering varies in shape
compared to the massive LO+PS contribution. In conclusion we do not
expect major retuning of the details of the cluster model to be
necessary as a result of the merging procedure.

\subsection{Reduction of Negative Weights}
\label{sec:redNeg}
One potential technical problem of unitarized merging may be the
appearance of negative weights.  Since the higher multiplicities are
clustered, weighted and subtracted, depending on the merging scale,
strong compensation effects and therefore negative contributions are
unavoidable.  A method to reduce the negative contributions is to
introduce an auxiliary cross section which cancels in the full result,
as it is done in NLO calculations.  Introducing such helper weights may
also serve as a strong check for the consistency of the implementation.
In \cite{Nagy:2003tz} a cut on the dipole phase space for NLO
calculations performed with CS dipoles was introduced to speed up the
calculation.  We can use these expressions to get the analytically
integrated dipole phase space above the cuts imposed in
\cite{Nagy:2003tz} and subtract the same regions from the differential
clustered contributions. Hereby one contribution is subtracted from the
clustered higher multiplicity and the compensating piece is added to the
lower multiplicity.  

To achieve the same result, we change the algorithm for LO merging.  The
first backward clustering is made randomly and without phase space
restrictions instead of clustering with the clustering algorithm
described in Sec.~\ref{sec:clusterprob}.  In addition we cluster the
first step with the cluster algorithm of Sec.~\ref{sec:clusterprob} and calculate 
the weight of the LO
cross section only if the same underlying contribution was picked. For
the chosen channel the dipole is calculated if the phase space point is
above the phase space cut.  The sum of dipole and, if picked, clustered
and negative LO contributions is multiplied by the number of dipoles,
which compensates for the random choice of the underlying
configurations. The history weight is then calculated for the dipole up
to the underlying contribution and for the LO contribution with one
additional step.  With these changes we integrate the dipole parts above
the cut and weighted with the history weights to the underlying
configuration.
 
The integrated counterparts, known in the analytic form can now be added
to the points calculated for the $\dd \sigma_{n-1}$ Born
configurations. As these parts are negative the $\dd \sigma_{n-1}$ is
suppressed and parts of the no emission probability produced by
clustering and subtracting of the higher multiplicity are compensated.

Varying the cut not only changes the proportion of negative event
weights, but also checks the framework.  For $e^+e^-$ we checked the
differential contributions and get a reduction of negative weights as
shown in the following table,
\begin{center}
  \begin{tabular}{ l  l  l }
    \hline\hline
    cut as in \cite{Nagy:2003tz}& positive events & negative events \\ \hline
    1 & 6981 & 3019 \\ 
    0.1 & 7218 & 2782 \\ 
    0.05 & 7313 & 2687 \\
    \hline\hline
  \end{tabular}
\end{center}
Note that although the proportion of events with negative weight is
reduced the unweighting of final event samples may still not be improved
much.  The multiplication of the number of dipoles and therefore
enhanced maximum weights may reduce performance.  We therefore only use
this for testing purposes. 

\subsection{Expansion Schemes and Scale Variations}
\label{sec:expansionSchemes}

\begin{figure}[t]
  \centering
\scalebox{.75}{\includegraphics{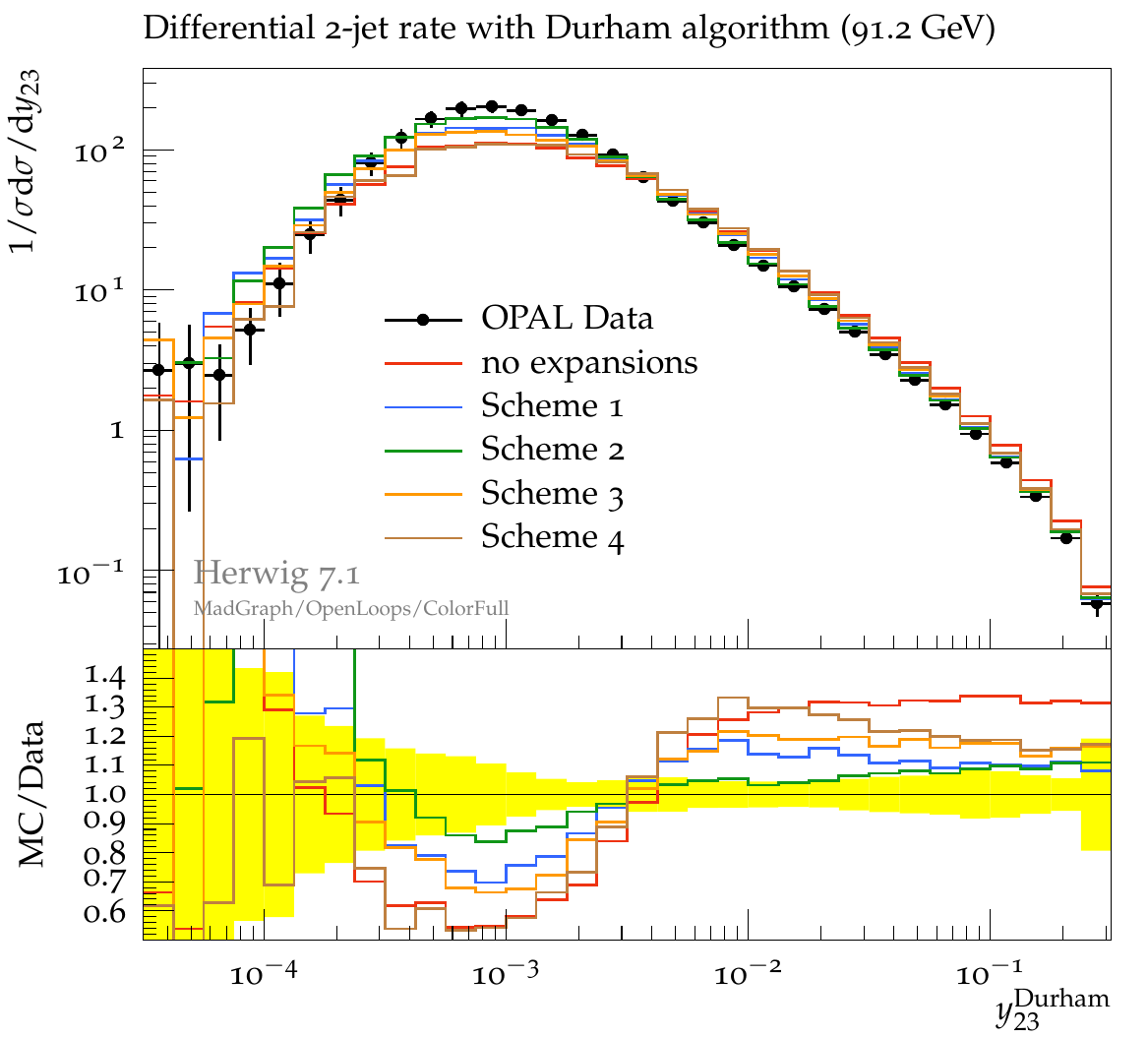}}
\caption{ To show the effects of choosing different schemes of history
  weight expansions as described in Sec.~\ref{sec:varyingexpansion}. We
  compare to data from \cite{Pfeifenschneider:1999rz}. }
  \label{fig:LEPschemesA}
\end{figure}

\begin{figure}[t]
  \centering
\scalebox{.75}{\includegraphics{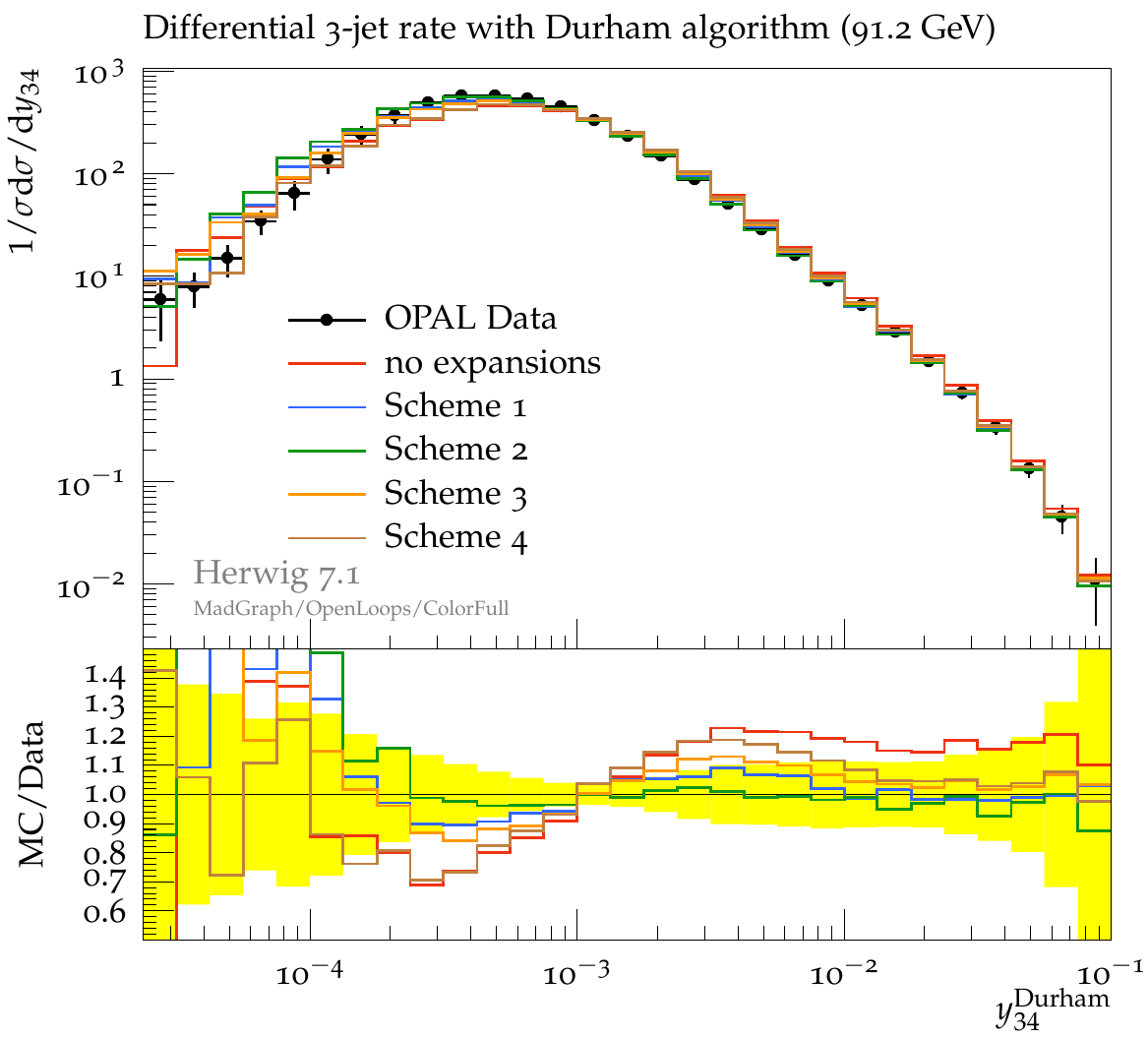}}
\caption{ To show the effects of choosing different schemes of history
  weight expansions as described in Sec.~\ref{sec:varyingexpansion}. We
  compare to data from \cite{Pfeifenschneider:1999rz}.}
  \label{fig:LEPschemesB}
\end{figure}

\begin{figure}[t]
  \centering
  \scalebox{.75}{\includegraphics{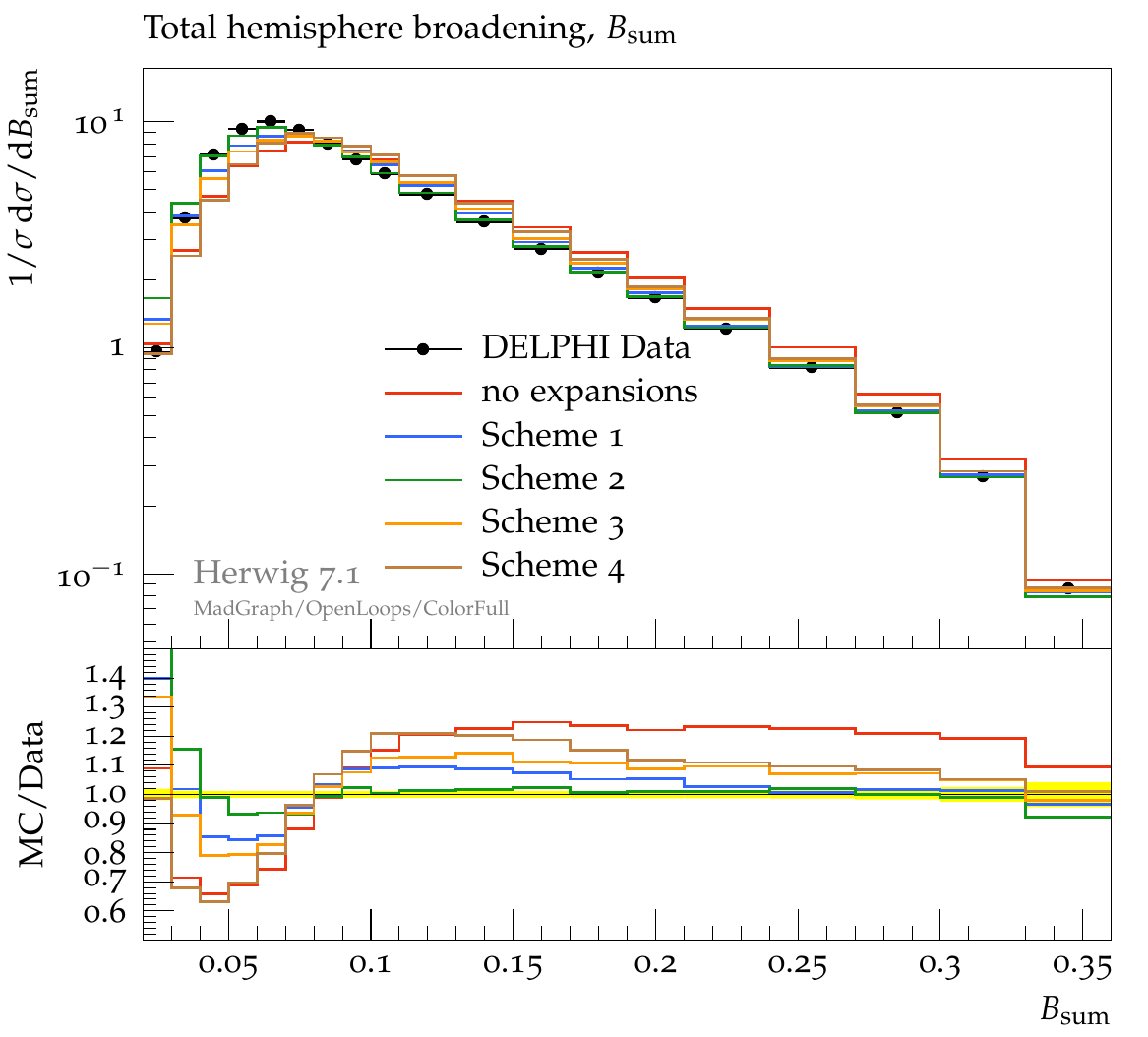}}
  \caption{To illustrate the effects of choosing different schemes of history
    weight expansions as described in Sec.~\ref{sec:varyingexpansion}.
    We compare to data from \cite{Abreu:1996na}. }
  \label{fig:LEPschemesC}
\end{figure}

As described in Sec.~\ref{sec:varyingexpansion} various schemes can be
constructed to include the shower history expansion. In
Figs. \ref{fig:LEPschemesA}, \ref{fig:LEPschemesB} and
\ref{fig:LEPschemesC} we choose a rather small merging scale of
$\rho=4\,\text{GeV}$ and show various choices of these schemes for jet
production at LEP. All lines are variations of the merging of three LO
cross sections, where the production process and the first emission
process are corrected with NLO contributions.  We set
$\alpha^{\overline{{\rm MS}}}_S (M_Z)=0.118$. While the red line,
corresponding to no expansion of the history weights, clearly
overestimates the regions of high emission scales, the schemes with
expansion tend to describe the data measured at LEP better.  In the
simulation the CMW modified strong coupling was used according to
Sec.~\ref{sec:CMW}. The expansion of the shower $\alpha_S$ compared to
the $\overline{\text{MS}}$ coupling suppresses the emission
contribution, which leads to the observed behaviour.  In addition the
NLO correct schemes, that are all of the same accuracy, are performing
rather differently in the comparison to data.  While scheme 1 (all shower
expansions weighted with the full reweights), scheme 3 (all expansions
weighted with Sudakov suppression weights\footnote{Note that at LEP no
  PDF reweighting is needed.})  or scheme 4 ($\alpha_S$-ratio expansion
only weighted with Sudakov suppression) tend to overestimate the data in
the softer region, the choice of scheme 2 (Sudakov expansion weighted
only with Sudakov suppression) is performing well over the full range of
energies. The expansion of the Sudakov form factor is producing a
squared logarithmic contribution, which is suppressed by scheme 2.  This
leads us to make scheme 2 our preferred choice albeit the other schemes
are formally of the same accuracy.  We propose to take all schemes into
account as an evaluation of theoretical uncertainties.

\begin{figure}[t]
  \centering
\scalebox{.75}{\includegraphics{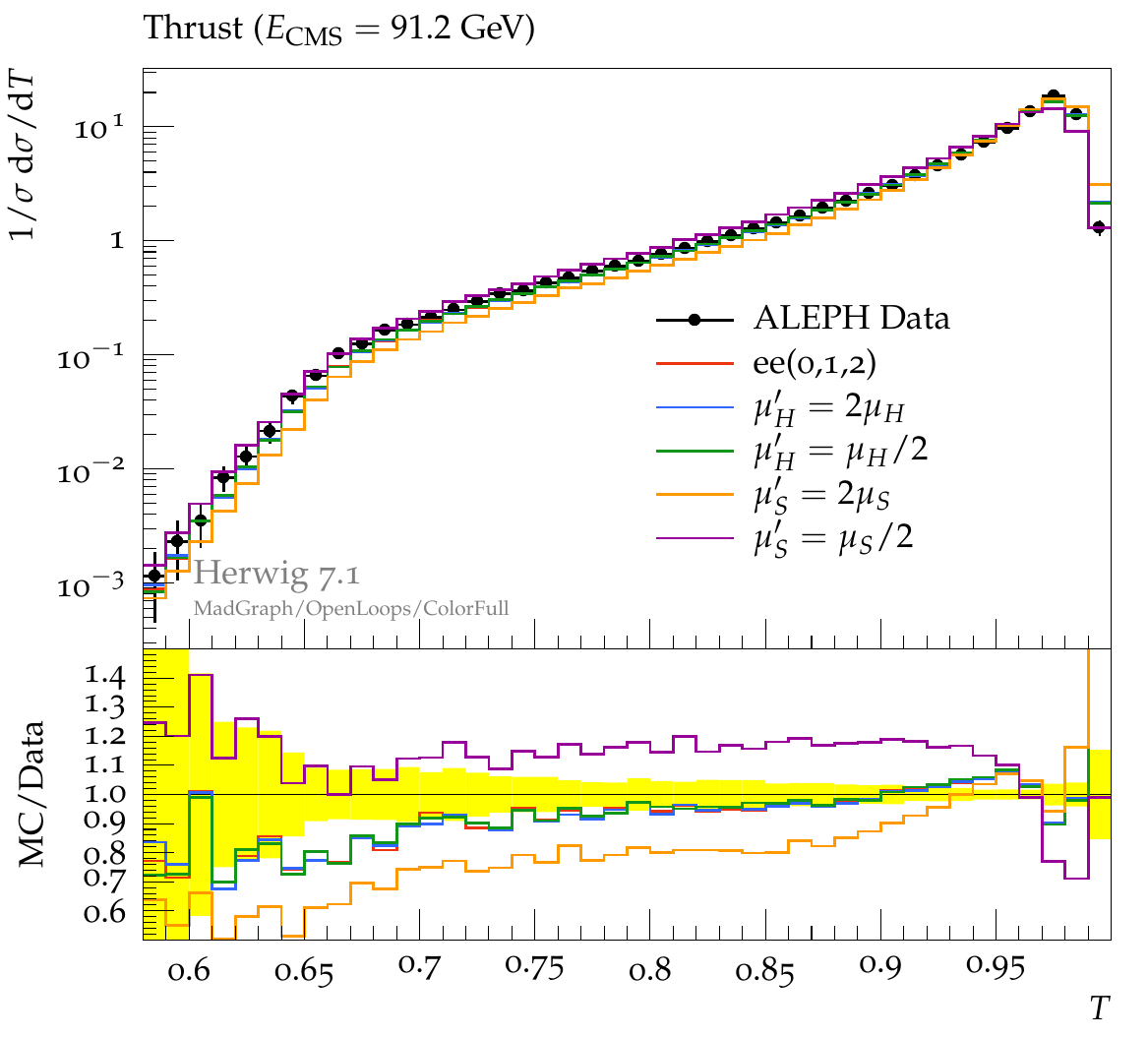}}
\caption{ Scale variation for an $ee(0,1,2)$ merged
  sample, compared with data \cite{Heister:2003aj}. 
  Since the LO is independent of the renormalization scale the
  variation of the hard scale argument does not affect the
  contribution. Variation of the argument of the coupling constant used
  in the showering process however leads to more spherical events by
  lowering the scale and more pencil-like events by enhancing the scale
  argument. }
  \label{fig:LEPscalesA}
\end{figure}

\begin{figure}[t]
  \centering
\scalebox{.75}{\includegraphics{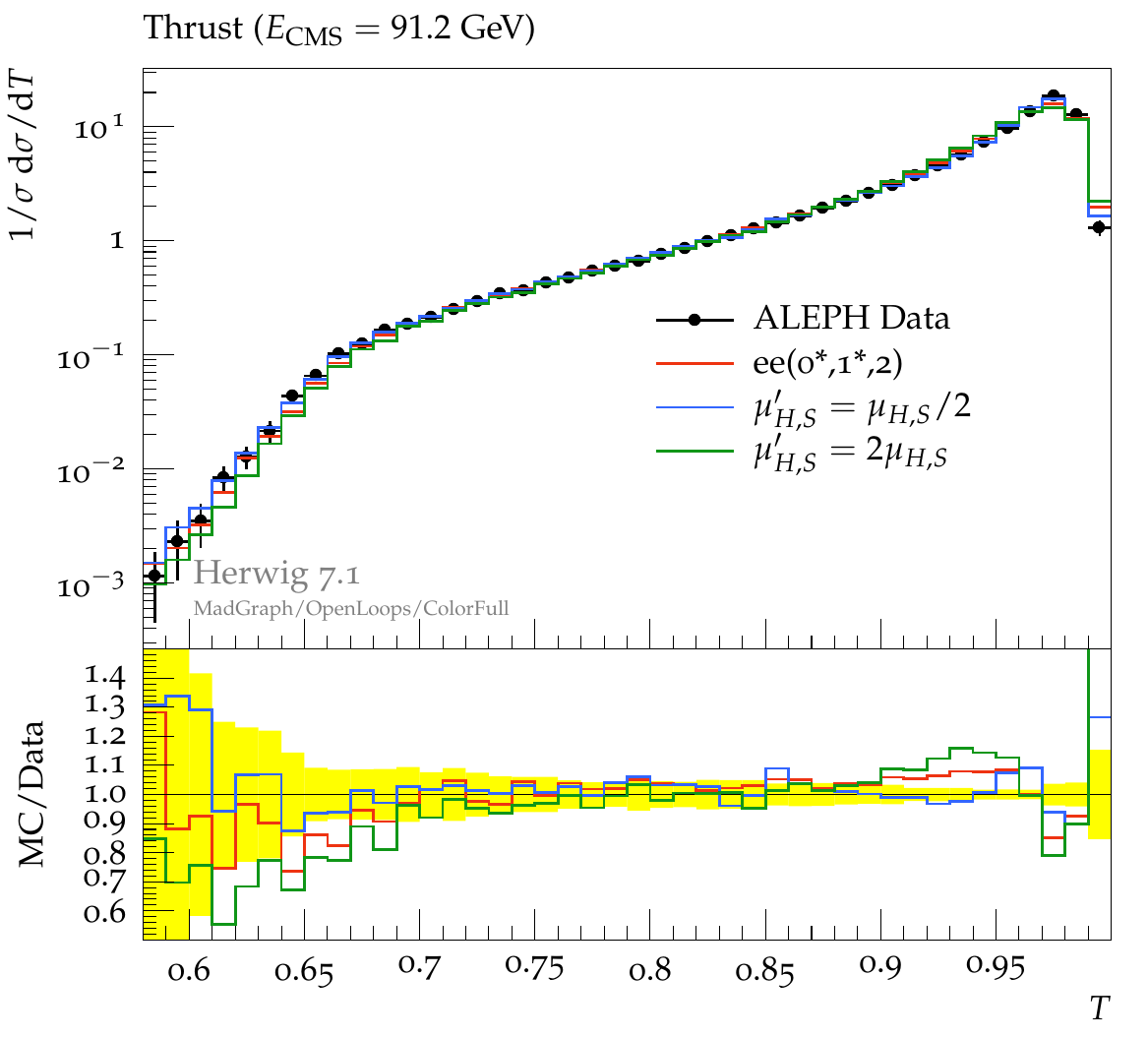}}
\caption{ Including NLO corrections for the production process and one
  additional emission process, $ee(0^*,1^*,2)$, reduces the scale
  variation of the thrust distribution.  Here we compare to data from
  ALEPH \cite{Heister:2003aj}.  The uncertainty bands do not cover the
  variations produced by changing the schemes described in
  Sec.~\ref{sec:expansionSchemes}.  Observables more sensitive to
  multiple emissions are still showing large error bands.  }
  \label{fig:LEPscalesC}
\end{figure}

In Figs. \ref{fig:LEPscalesA} and \ref{fig:LEPscalesC} we show the effect of 
scale variations in LO and NLO merged simulations. While scale variations of the 
scales used in shower emissions produce an uncertainty of roughly 40 \% in the LO samples,
the NLO corrections to the production process and the process with one additional emission
 compensate for the scales in the first shower emission.

\section{Results}
\label{sec:results}

In this section we collect a number of results for simulations with the
merging methods set up previously.  We describe the simulation setup 
before we present the results.  We begin
with jet-production from $e^+e^-$ annihilation at LEP before going
towards $W$ and $Z$ boson production with additional jets at the LHC.
Higgs boson production in association with additional jets has been
chosen as a special case as here the higher order corrections are known
to be particularly sizeable.  We finally present results for dijet
production at the LHC where the Born process already has all legs colour
charged.  We focus on the effect of merging more and more processes to
the Born setup, either with additional legs or with additional virtual
corrections.  We occasionally also vary the merging scale.

\subsection{Simulation setup}
\label{sec:setup}

We briefly summarize the parameters that we use in the
following simulations which are important for our discussion from the
point of view of perturbation theory.  All parameters are used
throughout unless explicitly stated. 

We use the MMHT2014nlo68cl \cite{Harland-Lang:2014zoa} NLO PDF set
interfaced to Herwig via LHAPDF6 \cite{Buckley:2014ana} for LO and NLO
MEs in order to have a common basis for all samples.  We use
an implementation of $\alpha_S$ with two-loop running and fixed
$\alpha_S(M_Z) = 0.118$.  $\alpha_S$ is modified with the CMW scheme,
cf.\ Sec.~\ref{sec:CMW}.  All LO MEs are obtained from
MadGraph\cite{Alwall:2014hca} via a dedicated interface.  In addition,
we use ColorFull \cite{Sjodahl:2014opa} for colour
correlations.  The merging scale for LEP is always
$\rho = 4\,\text{GeV}$, while for LHC we use $\rho = 10\,\text{GeV}$.

In order to demonstrate the performance of our implementation in
different situations, we simulate five different processes: jet
production in $e^+e^-$ annihilation at LEP and four more processes at
the LHC, namely $Z^0$ and $W^\pm$ production with additional jets, Higgs
production and finally dijet production.  The scales for these
production processes are chosen as follows:
\begin{tabbing}
\qquad \= LEP: \qquad \= $\mu_Q$=$\mu_R$=$E_{CM}$\\
\> $Z^0$:  \> $\mu_Q$=$\mu_R$=$\mu_F$=$M_{ll}$\\
\> $W^{\pm}$:  \>$\mu_Q$=$\mu_R$=$\mu_F$=$M_{l\nu}$\\
\> Higgs:  \> $\mu_Q$=$\mu_R$=$\mu_F$=$M_H$\\
\> Dijet: \>$\mu_Q$=$\mu_R$=$\mu_F$=$\max(p^T_j)$
\end{tabbing}
Here, $\max(p^T_j)$ is the maximal transverse momentum of anti-$k_T$
jets with a cone size of $R=0.7$.  We smear the merging scale as
described in Sec.~\ref{sec:smearing} with $\delta=0.1$ to get a 10 \%
variation of the merging scale.  

\subsection{LEP}

\begin{figure}[t]
  \centering
\scalebox{.75}{\includegraphics{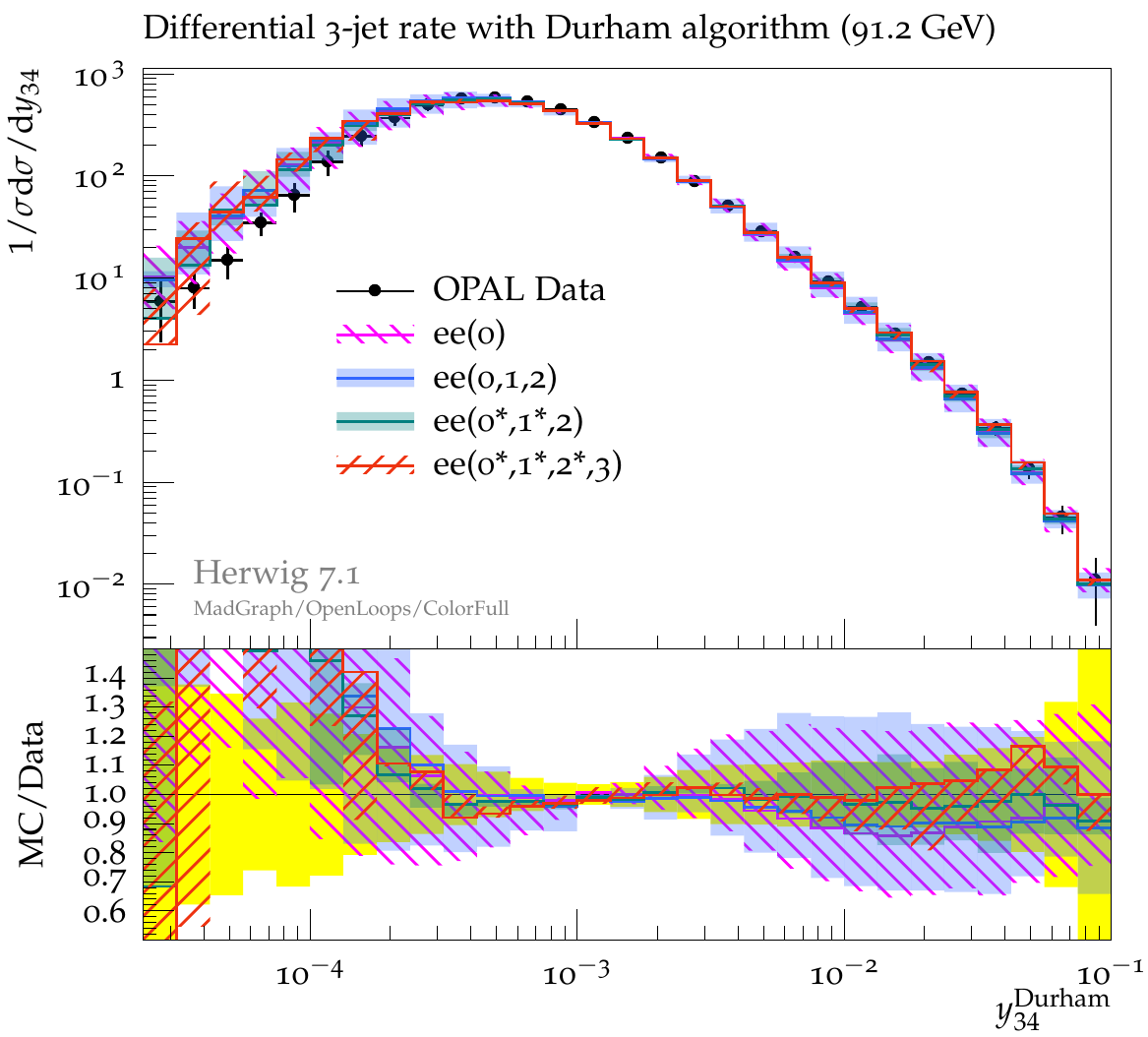}}
\caption{Comparison to OPAL data from
  \cite{Pfeifenschneider:1999rz}. The scale uncertainties of the LO+PS
  $ee(0)$ are not reduced by merging with higher
  multiplicities. Inserting NLO corrections to the first and second
  emission reduces the bands from scale variations significantly. The
  observable measures the transition from a three to a four jet
  configuration and is therefore sensitive to the second emission. The
  NLO corrections to the second emission included in $ee(0^*,1^*,2^*,3)$
  reduce the uncertainty band even more.}
  \label{fig:Pfeifenschneider:1999rz}
\end{figure}
\begin{figure}[t]
  \centering
\scalebox{.75}{\includegraphics{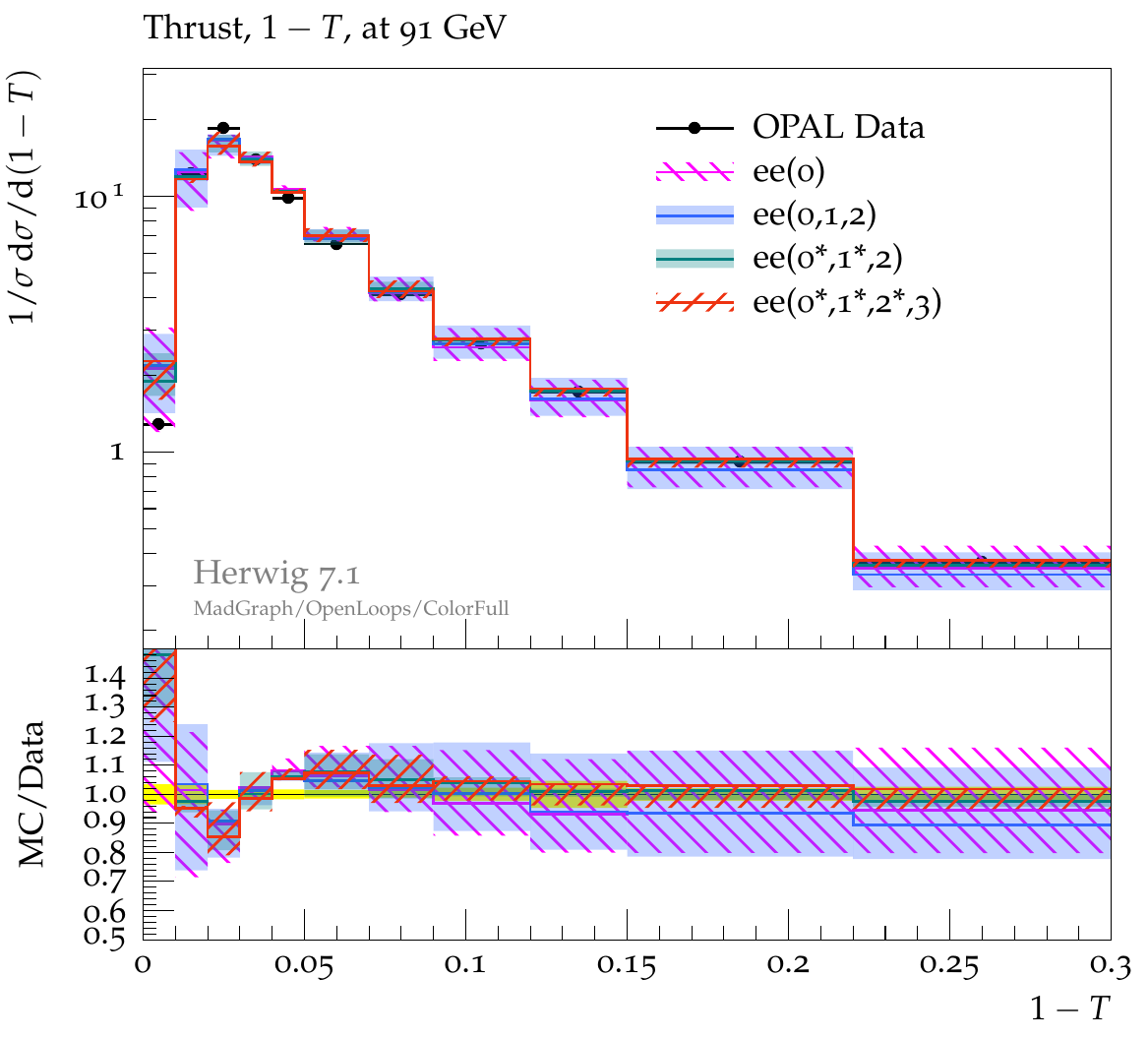}}
\caption{The thrust $1-T$ as measured at LEP \cite{Abbiendi:2004qz} compared
  to various simulations with increasing inclusion of NLO corrections to
  higher multiplicities.  }
  \label{fig:Abbiendi:2004qz}
\end{figure}
\begin{figure}[t]
  \centering
\scalebox{.75}{\includegraphics{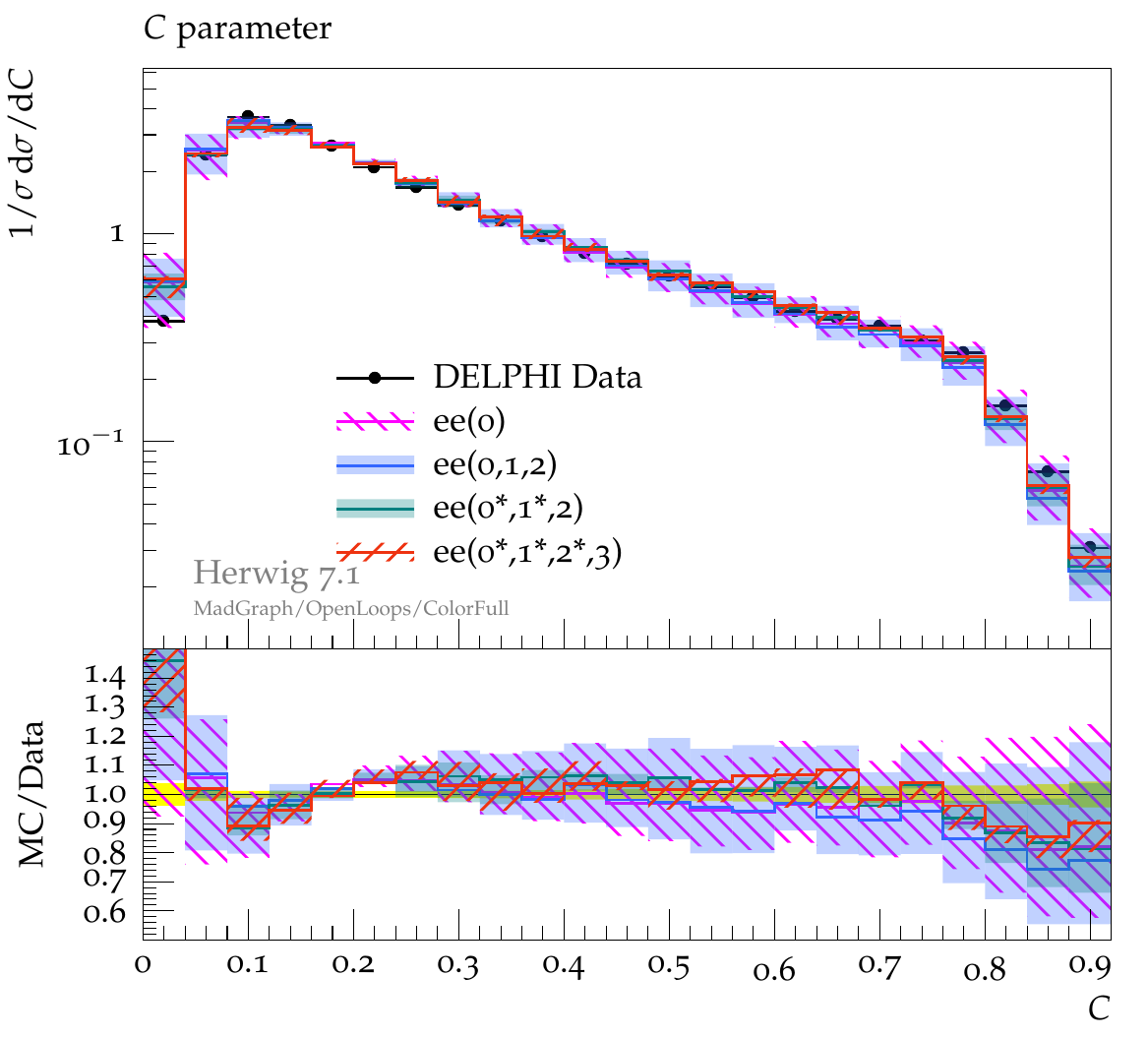}}
\caption{Data \cite{Abreu:1996na} comparison to the $C$-parameter which
  is sensitive to the first emission. NLO corrections to the first and
  second emission reduce the scale uncertainties. }
  \label{fig:Abreu:1996na-C}
\end{figure}
\begin{figure}[t]
  \centering
\scalebox{.75}{\includegraphics{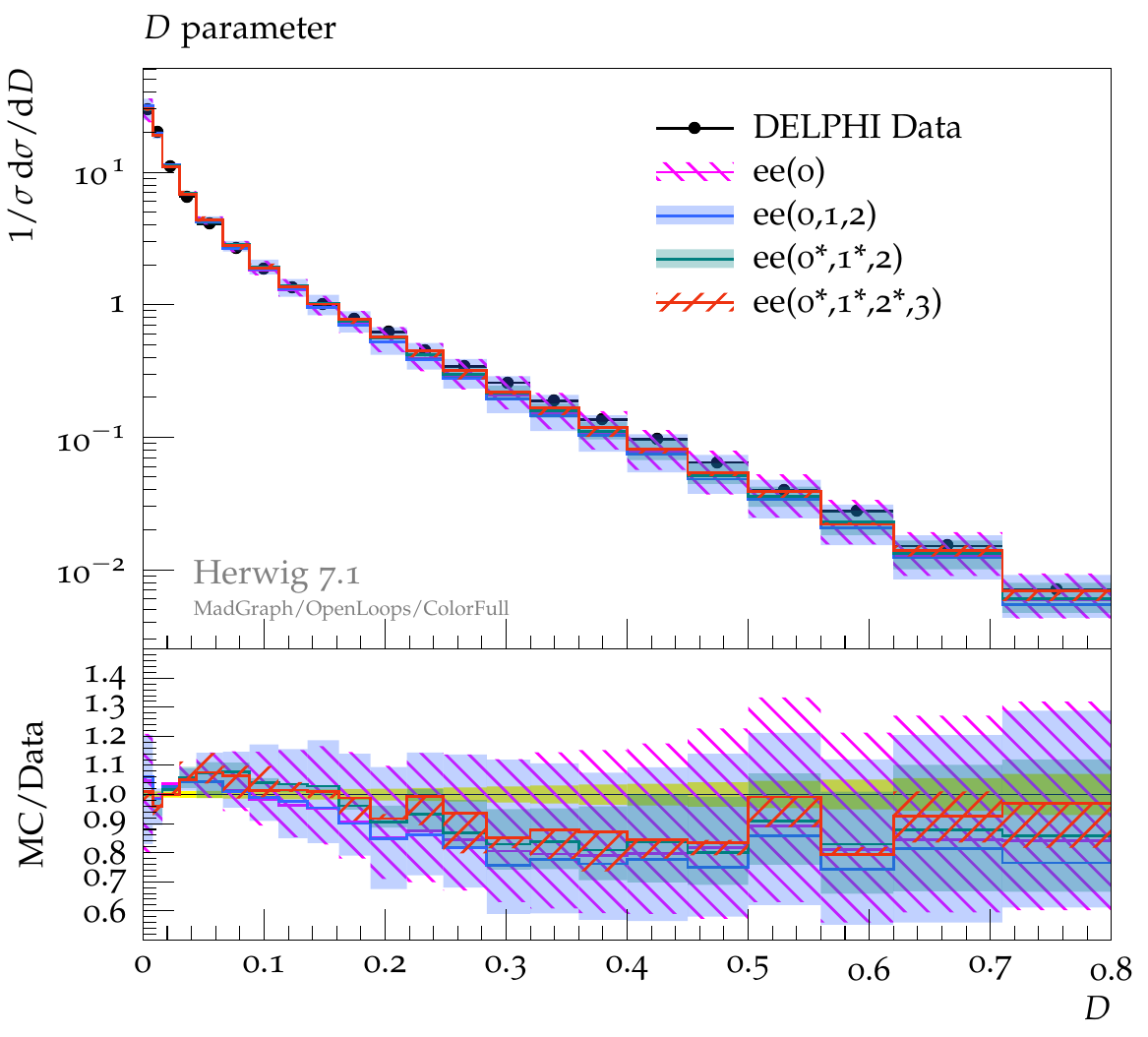}}
\caption{Data \cite{Abreu:1996na} comparison to the $D$-parameter
  observable, which is sensitive to the second emission.}
  \label{fig:Abreu:1996na-D}
\end{figure}

When comparing to data, we first consider data taken from
hadron production processes in $e^+e^-$ annihilation at LEP.  Here, we
find the cleanest environment regarding the development of QCD cascades.
When comparing results from our new simulations to the results at LEP
but also relative to the LO only simulation we have to be aware of some
caveats.  As all other event generators, a large part of the simulation
in \textsc{herwig} has been developed with LEP results as the first benchmark.
Hence, a large part of the modeling, particularly of hadronic final
states, has been adjusted with LEP data as the most important benchmark.
Therefore, when we encounter a worsening of our description at the first
sight we must not necessarily be surprised.  We would expect an
improvement of the description of many observables with our improved
approach, particularly in regions where they should be dominated by
perturbative physics.  If this is not the case it might well be that the
non-perturbative components of the program had previously been adjusted
to compensate for shortcomings in the perturbative description of
observables.  

In this paper we will focus on the relative improvement of results when
more and more perturbative information has been added to the simulation
and leave the discussion regarding non-perturbative parameters to a
re-tuning in conjunction with a new release of the program.  In order to
achieve a comparison of the different components of the program we leave
the hadronization model as it is and stick to an $\alpha_S(M_Z)$ close
to the world average\cite{Olive:2016xmw}, see Sec.~\ref{sec:setup}.

Fig.~\ref{fig:Pfeifenschneider:1999rz} shows the differential three-jet
rate with the Durham jet algorithm as it was measured by the OPAL
experiment at LEP.  This observable measures the hardness of the second
emission from a dijet system.  We show the pure LO result $ee(0)$ as
well as a result with two extra emissions $ee(0, 1, 2)$.  Additional
loop corrections are shown in the results $ee(0^*, 1^*, 2)$ and
$ee(0^*, 1^*, 2^*, 3)$, where the latter is expected to describe even
observables related to the fourth jet in the system at NLO accuracy.  We
vary the renormalization scale used to calculate the ME and the scale of
the shower emissions only synchronized by factors of 2 up and
down. Because the LO merging does not reduce the scale uncertainties the
bands are overlapping at this level. Inclusion of NLO corrections to up
to the second additional jet, so up to the $2 \to 4$ process,
successively improves the scaling behaviour of the simulation and hence
reduces the differential uncertainty band from roughly 40 \% down to a
10 \% level. In this observable the two simulations with NLO merging
give relatively similar results with slight improvements from
$ee(0^*, 1^*, 2^*, 3)$ concerning the scale variations.
 
Using an enhanced ("tuned") $\alpha_S$ and correcting the additional
emissions in the $\bar{MS}$ scheme would lead to an over shooting in the
data description and tuning would then require a reduced $\alpha_S$
value.  The various contributions are produced by using scheme~2
as described in Sec.~\ref{sec:varyingexpansion}.

Further data comparisons are presented in
Figs.~\ref{fig:Abbiendi:2004qz}, \ref{fig:Abreu:1996na-C} and
\ref{fig:Abreu:1996na-D}. Here the trust ($1-T$) and the $C$ and $D$
parameters are shown compared to OPAL and DELPHI data. We find that
while the shape is hardly modified by the higher jet multiplicities, the
scale uncertainties shrink again as it was seen in the three-jet rate.
The $D$ parameter stands out as this is an observable that is sensitive
to the fourth, ``out-of plane'', emission of the system.  We find that
the NLO merged simulation with higher corrections,
$ee(0^*, 1^*, 2^*, 3)$, shows a significantly smaller scale dependence
than the one which only corrects the third jet, $ee(0^*, 1^*, 2)$.  The
overall discrepancy between data and simulation in the $D$ parameter in
the tail requires further investigation and may be related to the tuning
of non-perturbative parameters, as discussed above.  We should note that
the observables are normalized to unity and hence undershooting the data
in the tail, where we expect the strength of our approach, might only be
a result of an overshooting in the bulk region which is presumably more
subject to non-perturbative corrections.

\subsection{$\bf{Z^0}$ boson production at LHC}
\begin{figure}[t]
  \centering
\scalebox{.75}{\includegraphics{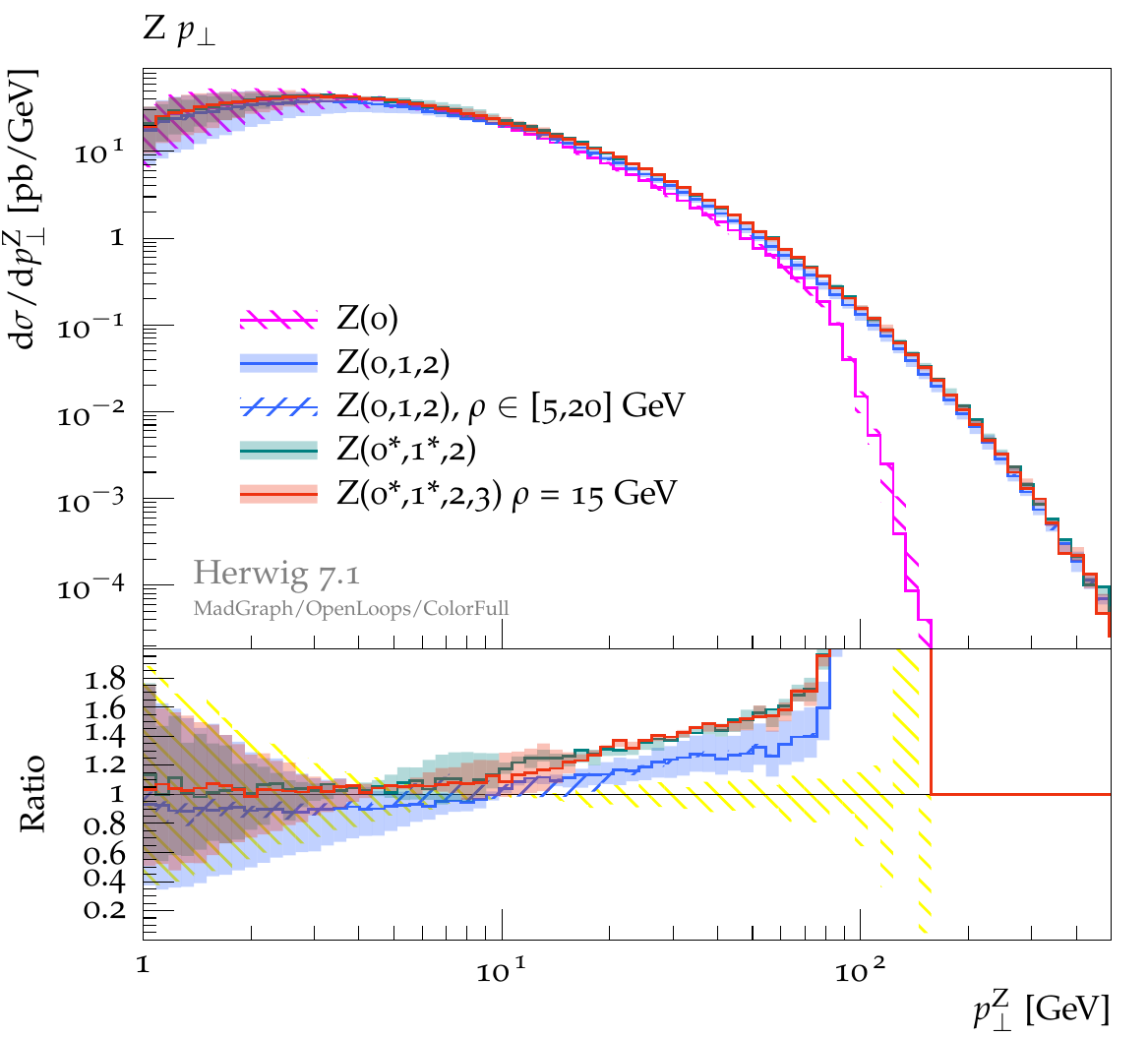}}
\caption{ Transverse momentum of an $e^+e^-$-pair close to the $Z^0$
  boson mass. While the LO distribution dies out at the Z mass due to
  phase space restrictions, the merged distributions fill the full phase
  space. The uncertainty bands are produced by synchronized variation of
  the renormalization and factorization scale in the shower and ME
  calculation. The NLO corrections to the first emission reduce the
  uncertainty estimate in the region above the merging scale.  }
  \label{fig:graphic17}
\end{figure}
\begin{figure}[t]
  \centering
\scalebox{.75}{\includegraphics{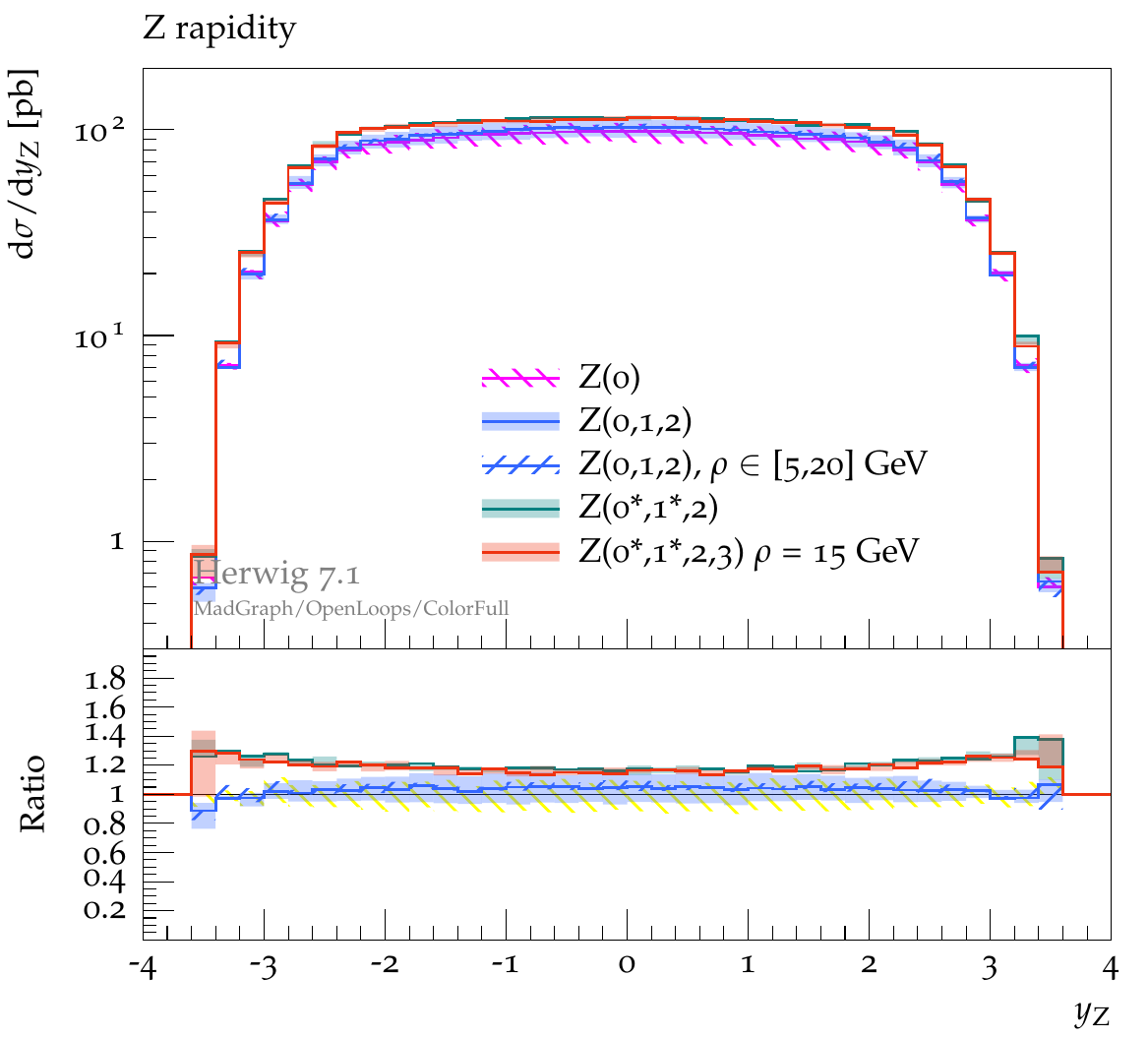}}
\caption{As an example for unitarized cross sections, we show the
  rapidity of the $Z^0$ boson for LO and NLO merged samples. The slight
  increase of the cross section with two additional hard jets is due
  to the non unitarization of unordered histories.  }
  \label{fig:graphic18}
\end{figure}
\begin{figure}[t]
  \centering
\scalebox{.75}{\includegraphics{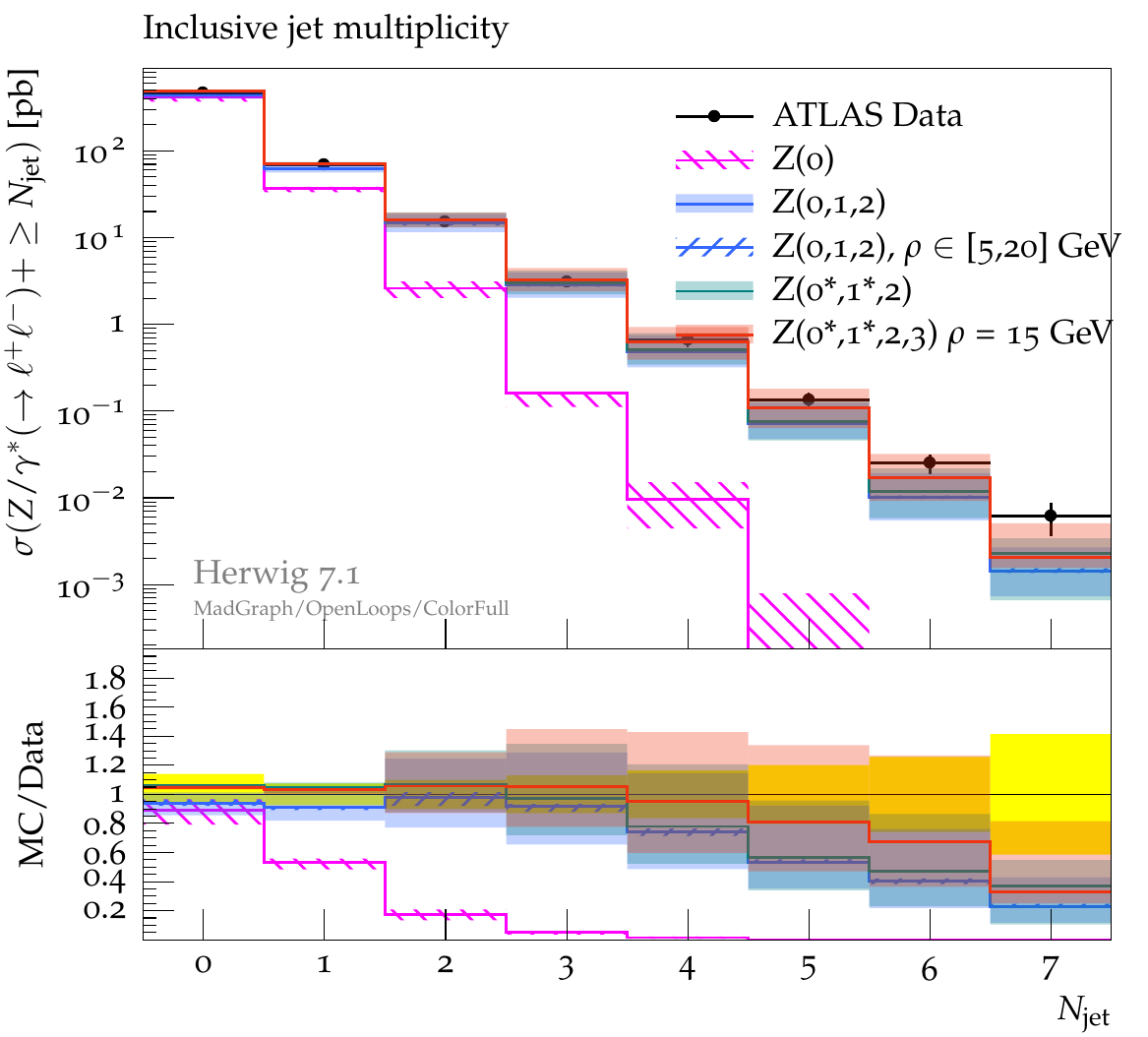}}
\caption{Jet multiplicities as measured by ATLAS \cite{Aad:2013ysa}
  compared to our simulation. The LO+PS $Z(0)$ as well as the not included
  $Z(0,1)$ fail to describe higher jet multiplicities. Inclusion of
  merged samples with at least two additional jets construct
  back-to-back configurations and open the phase space for emissions of
  multiple hard jets. The inclusion of MEs with three
  additional hard jets describe the higher multiplicities more
  appropriate, although the scale uncertainties are rather large here.
}
  \label{fig:Aad:2013ysa}
\end{figure}

In this section we describe the simulation results for $Z^0$ boson
production, i.e.\ we consider final states with a lepton pair in the
mass range $m_{ll}\in [66,116]\,\text{GeV}$, which we call a $Z^0$ boson in
the following.  We show three plots
(Figs.~\ref{fig:graphic17}, \ref{fig:graphic18} and
\ref{fig:Aad:2013ysa}) that each exhibit a specific property of the
unitarized merging.  Note, that in Fig.~\ref{fig:DIPforME} we already
show the transverse momentum of the $Z^0$ boson in a merged simulation
with two LO contributions, see Sec.~\ref{sec:DipforME}, with a
particular focus on the merging scale.  All figures show distributions
for $Z(0)$, $Z(0,1,2)$, $Z(0^*,1^*,2)$ and $Z(0^*,1^*,2,3)$ with scale
variation bands as described above (cf.\
Sec.~\ref{sec:scalevariations}).  For $Z(0,1,2)$ we also show the
variation of the merging scale, $5\,\text{GeV} < \rho < 20\,\text{GeV}$,
as a hashed blue band.

In Fig.~\ref{fig:graphic17} we show the transverse momentum of the
lepton pair.  As the $p^T$ of the $Z^0$ boson in this case is directly
linked to the ME region definition for the first emission the merging
scale variation directly shows the effects of the merging at the
boundaries of the ME region.  The synchronized renormalization and
factorization scale variation is strongly reduced above the merging
scale.  Since the transverse momentum of the $Z^0$ boson is rather
independent of the ME contributions of the third additional jet, the
scale variation of the two NLO merged samples $Z(0^*,1^*,2)$ and
$Z(0^*,1^*,2, 3)$ effectively probes the merging scale dependence.  This
manifests itself in rather large scale variations in the vicinity of the
merging scale.  To illustrate this point, we have chosen the merging
scale for the latter simulation as $\rho = 15\,\text{GeV}$ and find a
shift of the variation band below the merging scale.

The rapidity of the $Z^0$ boson, see Fig.~\ref{fig:graphic18} is an
inclusive observable with respect to parton shower effects. $Z(0,1,2)$
is flat compared to $Z(0)$ and receives contributions from unordered
histories above the hard scale of the shower, which leads to a slight
enhancement of the cross section. The contributions with NLO corrections
are enhanced due to the $K$ factor of the NLO $Z^0$ production process
and are more stable with respect to scale variations.

In Fig.~\ref{fig:Aad:2013ysa} we compare our simulation with the jet
multiplicity as it was measured in \cite{Aad:2013ysa}. Without the
contribution of the $pp\to Zjj$ cross sections the phase space for the
second parton shower emission, which is not corrected for in either
$Z(0,1)$ or in $Z^0$ production in NLO matching, is not capable of
describing higher jet multiplicities.  The recoiling of the two jets is
suppressed for shower emissions, leading to an undershooting in
back-to-back configurations with respect to the measured data.

\subsection{$\bf{W^\pm}$ boson production at LHC}

\begin{figure}[t]
  \centering
  \scalebox{.75}{\includegraphics{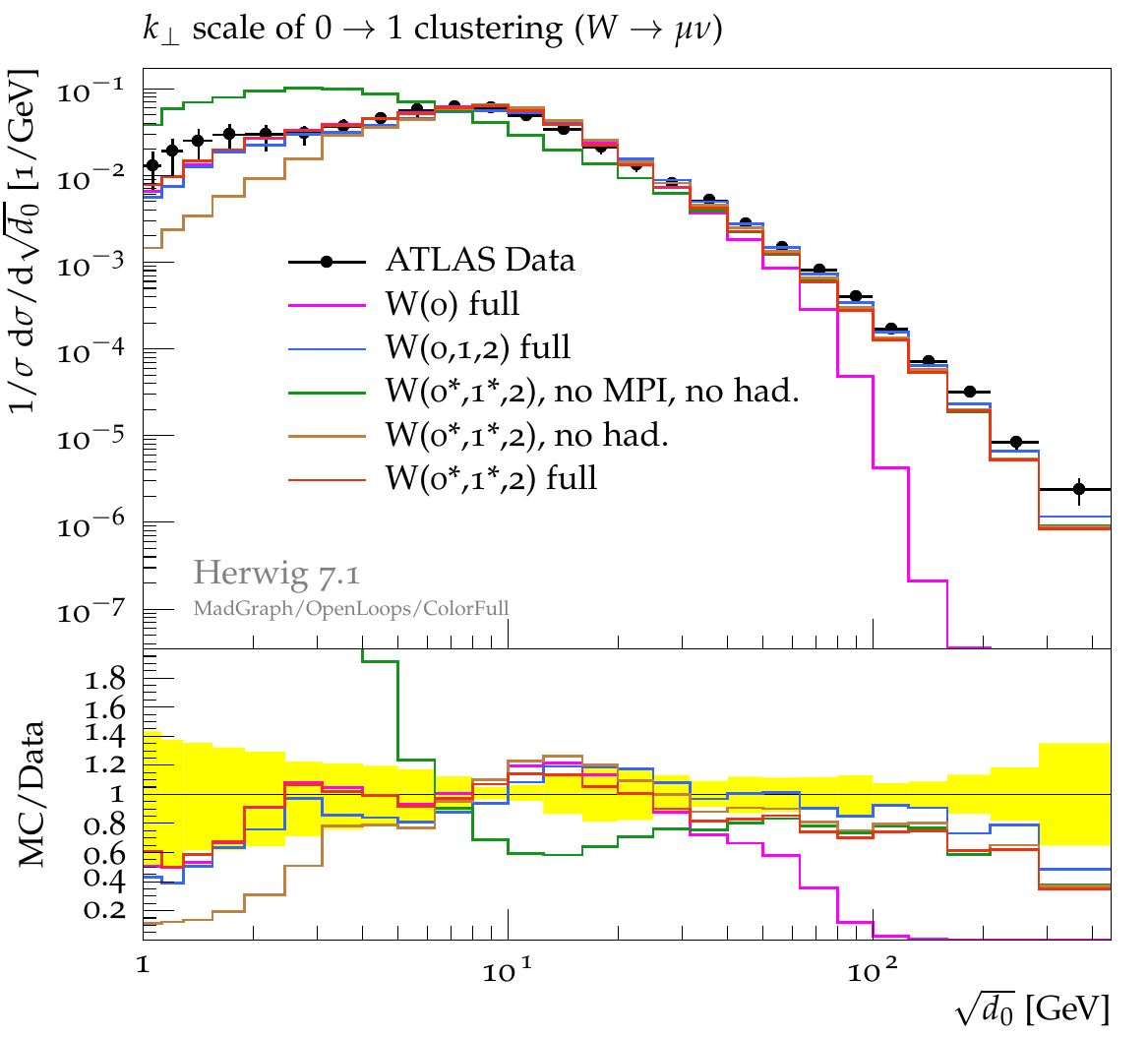}}
  \caption{$d_0$ is a measure for the hardness of the first emission
    which was measured at the ATLAS experiment \cite{Aad:2013ueu}. While
    the MEs are needed to fill the phase space of hard emissions, the
    MPI model and hadronization are needed to model the lower scales of
    the spectrum. }
  \label{fig:Aad:2013ueu}
\end{figure}
\begin{figure}[t]
  \centering
  \scalebox{.75}{\includegraphics{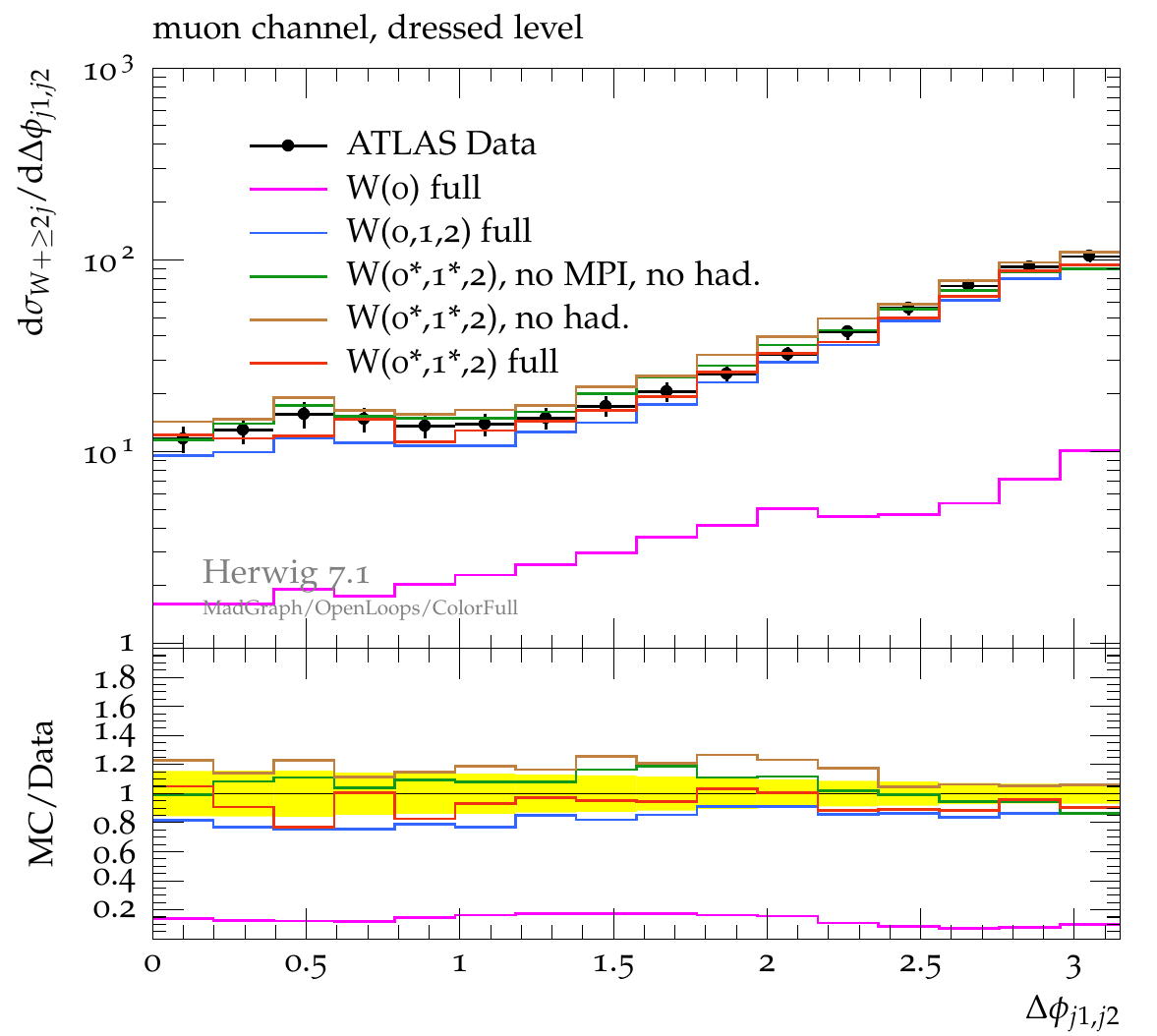}}
  \caption{Difference of the azimuthal angle between the two hardest
    jets in $W$ production as measured by ATLAS~\cite{Aad:2014qxa}. ME
    merging is needed to describe the second additional jet and with the
    NLO corrections the rate of the production is improved compared to
    LO merging.}
  \label{fig:Aad:2014qxa}
\end{figure}

Closely linked to $Z^0$ boson production is the production of a single
$W^\pm$ boson, which we consider in the leptonic channel.  As the
neutrino of the $W$ decay leaves the detector, the transverse momentum is
less well measured. Closely linked to the transverse momentum but in
this study more interesting are the splitting scales of the $k_T$
algorithm.  These are resolution scales $d_i$ of the $k_T$ algorithm at
which the event switches from an $i$ jet event to an $i+1$ jet event in
$W^\pm$ boson production. Fig.~\ref{fig:Aad:2013ueu} shows our simulated
result of the $\sqrt{d_0}$ distribution, compared to data from ATLAS
\cite{Aad:2013ueu}. We show $W(0)$, $W(0,1,2)$ and $W(0^*,1^*,2)$ with
the full event generator setup including MPI and hadronization effects.
In addition, we gradually switch off the non-perturbative effects for
$W(0^*,1^*,2)$.  With neither MPI nor hadronization included, the pure
QCD process tends to overshoot the data at low scales. Inclusion of MPI
 corrects for the medium range scales at approximately
$10\,\text{GeV}$ at LHC energies but undershoots the data for very low
scales.  Only the full simulation describes data from large scales down
to very small splitting scales of a few GeV.  Note that we do not show
the scale variations in this case for the sake of clarity.  The scale
variation is very similar to the scale dependence in the case of $Z^0$
production. 
 
As for the jet multiplicities in $Z^0$ boson production the correction
to the second emission is important in $W^\pm$ production as well in
order to fill the phase space available for two jet events.  A good
example of an observable for which this behaviour is important is the
azimuthal difference $\Delta \phi_{j1,j2}$ between the two hardest
jets. In Fig.~\ref{fig:Aad:2014qxa} our simulation of this is compared
to data measured by the ATLAS collaboration \cite{Aad:2014qxa}.  Once
again, we find the best simulation for $W(0^*,1^*,2)$ with inclusion of
all non-perturbative effects.  
 
\subsection{Higgs boson production in the LHC environment}

\begin{figure}[t]
  \centering
\scalebox{.75}{\includegraphics{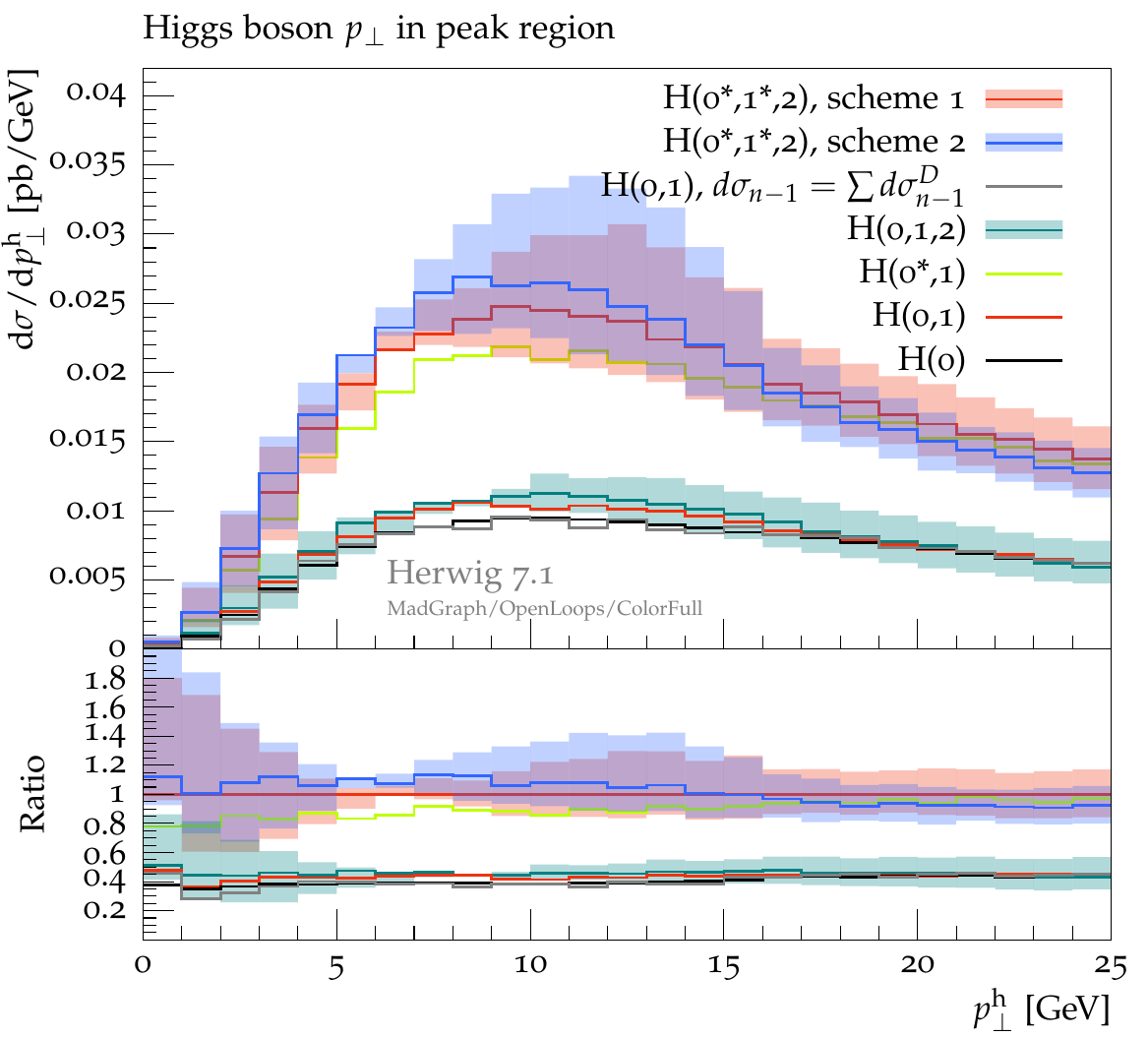}}
\caption{The enormous $K$-factor of the Higgs production process and the
  process with an additional jet in combination with a merging scale that
  is close to the Sudakov peak leads to a kink at the merging
  scale. Switching the scheme from the preferred scheme for LEP and Weak
  boson production to our scheme~1 shows a smoother contribution. Note
  that the choice of scheme is systematically beyond our accuracy. }
  \label{fig:graphic22}
\end{figure}

Higgs production is delicate due to the enormous NLO corrections to the
production process as well as for higher jet multiplicities. The
production is simulated via an effective ggH-vertex and has, due to the
gluon initial state and the color factor, a rather large emission
probability. The Sudakov peak is around $10\,\text{GeV}$. In spite of
the need for resummation at the Sudakov peak we choose the merging scale
to be of the same order as for the other processes at the LHC,
$\rho=15\,\text{GeV}$.  
 In Fig.~\ref{fig:graphic22} the contributions for the transverse
momentum of the Higgs boson are shown for LO 
 and NLO merging with up to two additional legs and loop corrections for
 two different schemes.  
 
 For LO we show four distributions. The pure parton shower in black which is 
 apart from statistical fluctuation identical to the merging with an additional 
 jet multiplicity if the ME are replaced with the dipole content $H(0,1)$ in gray, see Sec.~\ref{sec:DipforME}. 
The inclusion of the correct ME contributions for the first $H(0,1)$ or second emission $H(0,1,2)$, 
red and green respectively slightly change the behaviour in the ME region, and due to
 unitarization also the region below the merging scale.
 
The inclusion of NLO corrections to the production process, $H(0^*,1)$,
enhances the contribution without introducing a kink at the merging
scale $\rho$.  Further NLO corrections to the process with one
additional jet $H(0^*,1^*,2)$ are then scheme dependent, see
Sec.~\ref{sec:expansionSchemes}.  The preferred scheme~2 (in blue) for
dijet production at LEP introduces a enhancement below the merging
scale, which becomes visible only for low merging scales as it was
chosen here.  By using the alternative scheme~1, where the Sudakov
expansion is treated as the expansion of the $\alpha_S$ ratios, a
smoother transition at the Sudakov peak is produced. Note, that the 
choice of scheme is above the claimed accuracy and needs to be treated as
an uncertainty estimation.
Further more we want to point out that the corrections to the production process,
are allowed to emit into the full shower phase space for  $H(0^*,1)$, but are vetoed
for the $H(0^*,1^*,2)$ process. In the case of $H(0^*,1^*,2)$ the 
$\mathcal{O}(\alpha_S)$ contributions to the process with an additional emission are unitarized 
to the $H(0)$ phase space. The rather smooth transition at the merging scale is therefore 
due to a compensation of corrections of the production and the one-additional 
emission process.

\subsection{Dijet production at LHC}

\begin{figure}[t]
  \centering
\scalebox{.75}{\includegraphics{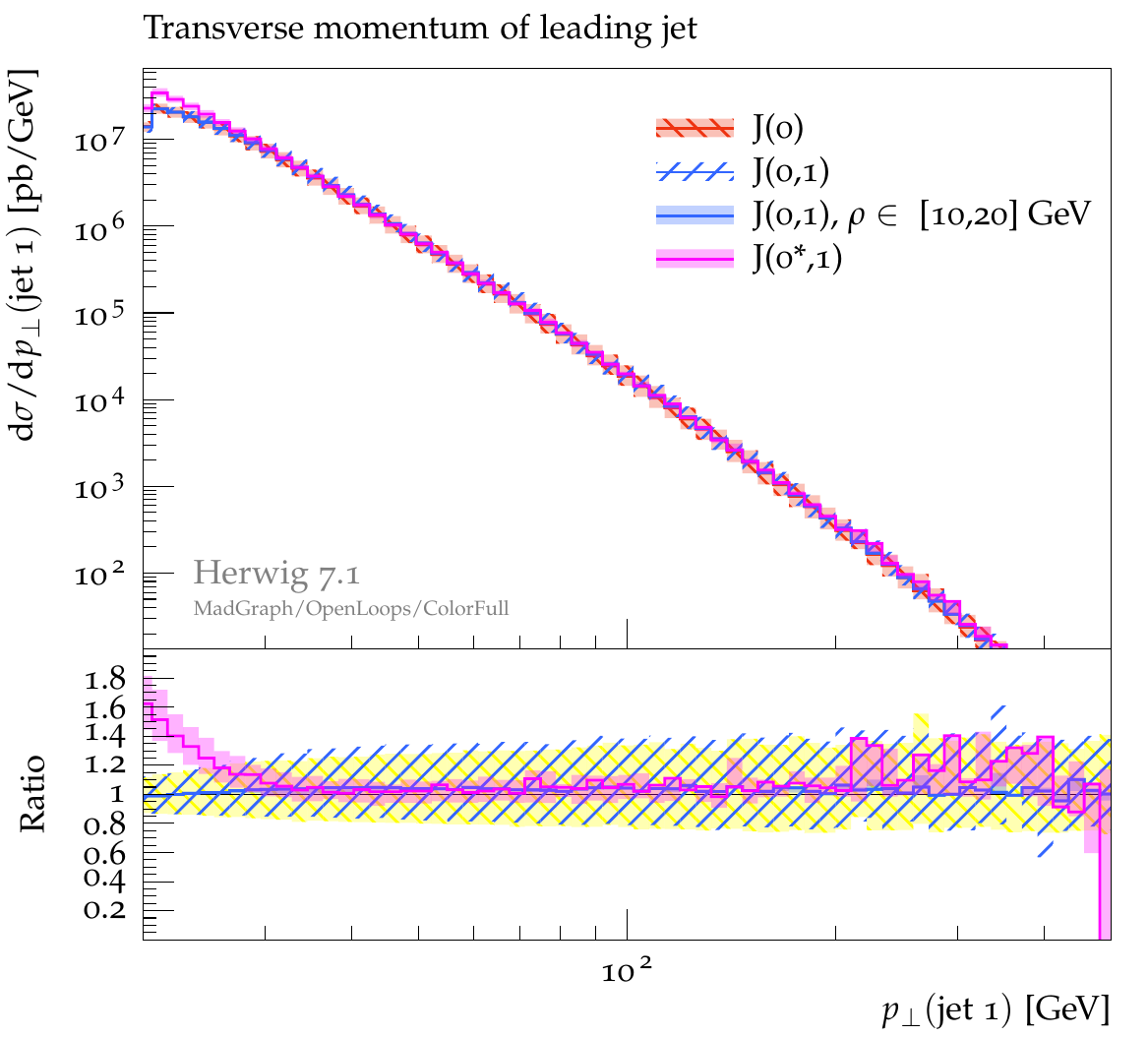}}
\caption{The transverse momentum of the hardest jet should hardly be
  altered by the merging algorithm as the observable is already
  described at LO. The uncertainty estimates from scale variation is not
  altered by the merging process.}
  \label{fig:graphic23}
\end{figure}

 \begin{figure}[t]
  \centering
\scalebox{.75}{\includegraphics{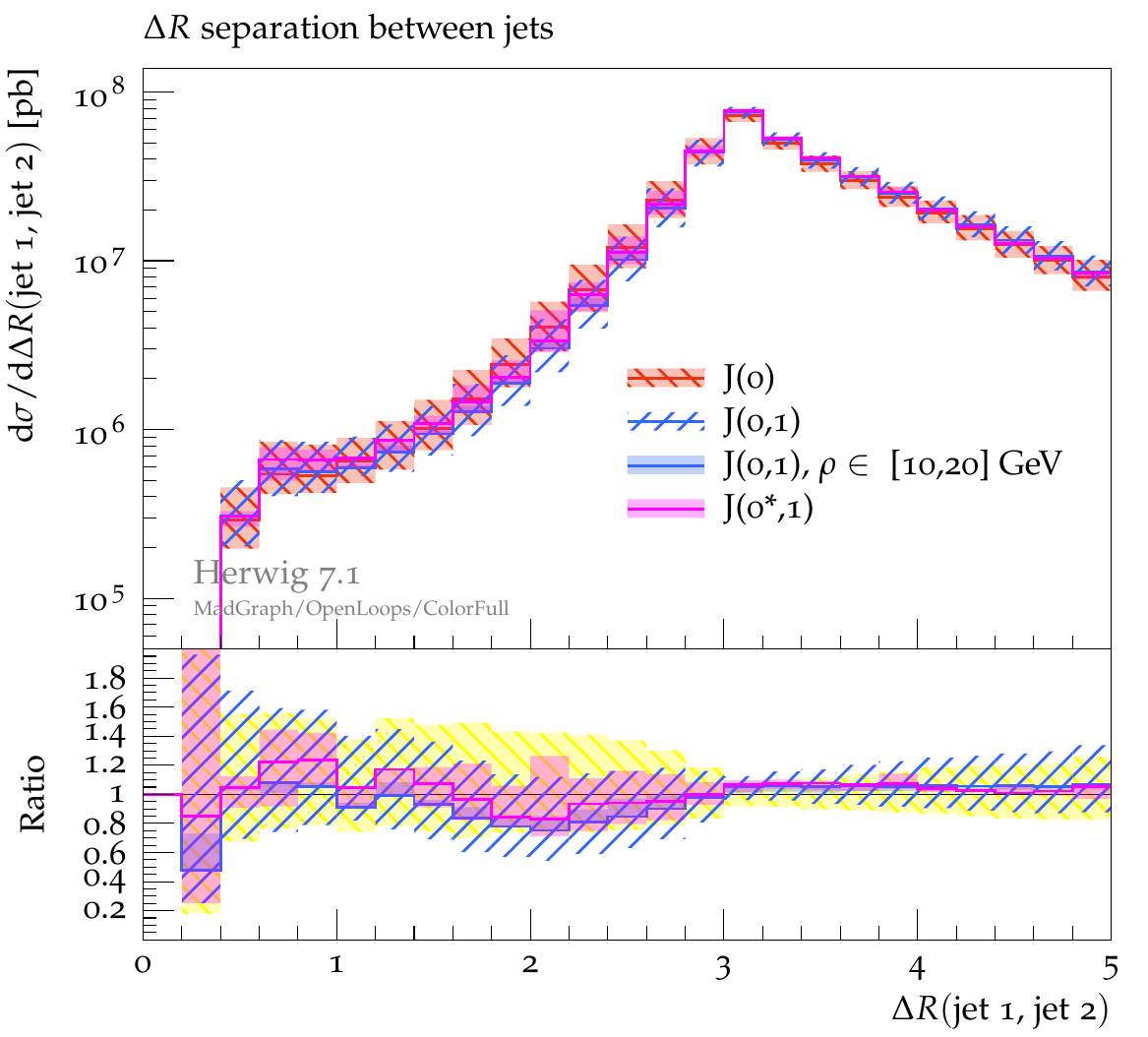}}
\caption{$\Delta R$ separation between the two hardest jets in dijet
  production. The inclusion of NLO contributions for the production
  process reduces the scale uncertainties in the region above
  $\Delta R=\pi$. The region below $\pi$ is filled by parton shower
  effects.  }
  \label{fig:graphic24}
\end{figure}

\begin{figure}[t]
  \centering
\scalebox{.75}{\includegraphics{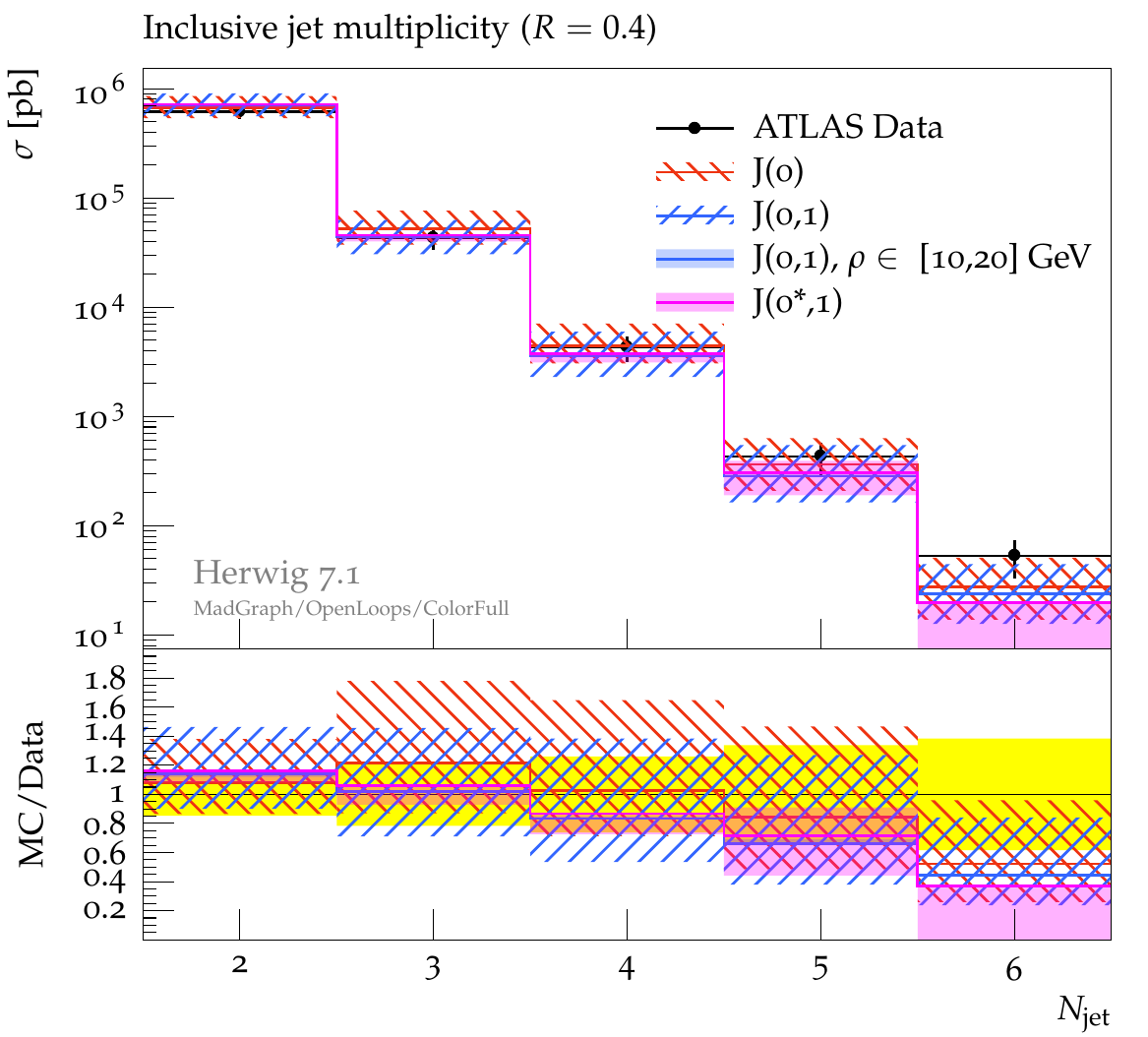}}
\caption{Comparing measured jet multiplicities \cite{Aad:2011tqa} to
  merged samples shows an overshooting of the three jet contribution of
  pure showering that is corrected by the merging. However the measured
  higher jet multiplicities are included in the uncertainty estimation,
  the simulation tends to undershoot this contributions.}
  \label{fig:Aad:2011tqa}
\end{figure}

The last process we consider in this paper is the production of di-jets
at a hadron collider.  In contrast to the production of a single vector
boson the scale of the production process is more ambiguous in this
case.  Scale choices like the invariant dijet mass $m_{jj}$, the scalar
sum of the transverse momenta $H_T$ and the transverse momentum of the
hardest jet are reasonable for the production process. For the results
shown here, we choose the transverse momentum of the hardest jet to be
the shower starting scale as well as the renormalization and
factorization scale. As there is only the dijet system in the first
place at LO this coincides with the $p_T$ of either one of the two
partons. We require a $p_T$ cut on a single inclusive jet of
$20\,\text{GeV}$ for the production process.  Here we only show samples
for $J(0)$, $J(0,1)$ and $J(0^*,1)$ while we leave a detailed study
with higher multiplicities for future work.

Fig.~\ref{fig:graphic23} shows the transverse momentum of the
leading jet for the various approximations. The uncertainty band is
produced by varying the renormalization and factorization scale in the
hard process and the shower synchronized by factors of two. The $J(0)$
contribution is hardly affected by merging with one additional jet,
$J(0,1)$. In both LO distributions, the uncertainty band covers up to
30~\% in both directions. Inclusion of the NLO correction to the
production process $J(0^*,1)$ then decreases the scale variation to a
10~\% level.  Below the jet cut of $20\,\text{GeV}$, the NLO correction
is enhanced.  This enhancement can be explained by the cut on the real
emission process, which requires that only the clustered process needs
to fulfil the cutting criterion.  Above the generation cut at
$20\,\text{GeV}$ for a single inclusive jet the NLO corrected
contribution $J(0^*,1)$ is similar in shape and size to the LO merged
sample $J(0,1)$.

The observable pictured in Fig.~\ref{fig:graphic24} reflects
the $R$ separation of the two leading jets and has two regions of
interest.  The first region is $\Delta R_{jj}\geq\pi$, that is already
present in fixed order dijet production at LO as $\Delta\phi_{jj}=\pi$
here. The region $\Delta R_{jj}<\pi$ can only be filled by additional
emissions, either the fixed order NLO or the parton shower.  The merged
samples hardly alter the region above $\pi$, while the contribution
below is modified by the inclusion of the cross section to the process
with an additional emission.  Note that the scale uncertainty band of
the NLO corrected contribution shrinks in the 'inclusive' region above
$\pi$ and shows larger variations below $\pi$.
 
In Fig.~\ref{fig:Aad:2011tqa} the previously described contributions are
compared to data measured by the ATLAS collaboration
\cite{Aad:2011tqa}. While the pure LO+PS contribution $J(0)$ tends to
overshoot the data for the third jet, the merged samples provide a
better description for this multiplicity. In all three cases the
description of higher jet multiplicities tend to undershoot the measured
data.

\section{Conclusion and Outlook}

In this paper we have presented a new algorithm to combine NLO QCD
calculations for the production of multiple jets together with a parton shower
using a modified unitarized approach. We have implemented this algorithm based
on the Matchbox NLO module and together with the dipole shower evolution as
available in the Herwig event generator to obtain a flexible and
fully-realistic simulation of collider final states.  The implementation
will become available with the 7.1 release of Herwig.  

Improving on just NLO matched simulations, the modified unitarization
procedure allows us to remove contributions which can lead to spurious
terms in inclusive cross sections at parametrically the same level as
NLO QCD corrections. At the same time this allows us to preserve finite
enhancements at higher jet multiplicity. A strict unitarization of these
contributions would otherwise have lead to unphysical predictions. In
order to arrive at this level of simulation it is crucial to identify
which contributions can lead to logarithmic enhancements, and which
momentum configurations are treated as contributing to an additional
hard subprocess.

We have performed a detailed comparison to available collider data. We
have investigated the impact of formally subleading ambiguities in order
to estimate the theoretical accuracy of the advocated procedure. We do
find the expected improvements, namely an overall improved description
of multiple jet emissions and a reduction of the associated scale
uncertainties.  We also find that remaining ambiguities from formally
higher order terms can be large and that our approach allows an estimate
of the size of these effects.  The method is expected to shed light on
dominant contributions at even higher orders, with NNLO QCD corrections
in reach for a combination with the modified unitarized merging
algorithm.

\section*{Acknowledgments}

The authors want to thank the other members of the Herwig collaboration for
continuous discussions and support. We especially thank Mike Seymour and Michael Rauch for a 
careful reading of the manuscript. Further we thank David Grellscheid for his support with 
technical aspects of the code. This work was supported by the European
Union Marie Curie Research Training Network MCnetITN, under contract
PITN-GA-2012-315877. SP acknowledges support by a FP7 Marie Curie Intra
European Fellowship under Grant Agreement PIEF-GA-2013-628739 during part of
this project, and the kind hospitality of the Erwin-Schr\"odinger-Institute at
Vienna while part of this work has been finalized.

\appendix
\section{Example Shower History}
\label{app:Example}
For an initial hard LO event with no emissions we have the weight $X_0$
and describe its state with the test function $u(\phi_0, Q_S)$.  In the
initial state we then have 
\begin{equation}
X_0 u(\phi_0,Q_S)=\alpha^m_S(\mu_R)f_0(\eta_0,\mu_F) \mathcal{W}_0(\phi_0) u(\phi_0,Q_S)\;.
\end{equation}
Here $\mathcal{W}_0(\phi_0)$ are the scale choice independent parts of
the weight assigned to a phase space point. The scale dependent
functions $\alpha_S(\mu_R)$ and the PDFs are extracted from the rest of
the weight.  $u(\phi_0,Q_S)$ is used to define the initial conditions of
the showering. $Q_S$ is the starting scale for parton showering. Usually
$\mu_R$, $\mu_F$ and $Q_S$ are functions of $\phi_0$. The PDF weight is
written in a condensed notation as
\begin{equation}
f_0(\eta_0,\mu_F) =  f^a_0(\eta^a(\phi_0),\mu_F)f^b_0(\eta^b(\phi_0),\mu_F)\;.
\end{equation}
The shower emission is then produced according  to,
\begin{align}
X^{\PS}_{1}&=\alpha_S(q^1)\frac{f_1(\eta_{1}(\phi^{\alpha_1}_{1}(\phi_0)),q^1))}{f_i(\eta_0(\phi_0),q^1)} w_N(Q_S|q^1) \\
&\times  P_{10}(\phi^{\alpha_1}_{1}(\phi_0))X_{0}  \;.
\end{align}
We again extracted only the scale dependent functions from the splitting
kernels.  Note that in the PDF ratio, if the momentum fraction of one (or
both) of the incoming legs is not changed, the ratio is one for this side
(both sides).

The weight $X_1$ of the next higher multiplicity, calculated at the
scales of the multiplicity $n$ is,
\begin{equation}
X_{1}=w^1_H \alpha^{m+1}_S(\mu_R)f_1(\eta_{1},\mu_F) \mathcal{W}_1(\phi^{\alpha_1}_{1}(\phi_0)) \;.
\end{equation}
Here $w^1_H$ is the history weight we need to apply to the $X_1$ weights
to mimic the scale handling of the shower.  Comparing $X_1$ and
$X_1^{\PS}$ with the approximation
\begin{equation}
P_{10}(\phi^{\alpha_1}_{1}(\phi_0))\mathcal{W}_0(\phi_0)\approx\mathcal{W}_1(\phi^{\alpha_1}_{1}(\phi_0))
\end{equation}
gives
\begin{equation}
w_H^1=\frac{\alpha_S(q^1)}{\alpha_S(\mu_R)}\frac{f_1(\eta_{1},q^1))}{f_1(\eta_{1},\mu_F)}\frac{f_0(\eta_{0},\mu_F)}{f_0(\eta_{0},q^1)}w_N(Q_S|q^1,\phi_0)\;.
\end{equation}
In the shower line, we get for $k$ additional emissions,
\begin{align}
X^{\PS}_{k}&=\\\nonumber\Bigg[&\prod_{i=1}^k \alpha_S(q^i)\frac{f_i(\eta_{i},q^i)}{f_{i-1}(\eta_{i-1},q^i)} w_N(q^{i-1}|q^i) 
 P^{\alpha_i}_{i,i-1}(\phi^{\alpha_i}_{i})\Bigg]X_{0}  \;.
\end{align}
Comparing to the direct evaluation
\begin{equation}
X_{k}=w^k_H \alpha^{m+k}_S(\mu_R)f_k(\eta_{1},\mu_F) \mathcal{W}_k(\phi^{\alpha_k}_{k}) 
\end{equation}
accordingly results in the weight 
\begin{equation}
w_H^k =   \frac{f_k(\eta_{k},q^k)}{f_{k}(\eta_{k},\mu_F)} \prod_{i=0}^{k-1}\frac{\alpha_S(q^i)}{\alpha_S(\mu_R)}\frac{f_i(\eta_{i},q^i)}{f_i(\eta_{i},q^{i+1})}  w_N(q^i|q^{i+1},\phi_i)\;.
\end{equation}

\bibliography{merging}

%% end contents %%
\end{document}